\documentclass[11pt]{article}

\usepackage{amsmath}
\usepackage{authblk}
\usepackage{enumitem}  
\usepackage[usenames, dvipsnames]{color} 
\usepackage{graphicx}     
\usepackage{subcaption}
\usepackage[utf8]{inputenc}
\usepackage{url}
\usepackage{cite}
\usepackage{amssymb}
\usepackage{mathrsfs}
\usepackage{comment}
\usepackage{amsthm}
\usepackage{float}
\usepackage{makecell}
\usepackage{multirow}

\usepackage[
  top=3cm,
  bottom=3cm,
  left=3cm,
  right=3cm
]{geometry}

\usepackage{physics}

\usepackage{overpic}

 \setlist[description]{font=\normalfont\space}

\def\beq{\begin{equation}}
\def\eeq{\end{equation}}

\newcommand{\ben}{\begin{enumerate}}
\newcommand{\een}{\end{enumerate}}
\newcommand{\be}{\begin{equation}}
\newcommand{\ee}{\end{equation}}

\usepackage{tikz}
\usetikzlibrary{positioning}
\usetikzlibrary{intersections}
\usetikzlibrary{fadings} 
\usetikzlibrary{arrows.meta} 
\usetikzlibrary{arrows}

\tikzfading[name=fade out,
inner color=transparent!0,
outer color=transparent!100]

\definecolor{cherryblossompink}{rgb}{1.0, 0.72, 0.77}
\definecolor{lightblue}{rgb}{0.68, 0.85, 0.9}

\usetikzlibrary{decorations.pathmorphing}
\usetikzlibrary{decorations.pathreplacing,decorations.markings}

\usetikzlibrary{backgrounds,automata}

\usepackage{hyperref}

\hypersetup{
    bookmarks=true,         % show bookmarks bar?
    unicode=false,          % non-Latin characters 
    pdftoolbar=true,        % show Acrobat
    pdfmenubar=true,        % show Acrobat 
    pdffitwindow=false,     % window fit to page when opened  
    pdfstartview={FitH},    % fits the width of the page to the window
    pdfauthor={Author},     % author
    pdfsubject={Subject},   % subject of the document  
    pdfcreator={Creator},   % creator of the document
    pdfproducer={Producer}, % producer of the document
    pdfkeywords={keyword1} {key2} {key3}, % list off keywords
    pdfnewwindow=true,      % links in new window
    linktocpage=true,
    linkcolor=blue,           % color of internal links (change box color with linkbordercolor)
     linkbordercolor=blue,
    citecolor=blue,        % color of links to bibliography
    filecolor=blue,      % color of file links
    urlcolor=blue         % color of external links
} 

\begin{document}
 
\numberwithin{equation}{section}
 
 \title{ \bf\LARGE Black hole equations of state and response functions \\[9mm]
}

\author{\large Silvester G.A. Borsboom\thanks{silvester.borsboom@ru.nl} }
\author{\large Manus  R.  Visser\thanks{manus.visser@ru.nl}} 

\vspace{10mm}

\affil{\textit{Institute for Mathematics, Astrophysics and Particle Physics,}\\
\textit{and Radboud Center for Natural Philosophy,}\\
\textit{Radboud University, 6525 AJ Nijmegen, The Netherlands}}

 \date{\today}
 
\maketitle

\begin{abstract}
 We systematically develop a framework for equilibrium thermodynamics in terms of thermodynamic representations, which are choices of thermodynamic potential and independent variables, one from each conjugate pair. In each representation, equations of state arise from first derivatives of the potential and response functions from its second derivatives. We apply this framework to ideal gases and to Schwarzschild black holes in a spherical cavity in various representations, interpreting the cavity area and its conjugate surface pressure as the thermodynamic volume and pressure of a holographically dual system. For this quasi-local Schwarzschild black hole, stability depends on the thermodynamic representation, or equivalently on which variables are held fixed under equilibrium perturbations. The large black hole branch is thermally stable at fixed volume but mechanically unstable under isothermal compression, while the system is mechanically stable under adiabatic compression everywhere. At fixed pressure, the black hole is thermally unstable throughout the physical state space. We also find that the thermal expansion coefficient is negative everywhere and show that isenthalpic expansion always cools the black hole. More broadly, the framework provides a systematic route to deriving equations of state and response functions for a wide class of black hole systems using quasi-local gravitational thermodynamics.

\end{abstract}

\thispagestyle{empty}

\newpage

 \tableofcontents

\section{Introduction}
Equations of state play a central role in thermodynamics. They relate the macroscopic variables that characterize equilibrium states and provide the starting point for describing how a system responds to external perturbations. Heat capacities, compressibilities, thermal expansion coefficients, and other response functions are obtained by differentiating the equations of state with respect to the appropriate thermodynamic control variables. Response functions are typically the quantities most directly accessible in experiments. They characterize how a system reacts to thermal and mechanical perturbations and provide key information about stability.

In   introductory treatments of thermodynamics (see e.g.~\cite{Landau:1980mil,schroeder2000introduction,blundell2010concepts}), the phrase ``equation of state'' is commonly  associated with a relation among pressure, temperature, volume, and particle number. The paradigmatic example is the ideal gas law,
$  
PV = NT.
$  
However, as emphasized by Callen~\cite{callen1985thermodynamics} in his textbook, this relation is only one equation of state among several,  even for an ideal gas. A more systematic formulation begins by specifying a thermodynamic representation (the thermodynamic analog of an ensemble in statistical mechanics), which corresponds to a choice of thermodynamic potential together with a set of independent thermodynamic variables. Once this choice has been made, the remaining conjugate variables in the thermodynamic state space become functions of the independent variables. These representation dependent functional relations are what we will call the equations of state.

For example, in the Helmholtz representation the thermodynamic potential is the Helmholtz free energy $F(T,V,N)$, whose independent variables are the temperature $T$, volume $V$, and particle number $N$. The conjugate variables (entropy $S$, pressure $P$, and chemical potential $\mu$, respectively) are obtained by taking partial derivatives of $F$ and hence become functions of the independent variables,
\begin{equation}
S=S(T,V,N)\,, \qquad
P=P(T,V,N)\,, \qquad
\mu=\mu(T,V,N)\,.
\end{equation}
These three relations are the equations of state in the Helmholtz representation. For an ideal gas, $P=NT/V$ is only one of them (the others are shown in Eq.~\eqref{eq:idealgas}). If instead temperature, pressure, and particle number are treated as independent variables, as in the Gibbs representation, then the expression for the volume itself becomes an equation of state, i.e. 
$
V=V(T,P,N).
 $

% Equations of state thus depend on the choice of representation. This dependence is not
% merely a matter of notation: a representation determines which variables are treated as
% thermodynamic control parameters, which conjugate variables are expressed as equations of
% state, and which response functions and stability criteria are relevant. \textcolor{black}{Accordingly, when
% we refer to representation-dependent stability, we mean that the stability criteria depend
% on the thermodynamic constraints, or equivalently on the class of allowed equilibrium
% perturbations, specified by the independent variables of the representation.} For example,
% the Helmholtz and Gibbs representations correspond to different choices of control
% variables and naturally single out the constant-volume and constant-pressure heat capacities,
% respectively. Consequently, different representations generally lead to different equations
% of state and different response theories for the same thermodynamic system. In this article
% we apply this framework to black hole thermodynamics (see \cite{Wald:1999vt} for a review).

 % }

Equations of state thus depend on the choice of representation. This dependence is not
merely a matter of notation: a representation determines which variables are treated as
thermodynamic control parameters, which conjugate variables are expressed as equations of
state, and which response functions and stability criteria are relevant. \textcolor{black}{
The choice of representation therefore specifies the thermodynamic constraints, and hence
the class of equilibrium perturbations that are allowed. We use the term
representation-dependent stability precisely in this sense: the relevant stability criteria
depend on which variables are controlled and which are allowed to vary. For instance, the
Helmholtz representation keeps the volume fixed, whereas the Gibbs representation keeps
the pressure fixed and allows the volume to respond. The corresponding thermal and
mechanical stability criteria are therefore different. } Consequently, different representations
generally lead to different equations of state and different response theories for the same
thermodynamic system. In this article \textcolor{black}{we systematically develop  this framework and} apply it to black hole thermodynamics
(see \cite{Wald:1999vt} for a review).

Stationary black holes are thermodynamic equilibrium systems and should therefore admit a description in terms of thermodynamic potentials, equations of state, and response functions. The first issue is to identify the relevant thermodynamic variables. In particular, the definitions of pressure and volume are not automatic in black hole thermodynamics \cite{Dolan:2012jh}. For a Schwarzschild black hole, the standard thermodynamic variational relation contains no pressure-volume term,
\begin{equation} \label{bhfirstlaw1}
dM = T_{\rm H} dS\,.
\end{equation}
Here, $M$ is the mass, \textcolor{black}{$T_{\rm H}= 1/ (8 \pi G M)$ the Hawking temperature \cite{hawking1975}, and $S= 4 \pi G M^2$ the Bekenstein  entropy \cite{Bekenstein:1973ur} of a  Schwarzschild black hole in four spacetime dimensions. Since these quantities are all determined by the mass}, there is only one independent thermodynamic variable. Hence, the equilibrium state space is one-dimensional. As a consequence, many notions familiar from ordinary thermodynamics cease to exist: there is no compressibility, no thermal expansion coefficient, and no concept of mechanical stability. Furthermore, without a well-defined thermodynamic volume there is no physical notion of extensivity, nor any meaningful large-system (thermodynamic) limit in which the size of the system is taken to infinity.

Several definitions of thermodynamic pressure and volume have been proposed in the black hole literature. For example, Padmanabhan \cite{Padmanabhan:2002sha} identified the pressure of a static black hole with the radial component of the stress-energy tensor and the volume with the Euclidean volume enclosed by the horizon. However, the resulting thermodynamic volume is not independent of the entropy, leading to a degenerate thermodynamic state space \cite{Hansen:2016wdg,Hansen:2016gud}.

Another proposal, known as extended black hole thermodynamics or black hole chemistry \cite{Kastor:2009wy,Dolan:2010ha,Cvetic:2010jb,Dolan:2011xt,Dolan:2012jh,Kubiznak:2014zwa} (see \cite{Kubiznak:2016qmn,Mann:2025xrb} for reviews), interprets the cosmological constant as a thermodynamic pressure,
\begin{equation}
P_\Lambda = -\frac{\Lambda}{8\pi G}\,,
\end{equation}
with conjugate thermodynamic volume $V_\Lambda$. The corresponding variational relation
\begin{equation}
dM = T_{\rm H} dS + V_\Lambda dP_\Lambda\,,
\end{equation}
is formally identical to the thermodynamic relation $dH = T dS + VdP$,  provided the black hole mass is interpreted as enthalpy $H$ rather than internal energy. Nevertheless, the pressure appearing in black hole chemistry differs conceptually from the mechanical pressure of ordinary thermodynamics. Rather than describing the response of a system to changes in its volume, it is associated with variations of the cosmological constant, which is a coupling constant of the gravitational theory. In this sense, varying  $P_\Lambda$ changes the theory itself rather than the thermodynamic state of a fixed system \cite{Mancilla:2024spp}. \textcolor{black}{
Moreover, for static black holes a degeneracy remains in the
entropy-volume sector. Although the enthalpy $H(S,P_\Lambda)$ is a
function of two independent variables, the thermodynamic volume
$V_\Lambda$ is completely determined by the entropy $S$. Hence $S$ and
$V_\Lambda$ cannot be varied independently, and $E(S,V_\Lambda)$ is not a genuine two-variable fundamental relation. 
} Another peculiar feature is that the black hole equation of state is typically written in terms of a specific volume proportional to the horizon radius \cite{Kubiznak:2012wp}, rather than the thermodynamic volume $V_\Lambda$
  appearing in the variational relation above. Finally, for asymptotically flat Schwarzschild black holes the cosmological pressure vanishes identically. Black hole chemistry therefore does not provide an adequate definition of pressure and volume   for asymptotically flat Schwarzschild black holes.

Recently, we proposed a different notion of pressure and volume based on quasi-local gravitational thermodynamics \cite{york1986,Whiting:1988qr,Martinez:1989hn,Brown:1989fa,Braden:1990hw,Brown:1990fk,BrownYork1993,Brown:1992bq} and its holographic interpretation \cite{Borsboom:2026ash} (see also~\cite{Banihashemi:2024yye}). York \cite{york1986} showed that for a Schwarzschild black hole enclosed by a spherical cavity of finite radius, the variational relation takes the form 
\begin{equation} \label{yorksrelationfirstlaw}
dE = T dS - s dA\,,
\end{equation}
where $E$ is the Brown-York quasi-local energy, $T$ is the Tolman temperature at the cavity boundary, $S$ is the Bekenstein entropy, $A$ is the area of the cavity boundary, and $s$ is the corresponding surface pressure. We  assume a holographic duality between the gravitational theory in the bulk spacetime and a quantum system living on the timelike boundary, whose constant time cross-sections coincide with the cavity walls.  \textcolor{black}{
We expect this boundary theory to arise as a flat space limit of the
finite-cutoff holographic theories associated with $T\bar T$-type
deformations \cite{McGough:2016lol}, or their higher-dimensional $T^2$
generalizations \cite{Taylor:2018xcy,Hartman:2018tkw}.\footnote{
\textcolor{black}{A complementary approach was developed by Banihashemi, Shaghoulian and
Shashi \cite{Banihashemi:2024yye}, who study flat-space Einstein gravity
  at finite cutoff with conformal boundary conditions and compare
its thermodynamics with a putative boundary thermal effective field theory.
Their construction is distinct from a flat space limit of a
$T\bar T$-type deformation, but addresses the same   question of a
finite-cutoff boundary dual for flat space gravity.}
}
} Interpreting the boundary area as a thermodynamic volume and the surface pressure as its conjugate pressure in the dual quantum system,
\begin{equation}
V \equiv A\,,
\qquad
P \equiv s\,,
\end{equation}
transforms York's variational relation \eqref{yorksrelationfirstlaw} into the standard thermodynamic form
\begin{equation}
dE = T dS - P dV\,.
\end{equation}
\textcolor{black}{
The cavity wall is part of the thermodynamic setup rather than an
intrinsic property of the isolated black hole. Accordingly, $P$ is a
quasi-local boundary pressure rather than a bulk pressure of the
Schwarzschild geometry. In the holographic interpretation, however, it
plays the role of the mechanical pressure conjugate to the spatial volume
$V$ of the boundary system. A quasi-static compression or expansion
corresponds to moving the cavity wall inward or outward, thereby changing
its area and producing the work term $-PdV$. Moreover, we emphasize 
that the quasi-local pressure-volume sector is physically distinct from the pressure-volume sector of black hole chemistry, and there is no   map between
$(P_\Lambda,V_\Lambda)$ and $(P,V)$.
}

The crucial point is that the entropy and volume are now independent variables. The Bekenstein entropy \textcolor{black}{$S= A_h/ 4G$} is determined by the horizon area \textcolor{black}{ $A_h=4 \pi r_h^2 = 16 \pi G^2 M^2$,  where $r_h =2 GM$ is the Schwarzschild radius in four spacetime dimensions,} whereas the volume is associated with the cavity boundary area. Consequently, the equilibrium state space becomes genuinely two-dimensional and admits a non-trivial pressure-volume sector. In the limit where the cavity is taken to spatial infinity, the quasi-local energy $E$ approaches the   mass $M$ while the pressure vanishes, see \eqref{eq:yorkE} and \eqref{eq:yorkP} below.   York's relation \eqref{yorksrelationfirstlaw} therefore reduces in that case to the standard Schwarzschild variational relation   \eqref{bhfirstlaw1}.

Once the equilibrium state space becomes two-dimensional, the thermodynamic description acquires a larger set of distinct representations. For a Schwarzschild black hole without a cavity the state space is one-dimensional, so there is no meaningful distinction between, for example, Helmholtz and Gibbs representations, because there is no independent pressure-volume sector. With a cavity, however, the presence of an independent pressure-volume sector distinguishes these representations.  Different thermodynamic representations then correspond to different
classes of equilibrium perturbations and organize different response
functions.

York's original analysis \cite{york1986} captures the canonical (Helmholtz) representation at fixed volume. In this representation there are two equilibrium branches: a small black hole branch with negative fixed-volume heat capacity and a large black hole branch with positive   heat capacity. Brown et al. \cite{Brown:1989fa} subsequently formulated a broader class of thermodynamic ensembles for gravitating systems using Euclidean actions with different boundary conditions, including the canonical pressure ensemble (in which temperature and pressure are held fixed, and which we identify with the Gibbs representation) and the microcanonical pressure ensemble (in which energy and pressure are held fixed, and which we do not consider further here). Comer \cite{Comer:1992pc} later analyzed the stability properties of thermal black holes in these ensembles and showed, in particular, that the canonical pressure ensemble is nowhere thermodynamically stable. This conclusion was obtained from the behavior of the corresponding reduced Euclidean action under the allowed thermodynamic variations and does not identify the specific response functions responsible for the instability. A systematic formulation of the equations of state and associated response functions in each thermodynamic representation has, to our knowledge, also not been developed.
 \textcolor{black}{In particular, a unified response theory 
  that makes explicit how thermal and mechanical response functions depend on
the thermodynamic variables held fixed is lacking.}

%What has been missing is a unified account of how such a black hole in a cavity responds when its size, pressure, temperature, or entropy is controlled, and of how these different responses fit together.

We therefore systematically formulate the quasi-local thermodynamics of Schwarzschild black holes in the energy, Helmholtz, Gibbs, and enthalpy representations and derive the associated response functions from the corresponding Hessian structures. The Helmholtz representation reproduces the familiar York picture in which the large black hole branch is thermally stable at fixed volume. The Gibbs representation, by contrast, reveals a qualitatively different stability structure. As already shown by York \cite{york1986}, the isothermal compressibility is positive on what, in the Helmholtz representation, is called the small black hole branch, but changes sign and becomes negative on the large black hole branch. Thus, the large black hole is mechanically unstable under isothermal compression despite being thermally stable in the canonical representation. The sign change occurs precisely when the cavity boundary crosses the photon sphere, i.e.
\(
r_B=\frac{3}{2}r_h=3GM
\)
in four spacetime dimensions.

Further, we  show that in the Gibbs representation the constant-pressure heat capacity and the thermal expansion coefficient are negative throughout the physical state space. Hence, the Schwarzschild black hole is thermally unstable at fixed pressure, and increasing the temperature at fixed pressure decreases the thermodynamic volume rather than increasing it. In the enthalpy representation, by contrast, the Schwarzschild black hole is mechanically stable under adiabatic compression, demonstrating that mechanical stability depends crucially on which thermodynamic variables are held fixed. Finally, we find the Joule-Thomson coefficient is positive throughout the state space, implying that Schwarzschild black holes in a spherical cavity always cool under isenthalpic expansion and possess no finite inversion curve.

These results demonstrate that thermal and mechanical stability are distinct notions whose behavior depends on the thermodynamic representation and therefore on the class of equilibrium perturbations under consideration. In particular, the branch that is thermally stable in the canonical representation becomes mechanically unstable in the Gibbs representation. Schwarzschild black holes  \textcolor{black}{in a cavity} consequently exhibit a considerably richer equilibrium structure than is visible in the standard asymptotically flat treatment.

The remainder of this paper is organized as follows. In Section~\ref{sec:eqnofstateinTD} we review thermodynamic representations, equations of state, response functions, and the relation   to Hessian structures in equilibrium thermodynamics. In Section~\ref{sec:eqnstatenow} we illustrate this framework for a classical ideal gas. This provides a simple setting in which the representation-dependence of equations of state and response functions can be seen explicitly, while also establishing the notation and conventions used throughout the rest of the paper. In Section~\ref{sec:eqnstateschw} we derive the equations of state for Schwarzschild black holes in a cavity in the energy, Helmholtz, Gibbs, and enthalpy representations. Section~\ref{sec:responseschw} develops the corresponding response theory in these representations and analyzes the resulting thermal and mechanical stability properties. Here, we also plot the
  fundamental relations, equations of state, and response
functions for the four different representations.  We conclude in Section~\ref{sec:conclusion} with a discussion of the broader implications of these results and some directions for future work.  
Further, we summarize the behavior of the response functions for the
Schwarzschild black hole in a cavity across the different thermodynamic
representations in a table at the end of the main text.

Finally, the paper is supplemented by two appendices. In Appendix~\ref{appA} we summarize various standard thermodynamic representations and the associated equations of state and response functions.  In Appendix~\ref{appendixContactGeom} we review the contact   geometric description of the thermodynamic state space.  
Regarding our choice of units, we set $\hbar=c=k_B=1$ but we keep factors of Newton's constant $G$ throughout the paper.

\section{Equilibrium thermodynamics}\label{sec:eqnofstateinTD}

\subsection{Thermodynamic representations and equations of state}

In equilibrium thermodynamics, the macroscopic state of a system is characterized by a thermodynamic potential $\Phi$ together with $n$ conjugate pairs of thermodynamic variables $(y_a,x^a)$, with $a=1,\dots,n$.  Common examples of conjugate pairs include $(T,S)$,  $(\mu_i,N^i)$ and $(Y_j,\xi^j)$, where $T$ is the temperature, $S$ is the entropy, $N^i$ are internal state variables (such as particle numbers),  $\mu_i$ are the associated chemical potentials, $\xi^j$ are deformation coordinates (such as volume or area), and $Y_j$ are the corresponding generalized forces. A \emph{thermodynamic representation} is specified by choosing a thermodynamic potential together with exactly one variable from each conjugate pair as an independent variable \cite{callen1985thermodynamics}. The \emph{fundamental relation} then determines the thermodynamic potential as a function of those independent variables \cite{callen1985thermodynamics,landsberg1978thermodynamics}. Physically, the choice of independent variables corresponds to a choice of
thermodynamic control parameters and therefore determines which variables are
held fixed under infinitesimal perturbations to nearby equilibrium states.
The   variables conjugate to the chosen independent variables characterize how the thermodynamic potential changes
under infinitesimal variations of these control parameters.

More precisely, if the independent variables are chosen to be
\begin{equation}
\label{independentvariables}
X^A=(x^1,\dots,x^k,y^{k+1},\dots,y^n)\,,
\end{equation}
then the fundamental relation in this thermodynamic representation takes the form
\begin{equation}
\Phi=\Phi(X^A)\,,
\label{eq:general-fundamental-relation}
\end{equation}
with differential
\begin{equation} 
d\Phi
=
\sum_{i=1}^{k} y_i\,dx^i
-
\sum_{j=k+1}^{n} x_j\,dy^j\, .
\label{eq:firstlaw-general-representation}
\end{equation}
\textcolor{black}{
Here, upper indices label the thermodynamic variables chosen as
independent, while the corresponding conjugate variables carry lower
indices. This is purely a notational convention; no metric or index
raising or lowering is implied.
The conjugate variables are obtained by partially differentiating the thermodynamic potential,}
\begin{align}
y_i
&=
\frac{\partial\Phi}{\partial x^i}\,,
\qquad \quad  i=1,\dots,k\,,
\label{eq:yasdifferentialofx}
\\
x_j
&=
-
\frac{\partial\Phi}{\partial y^j}\,,
\qquad  j=k+1,\dots,n\,.
\label{eq:xasdifferentialofy}
\end{align}
The resulting functional relations
\begin{equation}
y_i=y_i(X^A)\,,
\qquad
x_j=x_j(X^A)\,,
\label{eq:general-eos}
\end{equation}
are the \emph{equations of state} in the chosen thermodynamic representation. Thus every thermodynamic representation contains $n$ equations of state, one for each conjugate pair. We emphasize that an equation of state can never be e.g. $E(P,V)$, because here a quantity depends on both variables in a conjugate pair. For a thermodynamic representation to be well defined, the independent variables must all come from different conjugate pairs. This means that the well-known conformal fluid  relation
\(E=(D-1)PV\), which follows from the tracelessness of the  stress-energy tensor for a global thermal equilibrium state on a spatially homogeneous geometry, is not  truly an equation of state, but rather a \textcolor{black}{different kind of} thermodynamic identity.

Different thermodynamic representations are related by Legendre transformations of the thermodynamic potential, which exchange one or more independent variables for their conjugates. In the transformed representation, the equations of state are again obtained by differentiating the transformed thermodynamic potential with respect to the new independent variables and they thereby express the remaining conjugate variables as functions of those independent variables. Hence, although the underlying thermodynamic system remains unchanged, the equations of state depend on the choice of  thermodynamic representation (see Appendix \ref{appendixContactGeom} for a   geometric exposition of this point).

As an example, consider a   thermodynamic system with a single deformation coordinate~$V$ (volume), with conjugate pressure $P$, and a single particle number $N$, conjugate to the chemical potential $\mu$. In the \emph{energy representation}, the thermodynamic potential is the internal energy $E$ and the independent variables are $(S,V,N)$. The fundamental relation  and its differential  are
\begin{equation}
E=E(S,V,N)\,,
\qquad
dE=TdS-PdV+\mu dN\,,
\end{equation}
with conjugate pairs $(T,S)$, $(P,V)$, and $(\mu,N)$. The equations of state are
\begin{equation}
T
=
\frac{\partial E}{\partial S}\,,
\qquad
P
=
-
\frac{\partial E}{\partial V}\,,
\qquad
\mu
=
\frac{\partial E}{\partial N}\,.
\label{energyrepresentationequationsofstate}
\end{equation}
Hence, the equations of state determine the temperature, pressure, and chemical potential   as functions of the independent variables $(S,V,N)$. Choosing a different thermodynamic potential together with a different set of independent variables defines a different thermodynamic representation and therefore leads to different equations of state. Several standard thermodynamic representations and their associated equations of state are summarized in   Appendix~\ref{appA}.

\subsection{Thermodynamic response functions and   metrics}\label{sec:responsandmetrics}

Thermodynamic response functions characterize how equilibrium states respond to
infinitesimal variations of the thermodynamic control parameters. They follow from differentiating the equations of state, or equivalently by taking
second derivatives of the thermodynamic potential. Response functions encode the local second-order differential structure associated with a given thermodynamic representation and characterize the response of the system to infinitesimal perturbations. They determine thermal and mechanical stability and describe properties such as compressibility and thermal expansion.

Concretely, for a thermodynamic representation with independent variables
$X^A$ and thermodynamic potential $\Phi(X^A)$, the \emph{response
functions} are obtained by differentiating the equations of state
\eqref{eq:general-eos} with respect to the independent variables. We define the
response coefficients with signs chosen such that they coincide directly with
the Hessian matrix of the thermodynamic potential:
\begin{align}
\chi_{iA}
&=
\frac{\partial y_i}{\partial X^A}
=
\frac{\partial^2\Phi}{\partial x^i\partial X^A}\,,
\qquad
i=1,\dots,k\,,
\\
\tilde{\chi}_{jA}
&=
-
\frac{\partial x_j}{\partial X^A}
=
\frac{\partial^2\Phi}{\partial y^j\partial X^A}\,,
\qquad
j=k+1,\dots,n\,.
\label{generalresponsefunctions}
\end{align}
Hence the response coefficients are components of the Hessian matrix
\begin{equation}
\mathcal H_{AB}^{(\Phi)}
=
\frac{\partial^2\Phi}{\partial X^A\partial X^B}\,.
\end{equation}
The Hessian matrix is symmetric by the symmetry of partial derivatives. The corresponding relations
between different response coefficients are the Maxwell relations. For a
thermodynamic state space with $n$ conjugate pairs, there are $n$ equations of
state and therefore $n^2$ response coefficients. However, the symmetry of the
Hessian matrix implies that there are
only
 $ 
\frac12 n(n+1)
$
independent response coefficients.

The diagonal components of the Hessian characterize intrinsic thermal,
mechanical, or chemical stiffness, while the off-diagonal components encode
couplings between different thermodynamic sectors. 
As an explicit example, consider the energy representation with two independent
variables,
\begin{equation}
E=E(S,V)\,,
\qquad
dE=TdS-PdV \,.
\end{equation}
The corresponding Hessian matrix is
\begin{equation}
\mathcal{H}^{(E)}
=
\begin{pmatrix}
\mathcal{H}^{(E)}_{SS} & \mathcal{H}^{(E)}_{SV}\\
\mathcal{H}^{(E)}_{VS} & \mathcal{H}^{(E)}_{VV}
\end{pmatrix}
=
\begin{pmatrix}
\displaystyle
\left(\frac{\partial T}{\partial S}\right)_V
&
\displaystyle
-\left(\frac{\partial P}{\partial S}\right)_V
\\[1.2em]
\displaystyle
\left(\frac{\partial T}{\partial V}\right)_S
&
\displaystyle
-\left(\frac{\partial P}{\partial V}\right)_S
\end{pmatrix}\,.
\label{energyhessian}
\end{equation}
The diagonal components define the thermal and mechanical stiffness
coefficients
\begin{equation}\label{stiffnessenergyrep}
K_T^{(E)}
=
\left(\frac{\partial T}{\partial S}\right)_V,
\qquad
K_P^{(E)}
=
-
\left(\frac{\partial P}{\partial V}\right)_S,
\end{equation}
which measure the resistance of the system to infinitesimal thermal and
mechanical deformations, respectively. The off-diagonal components\textcolor{black}{, denoted $\chi_{SV}^{(E)}$,} encode
thermal-mechanical coupling and satisfy the Maxwell relation
\begin{equation}
\left(\frac{\partial T}{\partial V}\right)_S
=
-
\left(\frac{\partial P}{\partial S}\right)_V.
\end{equation}
Further, other physically relevant
thermodynamic response functions are often obtained from the  fundamental
response coefficients (i.e. Hessian components) by normalization with thermodynamic variables, inversion,
or by combining different response coefficients.

For instance, in the \emph{Helmholtz representation} \(F(T,V,N)\), where $F=E-TS$ is the Helmholtz free
 energy,  the constant-volume heat capacity and the isothermal mechanical stiffness are
\begin{align}
C_V
&=
T\left(\frac{\partial S}{\partial T}\right)_{V,N}
=
- T\,\mathcal H^{(F)}_{TT}\,,
\\
K_P^{(F)}
&=
-\left(\frac{\partial P}{\partial V}\right)_{T,N}
=
\mathcal H^{(F)}_{VV}\,.
\end{align}
The corresponding isothermal bulk modulus is the volume-normalized stiffness,
\begin{equation}
B_T
=
V K_P^{(F)}
=
- V\left(\frac{\partial P}{\partial V}\right)_{T,N}
\,.
\end{equation}
Similarly, in the \emph{Gibbs representation} $G(T,P,N)$, where $G=E-TS+PV$ is the Gibbs free
energy, the constant-pressure heat capacity, 
thermal expansion coefficient, and isothermal compressibility  are, respectively,
\begin{align}
C_{P}
&=
T
\left(
\frac{\partial S}{\partial T}
\right)_{P,N}
=
-
T\,\mathcal H_{TT}^{(G)}\,,
\\
\alpha
&=
\frac1V
\left(
\frac{\partial V}{\partial T}
\right)_{P,N}
=
\frac1V
\mathcal H_{TP}^{(G)}\,,\\
\kappa_T
&=
-
\frac1V
\left(
\frac{\partial V}{\partial P}
\right)_{T,N}
=
-
\frac1V
\mathcal H_{PP}^{(G)}\,.
\end{align}
The isothermal
compressibility is the inverse of the isothermal bulk modulus, $\kappa_T = 1/B_T.$ More generally, heat capacities and
compressibilities characterize thermal and mechanical susceptibilities,
respectively, while the corresponding inverse quantities, namely stiffness
coefficients and bulk moduli, characterize thermal and mechanical rigidity.
Several further thermodynamic response functions in different representations
are reviewed in Appendix~\ref{appA}.

Local thermal and mechanical stability
is determined by the convexity or concavity properties of the thermodynamic
potential in the chosen representation, or equivalently by the definiteness
properties of the associated Hessian matrix. A thermodynamic system is locally
stable when the Hessian satisfies the positivity or negativity conditions
appropriate to the thermodynamic potential under infinitesimal perturbations to
nearby equilibrium states consistent with the chosen control variables. For example, local stability in the energy representation requires the full Hessian
\eqref{energyhessian} to be positive definite, implying for instance
\begin{equation}
K_T^{(E)}>0\,,
\qquad
K_P^{(E)}>0\,.
\end{equation}
\textcolor{black}{
Positivity of an individual Hessian component is only a necessary
condition for full local stability and characterizes stability with
respect to the corresponding restricted class of perturbations. For
example, $K_T^{(E)}>0$ implies stability under entropy variations at
fixed volume, $\delta V=0$, but does not by itself imply positive
definiteness of the full Hessian.
}

\textcolor{black}{
Throughout this work, stability refers to local response stability under
infinitesimal equilibrium perturbations compatible with the chosen
thermodynamic constraints. This should be distinguished from global
thermodynamic preference between distinct equilibrium phases or saddles,
which requires a comparison of their thermodynamic potentials or Euclidean
actions. Such global comparisons between competing equilibrium saddles are not considered here.
}

The Hessian matrix also defines a  \emph{thermodynamic metric} on
thermodynamic state space. In the energy representation this yields the
Weinhold metric \cite{Weinhold:1975xej}
\begin{equation}
g^{\rm W}_{AB}
=
\frac{\partial^2E}{\partial X^A\partial X^B}\,,
\end{equation}
while in the entropy representation one obtains the Ruppeiner metric \cite{Ruppeiner:1979bcp,Ruppeiner:1995zz}
\begin{equation}
g^{\rm R}_{AB}
=
-
\frac{\partial^2S}{\partial Y^A\partial Y^B}\,.
\end{equation}
More generally, every thermodynamic representation determines a Hessian metric
associated with the corresponding thermodynamic potential and choice of
independent variables. Different thermodynamic representations therefore define
different local thermodynamic geometries, corresponding to different classes
of equilibrium perturbations and response functions. Degeneracies of the
corresponding Hessian metrics signal vanishing or divergent response functions. They may therefore indicate
changes of local thermal or mechanical stability, branch mergers, or critical
behavior. 

In the following   we illustrate this general framework by computing
explicit equations of state and response functions across various representations for a classical ideal gas and a Schwarzschild black hole in a cavity. 
\textcolor{black}{The contact geometric viewpoint reviewed in Appendix~\ref{appendixContactGeom} further clarifies the
distinction between the underlying manifold of equilibrium states and the choice of
thermodynamic control variables. The thermodynamic potentials $E$, $F$,
$G$, and $H$ are different generating functions for the same equilibrium
Legendre submanifold (see Eq.~\eqref{eq:legendresubmfd}), so a Legendre transformation
does not change the underlying equilibrium system. What does change
between thermodynamic representations is the choice of independent variables, with an accompanying change of generating function for the Legendre submanifold.}

% \textcolor{black}{Our system is inspired by the formalism of contact geometry, explained in Appendix \ref{appendixContactGeom}, which cleanly separates two notions that are sometimes conflated. The thermodynamic potentials \(E,F,G,H\) provide different generating functions of the same equilibrium Legendre submanifold (see Eq. \eqref{eq:legendresubmfd}). A Legendre transformation therefore does not by itself change the
% underlying equilibrium system. A representation does, however, identify a set of thermodynamic control variables and hence a  class of allowed fluctuations. The Hessian of the corresponding potential is the Jacobian of the equations of state in that chart, and its restriction to the allowed fluctuation directions determines local response stability.}

\section{Equations of state and response functions  for an ideal gas}
\label{sec:eqnstatenow}

Here we apply the general framework developed above to a classical ideal gas. Although the ideal gas law
\begin{equation}
PV = NT,
\label{eq:ideal-gas-law}
\end{equation}
is usually referred to as \emph{the} equation of state of the ideal gas, it is in fact only one of multiple: since the ideal gas possesses three conjugate thermodynamic pairs, namely $(T,S)$, $(P,V)$, and $(\mu,N)$, there are in total three equations of state in every thermodynamic representation. \textcolor{black}{Useful references for the thermodynamic and statistical mechanical
description of the ideal gas used in this section are
\cite{Landau:1980mil,Pathria:1996hda,Lavis2025-LAVTFO-3}.}

\subsection{Helmholtz representation}

In the Helmholtz representation, where the independent variables are $(T,V,N)$, the Helmholtz free energy of a classical ideal gas is
\begin{equation}
F(T,V,N)
=
-NT \ln \!\left(\frac{V}{N}\right)
-\frac{f}{2}NT \ln T
-N Tc \,,
\end{equation}
\textcolor{black}{where $f$ denotes the number of quadratic microscopic degrees of freedom
per gas particle and $c$ is a constant fixed by the microscopic partition
function. For a monatomic ideal gas in three spatial dimensions, $f=3$ and
$c=\frac{3}{2}\ln(2\pi m/h^2)+1$.}
The equations of state follow from partial differentiation of the Helmholtz free energy,
yielding
\begin{align}\label{eq:idealgas}
S(T,V,N)
&=
N \ln \!\left(\frac{V}{N}\right)
+
\frac{f}{2} N \ln T
+
N\!\left(\frac{f}{2}+c\right)\,,
\nonumber
\\
P(T,V,N)
&=
\frac{NT}{V}\,,
\\
\mu(T,V,N)
&=
-
T \ln \!\left(\frac{V}{N}\right)
-
\frac{f}{2}T \ln T
+
T(1-c)\,.
\nonumber
\end{align}
Thus, the ideal gas law \eqref{eq:ideal-gas-law} is only one element of the complete set of equations of state associated with the Helmholtz representation.

Next, since $n=3$ there are   six independent response coefficients. They are given by the following diagonal and off-diagonal Hessian components
\begin{align}
\mathcal{H}^{(F)}_{TT}
&=
-\frac{fN}{2T}\,, \qquad 
\mathcal{H}^{(F)}_{TV}
=
\mathcal{H}^{(F)}_{VT}
=
-\frac{N}{V}\,,
\\
\mathcal{H}^{(F)}_{VV}
&=
\frac{NT}{V^2}\,, \qquad \,\,\,\mathcal{H}^{(F)}_{TN}
=
\mathcal{H}^{(F)}_{NT}
=
-
\ln\!\left(\frac{V}{N}\right)
-
\frac{f}{2}\ln T
+
1-\frac{f}{2}-c\,,
\\
\mathcal{H}^{(F)}_{NN}
&=
\frac{T}{N}\,, \qquad \,\,\,\,\,\,\,\mathcal{H}^{(F)}_{VN}
=
\mathcal{H}^{(F)}_{NV}
=
-\frac{T}{V}\,.
\end{align}
From the Hessian components one may construct the standard thermodynamic response functions. For example, the constant-volume heat capacity and isothermal bulk modulus are 
\begin{equation}
C_V
=
-T \mathcal{H}^{(F)}_{TT}
=
\frac{f}{2}N\,,
\qquad
B_T
=
V \mathcal{H}^{(F)}_{VV}
=
\frac{NT}{V}
=
P\,.
\end{equation}

\subsection{Energy representation}

Alternatively, in the energy representation the independent variables are $(S,V,N)$ and the fundamental equation for the internal energy is
\begin{equation}
E(S,V,N)
=
\frac{f}{2}N
\left(
\frac{e^{S/N-f/2-c}N}{V}
\right)^{2/f}\,.
\end{equation}
The equations of state follow from the partial derivatives \eqref{energyrepresentationequationsofstate} and are given by
\begin{align}
T(S,V,N)
&=
\left(
\frac{e^{S/N-f/2-c}N}{V}
\right)^{2/f}\,,
\nonumber\\
P(S,V,N)
&=
\left(
\frac{e^{S/N-f/2-c}N}{V}
\right)^{2/f}
\frac{N}{V}\,,
\\
\mu(S,V,N)
&=
\left(
\frac{e^{S/N-f/2-c}N}{V}
\right)^{2/f}
\left(
\frac{f}{2}
+1
-\frac{S}{N}
\right)\,.
\nonumber
\end{align}
In the energy representation the six independent response coefficients are given by the following   Hessian components,
\begin{align}
\mathcal{H}^{(E)}_{SS}
&=
\frac{2T}{fN}\,,
&
\mathcal{H}^{(E)}_{SV}
=
\mathcal{H}^{(E)}_{VS}
&=
-\frac{2T}{fV}\,, \nonumber
\\
\mathcal{H}^{(E)}_{VV}
&=
\frac{f+2}{f}\frac{NT}{V^2}\,,
&
\mathcal{H}^{(E)}_{SN}
=
\mathcal{H}^{(E)}_{NS}
&=
\frac{2T}{fN}
\left(
1-\frac{S}{N}
\right)\,, 
\\
\mathcal{H}^{(E)}_{NN}
&=
\frac{T}{N}
\left[
\frac{S}{N}
+
\frac{2}{f}
\left(
1-\frac{S}{N}
\right)
\left(
\frac{f}{2}+1-\frac{S}{N}
\right)
\right]\,,
&
\mathcal{H}^{(E)}_{VN}
=
\mathcal{H}^{(E)}_{NV}
&=
-\frac{T}{V}
\left[
1+\frac{2}{f}
\left(
1-\frac{S}{N}
\right)
\right]\,, \nonumber
\end{align}
where the temperature should be regarded as a function $T=T(S,V,N).$
The diagonal Hessian components are equal to the thermal, mechanical and chemical   stiffness coefficients in the energy representation, 
\begin{equation}
K_T^{(E)}
=
\mathcal{H}^{(E)}_{SS}
\,, \qquad 
K_P^{(E)}
=
\mathcal{H}^{(E)}_{VV}
\,, \qquad 
K_\mu^{(E)}
=
\mathcal{H}^{(E)}_{NN}\,.
\end{equation}
The mechanical stiffness coefficient in turn determines the adiabatic bulk modulus  
\begin{equation}
B_S
=
V \mathcal{H}^{(E)}_{VV}
=
\frac{f+2}{f}\frac{NT}{V}
=
\frac{f+2}{f}P \,.
\end{equation}

\subsection{Gibbs representation}

In the Gibbs representation, where the independent variables are $(T,P,N)$, the Gibbs free energy of a classical ideal gas is 
\begin{equation}
G(T,P,N)
=
NT\ln P
-
\left(1+\frac{f}{2}\right)NT\ln T
+
NT(1-c)\,.
\end{equation}
The equations of state are given by 
\begin{align}
S(T,P,N)
&=
N\left(1+\frac{f}{2}\right)\ln T
-
N\ln P
+
N\left(\frac{f}{2}+c\right)\,,
\nonumber
\\
V(T,P,N)
&=
\frac{NT}{P}\,,
\\
\mu(T,P,N)
&=
T\ln P
-
\left(1+\frac{f}{2}\right)T\ln T
+
T(1-c)\,,
\nonumber
\end{align}
and the six independent response coefficients are
\begin{align}
\mathcal{H}^{(G)}_{TT}
&=
-\frac{f+2}{2}\frac{N}{T}\,,
\qquad\,
\mathcal{H}^{(G)}_{TP}
=
\mathcal{H}^{(G)}_{PT}
=
\frac{N}{P}\,, \nonumber
\\
\mathcal{H}^{(G)}_{PP}
&=
-\frac{NT}{P^2}\,,
\qquad
\qquad \mathcal{H}^{(G)}_{TN}
=
\mathcal{H}^{(G)}_{NT}
=
\ln P
-
\left(1+\frac{f}{2}\right)\ln T
-
\frac{f}{2}
-c\,,
\\
\mathcal{H}^{(G)}_{NN}
&=
0\,,
\qquad
\qquad \qquad \mathcal{H}^{(G)}_{PN}
=
\mathcal{H}^{(G)}_{NP}
=
\frac{T}{P}\,. \nonumber
\end{align}
The vanishing of \(\mathcal H^{(G)}_{NN}\) is a consequence of the extensivity of the Gibbs free energy in $N$: 
\(
G(T,P,N)=Ng(T,P).
\)
From these Hessian components one obtains the well-known coefficients
\begin{equation}
C_P
=
-T\mathcal{H}^{(G)}_{TT}
=
\frac{f+2}{2}N\,,
\qquad
\alpha
=
\frac{1}{V}\mathcal{H}^{(G)}_{TP}
=
\frac{1}{T}\,, \qquad 
\kappa_T
=
-\frac{1}{V}\mathcal{H}^{(G)}_{PP}
=
\frac{1}{P}\,.
\end{equation}

\section{Equations of state for Schwarzschild black holes}
\label{sec:eqnstateschw}

We now determine the equations of state for a Schwarzschild black hole enclosed by a finite timelike boundary whose spatial cross-sections are spheres, i.e.\ a Schwarzschild black hole in a spherical cavity. The quasi-local thermodynamic system consists of the spacetime region between the event horizon and the cavity boundary. Its thermodynamic description and variational relation were reviewed in the Introduction. Following the holographic interpretation of quasi-local thermodynamics developed in \cite{Borsboom:2026ash}, the cavity area and surface pressure are identified, respectively, with the thermodynamic volume and pressure of the dual boundary system. The resulting thermodynamic state space is therefore two-dimensional, allowing the energy, Helmholtz, Gibbs, and enthalpy representations   to be constructed explicitly. In Figures \ref{fig:schwarzschild-energy-representation} to \ref{fig:schwarzschild-enthalpy-representation} we plot the fundamental relations and  equations of state  in these representations for a Schwarzschild black hole. 

In 3+1 dimensions the relevant quasi-local thermodynamic quantities are \cite{york1986}
\begin{align}
S=\frac{\pi r_h^2}{G}&\,,\;\;\;\;\;\; V=4\pi r_B^2\,, \label{eq:yorkSV}\\
E(r_h,r_B)&=\frac{r_B}{G}\left(1-\left(1-\frac{r_h}{r_B}\right)^{1/2}\right)\,,\label{eq:yorkE}
\\
P(r_h,r_B)
&=
\frac{1}{8\pi G r_B}
\left[
\left(1-\frac{r_h}{2r_B}\right)
\left(1-\frac{r_h}{r_B}\right)^{-1/2}
-1
\right]\,,\label{eq:yorkP}
\\
T(r_h,r_B)
&=
\frac{1}{4\pi r_h}
\left(1-\frac{r_h}{r_B}\right)^{-1/2}\,,\label{eq:yorkT}
\end{align}
where $r_h=2GM$ is the horizon radius and $r_B$ is the cavity radius. Here $S$ is the Bekenstein entropy, $E$ is the Brown-York quasi-local energy, $P$ is the holographic pressure, and $T$ is the Tolman temperature measured at the cavity boundary, i.e.\ the locally redshifted Hawking temperature seen by an observer at $r=r_B$.

\subsection{Energy representation}

In the energy representation the independent variables are $(S,V)$ and the thermodynamic potential is the quasi-local energy $E(S,V)$.
For a four-dimensional Schwarzschild black hole in a spherical cavity, the fundamental relation is obtained by combining   \eqref{eq:yorkSV} and \eqref{eq:yorkE}:
\begin{equation}
E(S,V)
=
\frac{\sqrt{V}}{2\sqrt{\pi}G}
\left[
1-
\sqrt{
1-
\left(
\frac{4GS}{V}
\right)^{1/2}
}
\right]\,,
\label{energyfundamental}
\end{equation}
defined for $4GS<V$, since the cavity should be outside the horizon. The equations of state, of which there are two since there are two independent variables, follow from   partial differentiation of the energy \eqref{energyrepresentationequationsofstate}, giving
\begin{align}
T(S,V)
&= 
\frac{1}{4\sqrt{\pi GS}}
\left[
1-
\left(
\frac{4GS}{V}
\right)^{1/2}
\right]^{-1/2}\,,
\label{SchwarzschildEOS1} \\
P(S,V)
&=
\frac{1}{4G\sqrt{\pi V}}
\left[
\frac{
1-\frac12
\left(
\frac{4GS}{V}
\right)^{1/2}
}{
\sqrt{
1-
\left(
\frac{4GS}{V}
\right)^{1/2}
}
}
-1
\right]\,.
\label{SchwarzschildEOS2}
\end{align}
The fundamental relation and equations of state are plotted in Figure \ref{fig:schwarzschild-energy-representation}.

\subsection{Helmholtz representation}
\label{sec:helmholtzrepr}
Unlike the energy representation, the Helmholtz or canonical representation is multivalued. For fixed temperature and volume, the entropy equation of state $S(T,V)$ admits two equilibrium branches above the minimum temperature
\begin{equation}
T_{\rm min}(V)
=
\frac{\sqrt{27}}{4\sqrt{\pi V}} \,.
\end{equation}
These correspond to the small and large   black hole branches, originally identified by York \cite{york1986}. 
% \textcolor{black}{The two branches merge at $T=T_{\min}$, below which no equilibrium
% Schwarzschild black hole exists inside the cavity. Geometrically, the
% projection of the equilibrium state space onto the Helmholtz variables
% $(T,V)$ develops a fold along $T=T_{\min}(V)$, where the two branches
% meet, as illustrated in Figure~\ref{fig:schwarzschild-canonical-representation}.}
\textcolor{black}{The two branches merge at $T=T_{\min}$, below which no equilibrium
Schwarzschild black hole exists inside the cavity. Geometrically, the
projection of the equilibrium state space onto the Helmholtz variables
$(T,V)$ becomes singular along $T=T_{\min}(V)$, where the two branches
meet, as illustrated in Figure~\ref{fig:schwarzschild-canonical-representation}. In Sec.~\ref{sec:helmholtzresponse} we show that this
singularity is a fold.}

Equivalently, the Helmholtz free energy becomes multivalued as a function of $(T,V)$. It is convenient to introduce the dimensionless variable
\begin{equation}\label{ratiorhrb}
x\equiv \left(\frac{4GS}{V}\right)^{1/2}
=\frac{r_h}{r_B},
\qquad 0<x<1,
\end{equation}
which measures the ratio of the horizon radius to the cavity radius. In terms of this variable, the two branches are described by
\begin{align} \label{smalllargebranches}
x_{\rm small}
=
\frac{
1+2\cos\left(\theta+\frac{4\pi}{3}\right)
}{3}\,,\qquad 
x_{\rm large}
=
\frac{
1+2\cos\theta
}{3}\,,
\end{align}
with
\begin{equation}
\theta(T,V)
=
\frac13
\arccos
\left(
1-\frac{27}{8\pi VT^2}
\right)\,,
\end{equation}
and the fundamental relation for the Helmholtz free energy is
\begin{equation}
F_{\rm small/large}(T,V)
=
\frac{\sqrt{V}}{2\sqrt{\pi}G}
\left[
1
-
\sqrt{1-x_{\rm small/large}}
-
\frac{x_{\rm small/large}}
{4\sqrt{1-x_{\rm small/large}}}
\right]\,.
\end{equation}
The canonical equations of state are then
\begin{align}
S_{\rm small/large}(T,V)
&=
\frac{V}{4G}
x_{\rm small/large}^2 \,,
\\
P_{\rm small/large}(T,V)
&=
\frac{1}{4G\sqrt{\pi V}}
\left[
\frac{
1-\frac12 x_{\rm small/large}
}{
\sqrt{1-x_{\rm small/large}}
}
-1
\right]\,.
\label{canonicalpressure}
\end{align}
The canonical representation therefore decomposes the Schwarzschild state space into two distinct equilibrium branches for $T>T_\text{min}$.
In earlier work \cite{Borsboom:2026ash} we showed that the two canonical
branches exhibit qualitatively different extensivity properties. In the
large-system limit $V\to\infty$ at fixed temperature, the Helmholtz free
energy of the large black hole branch becomes extensive, whereas the small
black hole branch remains non-extensive. \textcolor{black}{
The same asymptotic distinction is reflected in the response functions
derived below: in the fixed-temperature, large-volume limit, the  
thermal pressure coefficient $\gamma_V$ tends to zero on the small black
hole branch, while it approaches $1/(4G)$ on the large black hole branch, see \textcolor{black}{Eq. \eqref{eq:gammaVtozero}.}}
By contrast, in the energy representation, the quasi-local energy
$E(S,V)$ remains non-extensive in the large-volume limit at fixed entropy
density $S/V$. \textcolor{black}{
A further explanation of the branch structure in relation to the
canonical response functions is given in
Section~\ref{sec:helmholtzresponse}.
}

\subsection{Gibbs representation}

In the Gibbs representation the independent variables are $(T,P)$. The Euclidean action was calculated by imposing the correct boundary conditions in \cite{Brown:1989fa} and the stability properties were studied in \cite{Comer:1992pc} (under the name of ``canonical pressure ensemble''), but to our knowledge the equations of state in this representation have never been written down explicitly. 

It is more convenient to work with the horizon radius $r_h$ and cavity radius $r_B$, rather than the entropy and volume variables, $S= \pi r_h^2/ G$ and $
V=4\pi r_B^2 $.
Looking at the expression~\eqref{eq:yorkP} for the thermodynamic pressure, we introduce the dimensionless redshift parameter
\begin{equation}
y \equiv \sqrt{1-\frac{r_h}{r_B}}\,,
\qquad
0<y<1\,,
\label{eq:ydef}
\end{equation}
which measures the gravitational redshift between the horizon and the cavity wall.
The temperature relation~\eqref{eq:yorkT} immediately gives
\begin{equation}
r_h=\frac{1}{4\pi Ty}\,,
\label{eq:rhgibbs}
\end{equation}
which can be rearranged into
\begin{align}
    r_B=\frac{1}{4\pi Ty(1-y^2)}\,.\label{eq:rBgibbs}
\end{align}
Substituting these expressions for $r_h,r_B$ into the pressure equation yields
\begin{align}
    P(T,y)&=\frac{T}{4G}
(1-y)^3(1+y)\,.
\end{align}
Thus the pressure equation can be rewritten as a constraint on the redshift parameter:
\begin{equation}
(1-y)^3(1+y)=\frac{4GP}{T}\,.
\label{eq:gibbsconstraint}
\end{equation}
The function
$ 
g(y)=(1-y)^3(1+y)
$
is monotonically decreasing for $y\in(0,1)$, with $g(0)=1$ and $g(1)=0$.
Hence \eqref{eq:gibbsconstraint} admits a unique solution $y(T,P)\in(0,1)$ whenever
\begin{equation}
0<P<\frac{T}{4G} \,. \label{allowedregimenow2}
\end{equation}
which agrees with the physical domain derived in \cite{Brown:1989fa,Comer:1992pc}. 
Outside this wedge no equilibrium Schwarzschild black hole exists in the Gibbs representation. Unlike the Helmholtz representation, where two equilibrium branches exist above the minimum temperature, the Gibbs representation therefore possesses a single equilibrium branch throughout the allowed region~\eqref{allowedregimenow2}. After all, if one knows $T$ and $P$, then one also knows $y$ uniquely, and then $T$ and $y$ together can be used to find $r_h,r_B$ by means of equations \eqref{eq:rhgibbs} and \eqref{eq:rBgibbs}.

The solution of the quartic equation \eqref{eq:gibbsconstraint} may be written analytically using Ferrari's method, as follows:
\begin{equation}
y(T,P)
=
\frac12
-\frac12\sqrt{1+v}
+\frac12
\sqrt{
2-v+\frac{2}{\sqrt{1+v}}
}\,,
\label{eq:yexplicit}
\end{equation}
where
\begin{equation}\label{someuglyequation}
v=
\sqrt[3]{\frac{8GP}{T}+2\sqrt{\frac{(4GP)^2}{T^2}-16\frac{(4GP)^3}{27T^3}}}
+
\sqrt[3]{\frac{8GP}{T}-2\sqrt{\frac{(4GP)^2}{T^2}-16\frac{(4GP)^3}{27T^3}}}\,.
\end{equation}
The equations of state in the Gibbs representation are therefore
\begin{equation}
S(T,P)=\frac{\pi r_h(T,P)^2}{G}\,,
\qquad
V(T,P)=4\pi r_B(T,P)^2\,,
\label{eq:gibbseos}
\end{equation}
with $r_h(T,P)$ and $r_B(T,P)$ determined by \eqref{eq:rhgibbs}, \eqref{eq:rBgibbs},  \eqref{eq:yexplicit} and \eqref{someuglyequation}. The fundamental relation is simply the expression for the Gibbs free energy
\begin{align}
    G(T,P) = E - TS + PV
       = \frac{1}{8\pi G T\, y(1+y)}\,.
\end{align}
The fundamental relation and equations of state are plotted in Figure \ref{fig:schwarzschild-gibbs-representation}.
 
\subsection{Enthalpy representation}

In the enthalpy representation the independent variables are $(S,P)$ and the thermodynamic potential is
\begin{equation}
H=E+PV\,,
\qquad
dH=TdS+VdP\,.
\end{equation}
Physically, this representation is naturally adapted to adiabatic processes at fixed pressure. Since the entropy fixes the horizon radius directly,
the remaining task is to determine the cavity radius as a function of $(S,P)$.
As in the Gibbs representation, it is convenient to use the dimensionless redshift parameter $y$ in \eqref{eq:ydef}.
 In terms of $y$ one has
\begin{equation}
r_B=\frac{r_h}{1-y^2}\,.
\end{equation}
From equations \eqref{eq:rhgibbs} and \eqref{eq:gibbsconstraint} one finds
\begin{equation}
(1-y)^3(1+y)=16 \pi GPr_h y=16GP\sqrt{\pi GS}y\,.
\label{eq:enthalpyconstraint}
\end{equation}
The physical regime of the enthalpy representation is
\begin{equation}
S>0\,,
\qquad
P>0\,.
\end{equation}
For \(P<0\) the right-hand side of
\eqref{eq:enthalpyconstraint} is negative while the left-hand side is positive
for \(0<y<1\), so no physical solution exists. For \(P=0\), the solution is
\(y=1\), corresponding to the boundary limit \(r_B\to\infty\) at fixed $r_h$. Thus finite
cavity equilibria have \(P>0\).

For given values of \(S\) and \(P\), the left-hand side of
\eqref{eq:enthalpyconstraint} decreases monotonically from \(1\) to \(0\) as
\(y\) runs from \(0\) to \(1\), while the right-hand side is a linear function
of \(y\) with positive slope. The two sides therefore intersect exactly once,
yielding a unique physical solution \(y(S,P)\in(0,1)\).
Unlike the Helmholtz representation, but similar to the Gibbs representation, the enthalpy representation therefore contains only a single equilibrium branch. \textcolor{black}{The local regularity of the corresponding enthalpy and Gibbs projections
is discussed further in Sec.~\ref{sec:enthalpyresponse}.} 
Further, unlike the energy representation, whose physical domain is restricted by \(V>4GS\), the enthalpy representation covers the full positive \textcolor{black}{quadrant $S>0,P>0$.}

The quartic equation \eqref{eq:enthalpyconstraint} may be solved analytically
using Ferrari's method. The physical root is
\begin{equation}
y(S,P)
=
\frac12
-\frac12 w
+\frac12
\sqrt{
3-w^2-2w+2\sqrt{w^4-2w^2+5}
}\,,
\label{eq:y-enthalpy-explicit}
\end{equation}
where \(w=w(S,P)\) is given by
\begin{align}
&w(S,P)
=
\Bigg[
1
+
\sqrt[3]{
32GP\sqrt{\pi GS}
\left(
1+4GP\sqrt{\pi GS}
+
\sqrt{
1+\frac{184}{27}GP\sqrt{\pi GS}
+16\pi G^3P^2S
}
\right)
}
\nonumber\\
&+
\sqrt[3]{
32GP\sqrt{\pi GS}
\left(
1+4GP\sqrt{\pi GS}
-
\sqrt{
1+\frac{184}{27}GP\sqrt{\pi GS}
+16\pi G^3P^2S
}
\right)
}
\Bigg]^{1/2}\,.
\end{align}
Here the cube roots denote real cube roots.
The equations of state in the enthalpy representation then take the simple form
\begin{equation}
T(S,P)
=
\frac{1}{4\sqrt{\pi G S}\,y(S,P)}\,,
\qquad
V(S,P)
=
\frac{4GS}{\bigl(1-y(S,P)^2\bigr)^2}\,.
\end{equation}
The fundamental relation follows from $H=E+PV$.
Using
\begin{equation}
E=\frac{r_h}{G(1+y)}\,,
\qquad
PV=
\frac{r_h}{4G}\frac{1-y}{y(1+y)}\,,
\end{equation}
the enthalpy becomes
\begin{equation}
H(S,P)
=
\sqrt{\frac{S}{\pi G}}\,
\frac{1+3y(S,P)}{4y(S,P)\bigl(1+y(S,P)\bigr)}\,.
\label{eq:enthalpy-fundamental}
\end{equation}
Together with the explicit equation \eqref{eq:y-enthalpy-explicit}, this gives the fundamental relation in the enthalpy representation. This is plotted in Figure \ref{fig:schwarzschild-enthalpy-representation}.

\section{Response functions for Schwarzschild black holes}
\label{sec:responseschw}

Having derived the equations of state in the various thermodynamic representations, we now determine the corresponding response functions for a Schwarzschild black hole in a cavity. Since the thermodynamic state space is two-dimensional, there are three independent Hessian components in each representation. \textcolor{black}{Particular representatives of these three independent Hessian components  are plotted for four different representations in Figures \ref{fig:schwarzschild-energy-representation}, \ref{fig:schwarzschild-canonical-representation}, \ref{fig:schwarzschild-gibbs-representation} and \ref{fig:schwarzschild-enthalpy-representation}. Three derived response functions (i.e. not belonging to a particular representation) are plotted in Figure \ref{fig:schwarzschild-derived-response-functions}.  The plots are shown for $G=1$ in four bulk
spacetime dimensions. The shadows are included for visualization purposes
only. The behavior of all fifteen response functions is summarized in Table \ref{tab:schwarzschild-response-summary}.}

\subsection{Energy representation}

\begin{figure}[t]
    \centering
    \includegraphics[width=0.33\textwidth]{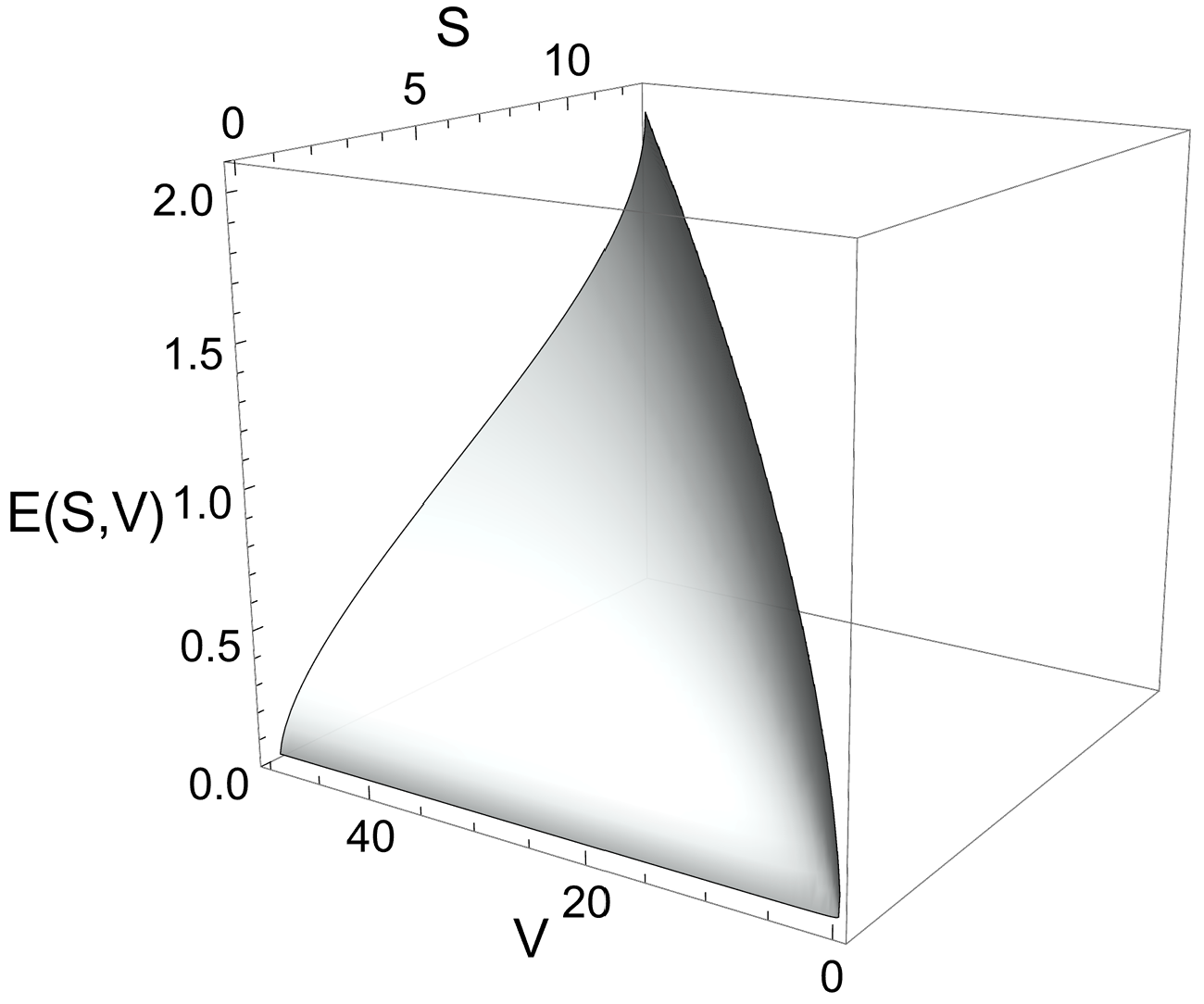}
    \includegraphics[width=0.35\textwidth]{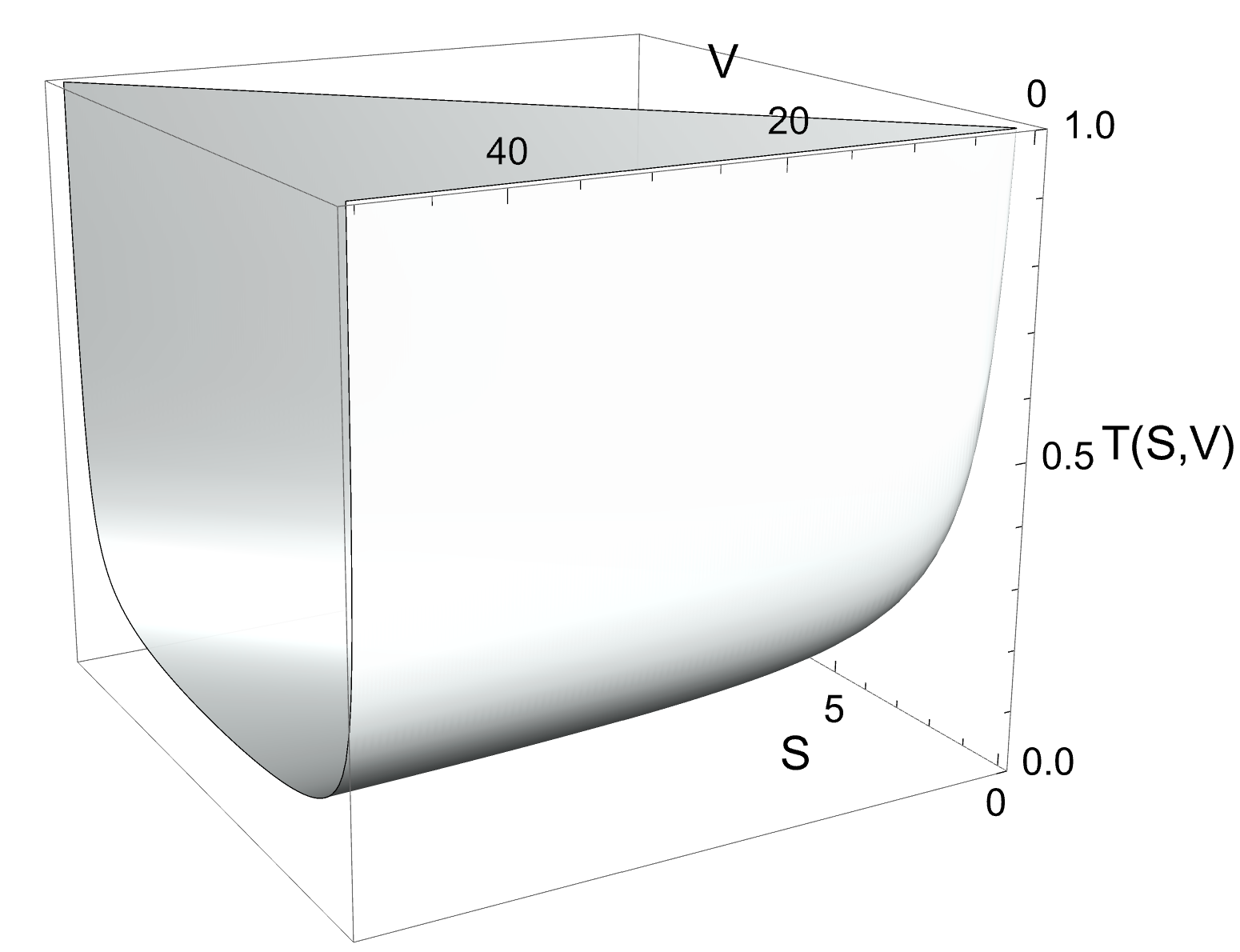}
    \includegraphics[width=0.29\textwidth]{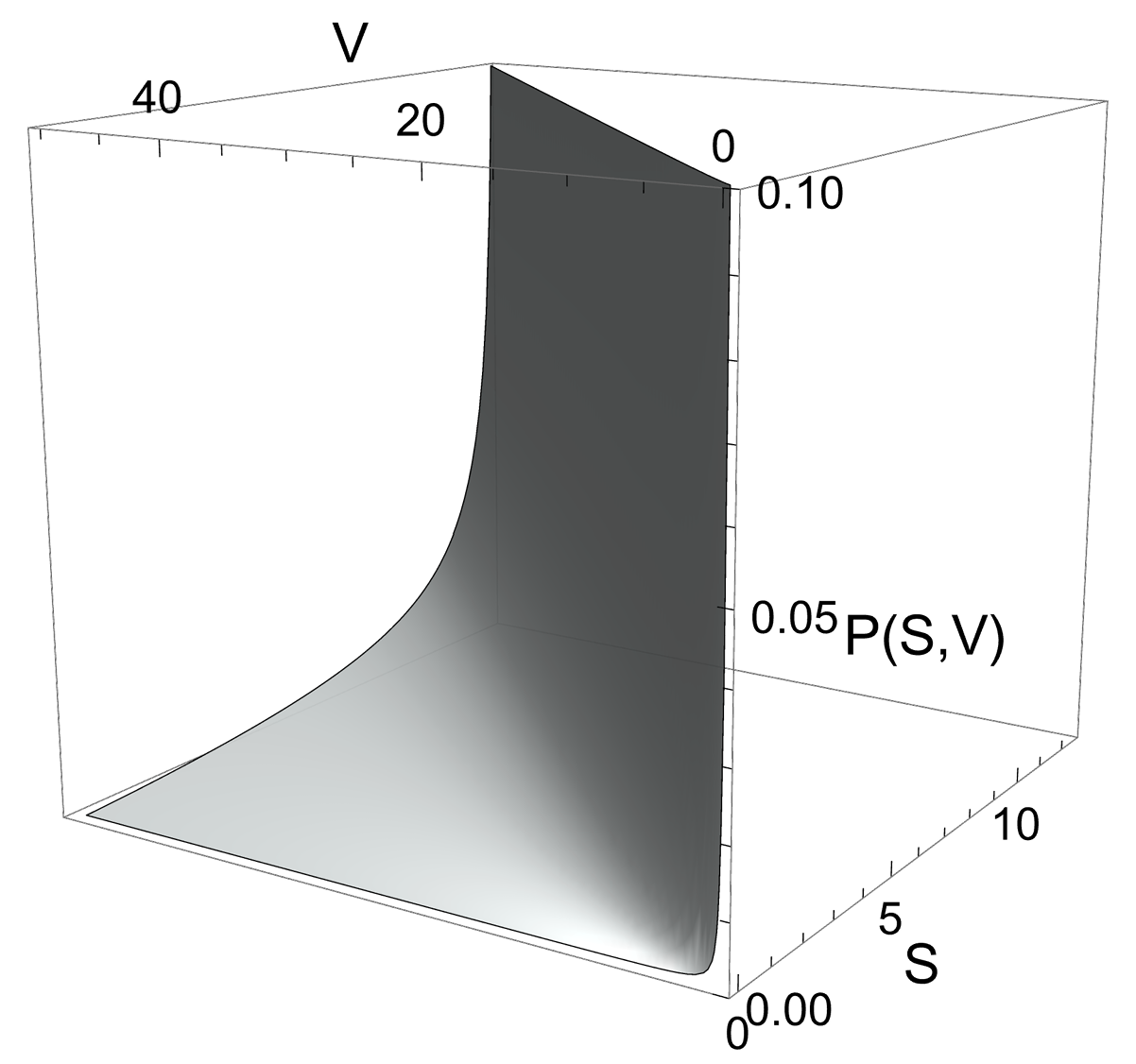}

    \includegraphics[width=0.33\textwidth]{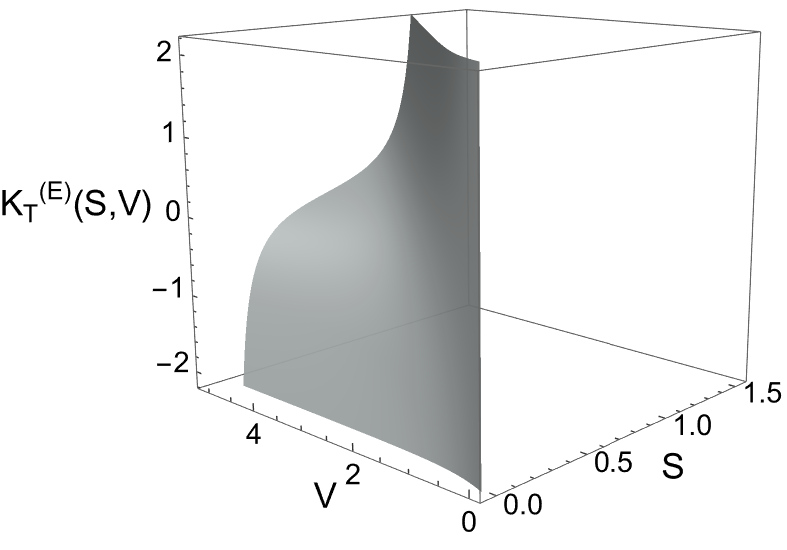}
    \includegraphics[width=0.33\textwidth]{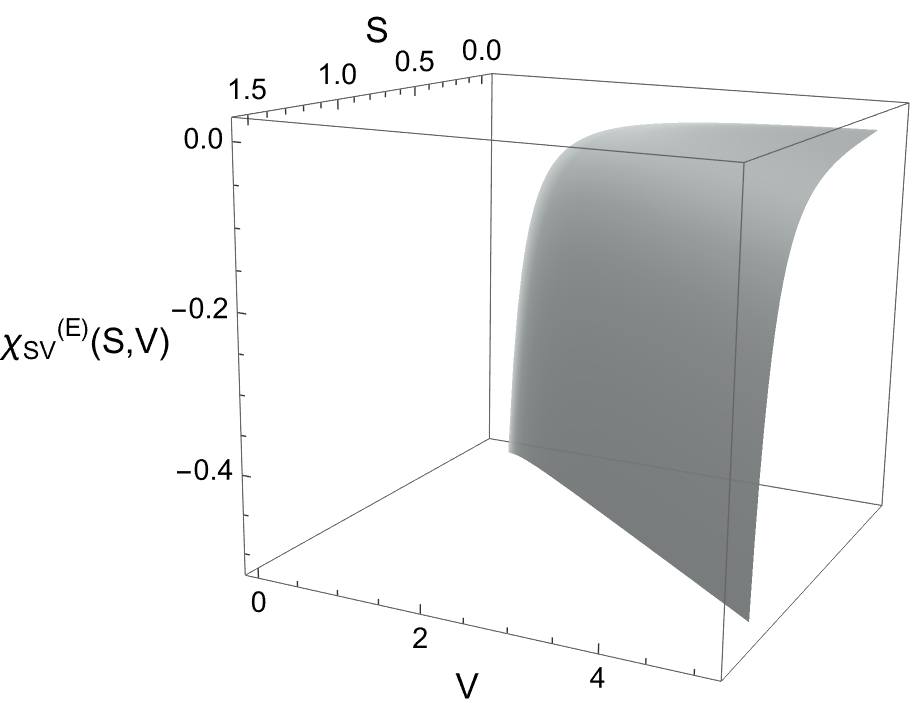}
    \includegraphics[width=0.32\textwidth]{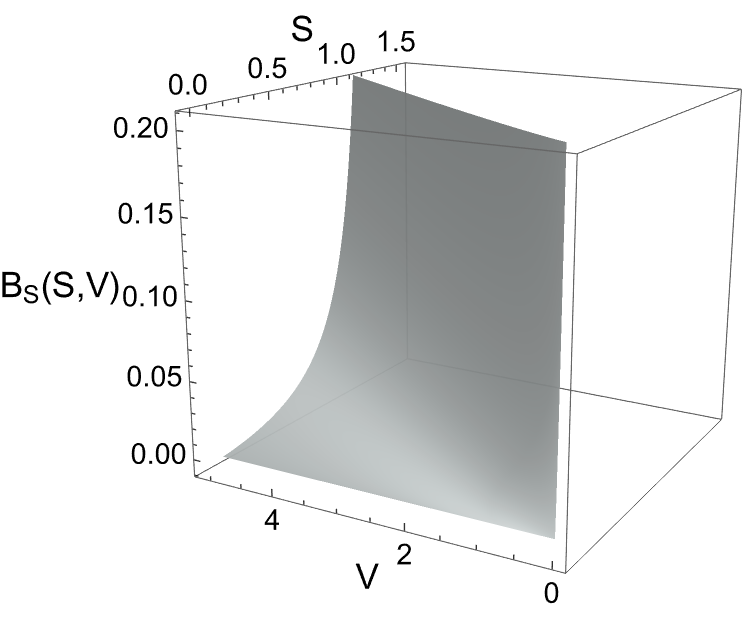}
    
    \caption{
   \emph{Energy representation.}
    Top row from left to right: the fundamental relation \(E(S,V)\) and the equations of state \(T(S,V)\) and \(P(S,V)\).
    The physical domain  is \(4GS<V\). \textcolor{black}{Bottom row: the isochoric thermal stiffness $K_T^{(E)}$, the isentropic thermo-mechanical coupling $\chi_{SV}^{(E)}$, and the adiabatic bulk modulus $B_S$.  The stiffness $K_T^{(E)}$ is negative for the small black hole branch, vanishes at the merger $x=2/3$, and is positive for the large black hole branch. Moreover, $\chi_{SV}^{(E)}<0$ and $B_S>0$ throughout the physical state space. The response coefficients become singular as the horizon approaches the cavity wall, $x\to1$.}
    }
    \label{fig:schwarzschild-energy-representation}
\end{figure}

In terms of the dimensionless variable $x$ \eqref{ratiorhrb},
the three independent Hessian components in the energy representation are
\begin{align}
\mathcal{H}^{(E)}_{SS}
&=
\left(\frac{\partial T}{\partial S}\right)_V
=
\frac{G(3x-2)}
{2\sqrt{\pi}\,V^{3/2}x^3(1-x)^{3/2}}\,, \label{firstequation51}
\\
\mathcal{H}^{(E)}_{SV}
=
\mathcal{H}^{(E)}_{VS}
&=
\left(\frac{\partial T}{\partial V}\right)_S
=
-\left(\frac{\partial P}{\partial S}\right)_V
=
-\frac{1}
{8\sqrt{\pi}\,V^{3/2}(1-x)^{3/2}}\,,
\\
\mathcal{H}^{(E)}_{VV}
&=
-\left(\frac{\partial P}{\partial V}\right)_S
=
\frac{
3x^2+4x\sqrt{1-x}-6x-4\sqrt{1-x}+4
}
{32\sqrt{\pi}G\,V^{3/2}(1-x)^{3/2}} \,.
\end{align}
The diagonal Hessian components $\mathcal{H}^{(E)}_{SS}$ and
$\mathcal{H}^{(E)}_{VV}$ define the thermal and mechanical stiffness
coefficients \eqref{stiffnessenergyrep}, respectively, while the
off-diagonal component describes thermo-mechanical coupling.

The isochoric thermal stiffness $K_T^{(E)}=\mathcal{H}^{(E)}_{SS}$ changes sign and vanishes at
$x=2/3$, i.e. at the photon sphere, which coincides with the point where
the temperature reaches its minimum. For $x<2/3$ one has
$\mathcal{H}^{(E)}_{SS}<0$, whereas for $x>2/3$ one finds
$\mathcal{H}^{(E)}_{SS}>0$, implying positive thermal stiffness. Thus,
the same transition that separates the small and large black hole
branches in the Helmholtz representation appears in the energy
representation as a change in sign of the thermal stiffness.

The mechanical stiffness $K_P^{(E)}
=\mathcal{H}^{(E)}_{VV}$ is positive for all allowed values of $S$ and
$V$,   and therefore the adiabatic bulk modulus,
\begin{equation}
    B_S=V\mathcal{H}^{(E)}_{VV}\,,
\end{equation}
is positive everywhere. Hence, the Schwarzschild black hole in a
spherical cavity is mechanically stable under adiabatic compression
throughout the state space.

\textcolor{black}{
These thermal and mechanical stability statements refer to the
corresponding restricted classes of perturbations. Indeed, the full
Hessian of $E(S,V)$ is indefinite throughout the physical state space,
with
\begin{equation}
\det \mathcal{H}^{(E)}
=
-\frac{
\left(1-\sqrt{1-x}\right)^2
\left(1+2\sqrt{1-x}\right)
}{
16\pi V^3 x^3(1-x)^{3/2}
}
<0\,,
\end{equation}
implying it has one positive and one negative eigenvalue. The fundamental relation $E(S,V)$ is therefore neither convex nor
concave as a function of both variables. Positivity of either diagonal
stiffness separately should thus not be interpreted as full
two-variable thermodynamic stability.
}

\textcolor{black}{
The off-diagonal Hessian component describes thermo-mechanical coupling.
In particular, $\chi_{SV}^{(E)}=\mathcal H_{SV}^{(E)}
=(\partial T/\partial V)_S$ is the isentropic thermo-mechanical coupling coefficient. This is negative throughout the
physical state space, so increasing the cavity volume at fixed
entropy lowers the wall temperature. It diverges as $x\to1$, when the
horizon approaches the cavity wall. The response functions are
visualized in the bottom row of
Figure~\ref{fig:schwarzschild-energy-representation}.
}

\subsection{Helmholtz representation}\label{sec:helmholtzresponse}

In the Helmholtz representation the independent variables are $(T,V)$, and one should regard
$x=x(T,V)$ as either the small or large canonical branch, cf. \eqref{smalllargebranches}. The Hessian matrix components of the
Helmholtz free energy are
\begin{align}
\mathcal{H}^{(F)}_{TT}
&=
-\left(\frac{\partial S}{\partial T}\right)_V
=
-\frac{2\sqrt{\pi}V^{3/2}x^3(1-x)^{3/2}}{G(3x-2)}=-\frac{1}{4\pi G T^3(3x-2)}\,,
\\
\mathcal{H}^{(F)}_{TV}
=
\mathcal{H}^{(F)}_{VT}
&=
-\left(\frac{\partial S}{\partial V}\right)_T
=
-\left(\frac{\partial P}{\partial T}\right)_V
=
-\frac{x^3}{4G(3x-2)}\,,
\\
\mathcal{H}^{(F)}_{VV}
&=
-\left(\frac{\partial P}{\partial V}\right)_T
=
-\frac{2(1-x)^{3/2}+3x-2}
{8\sqrt{\pi}G\,V^{3/2}(3x-2)} \,.
\end{align}
The constant-volume heat capacity follows from
\begin{equation}
C_V
=
-T\mathcal{H}^{(F)}_{TT}=\frac{T}{K_T^{(E)}}=\frac{1}{4 \pi G T^2(3 x-2)}\,.
\end{equation}
Similarly, the isothermal bulk modulus is
\begin{equation}
B_T
=
V\mathcal{H}^{(F)}_{VV} =
-\frac{2(1-x)^{3/2}+3x-2}
{8\sqrt{\pi}G\,\sqrt{V}(3x-2)} \,.
\end{equation}
\textcolor{black}{Thirdly, the thermal pressure coefficient equals \begin{equation}\label{gammavschw}\gamma_V=-\mathcal{H}_{TV}^{(F)}= \frac{x^3}{4G(3x-2)}\,.\end{equation}} 
\textcolor{black}{
This response function also reflects the different extensivity properties
of the two Helmholtz branches discussed in Sec.~\ref{sec:helmholtzrepr} and Ref.~\cite{Borsboom:2026ash}.
At fixed temperature and $V\to\infty$, the canonical relation
\begin{equation}
x^2(1-x)=\frac{1}{4\pi VT^2}
\end{equation}
implies
\begin{equation}
x_{\rm small}\sim \frac{1}{2\sqrt{\pi}\,T\sqrt{V}},
\qquad
1-x_{\rm large}\sim \frac{1}{4\pi VT^2}.
\end{equation}
Consequently,
\begin{equation}\label{eq:gammaVtozero}
\gamma_V^{\rm small}
\sim
-\frac{1}{64\pi^{3/2}GT^3V^{3/2}}
\longrightarrow 0,
\qquad
\gamma_V^{\rm large}
\longrightarrow \frac{1}{4G}.
\end{equation}
Thus, on the large black hole branch, which becomes extensive in the
Helmholtz representation, the thermal pressure coefficient approaches a
finite, volume-independent value, whereas on the small black hole branch,
which remains non-extensive in this representation, it vanishes.} 

\begin{figure}[t]
    \centering
    \includegraphics[width=0.3\textwidth]{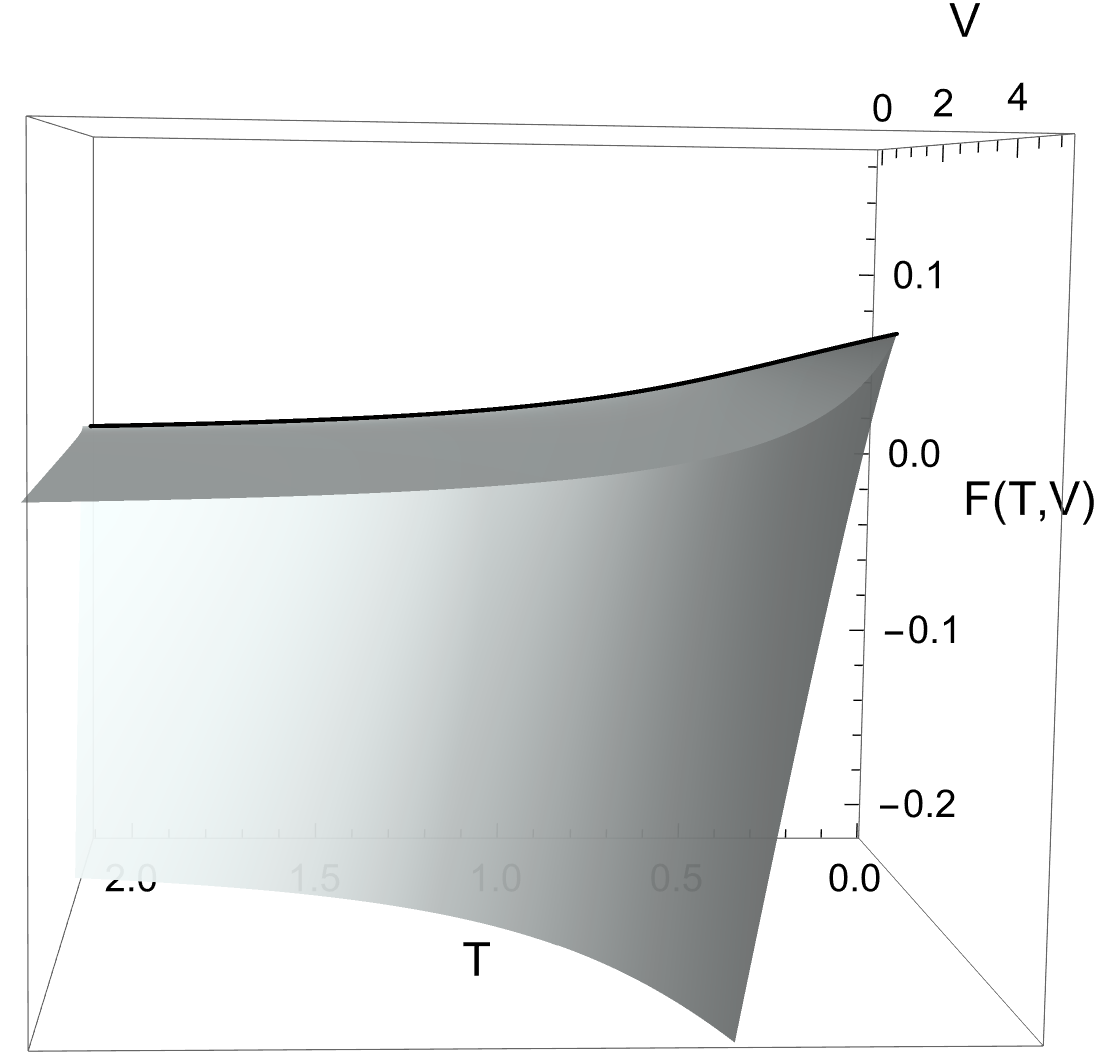}
    \includegraphics[width=0.36\textwidth]{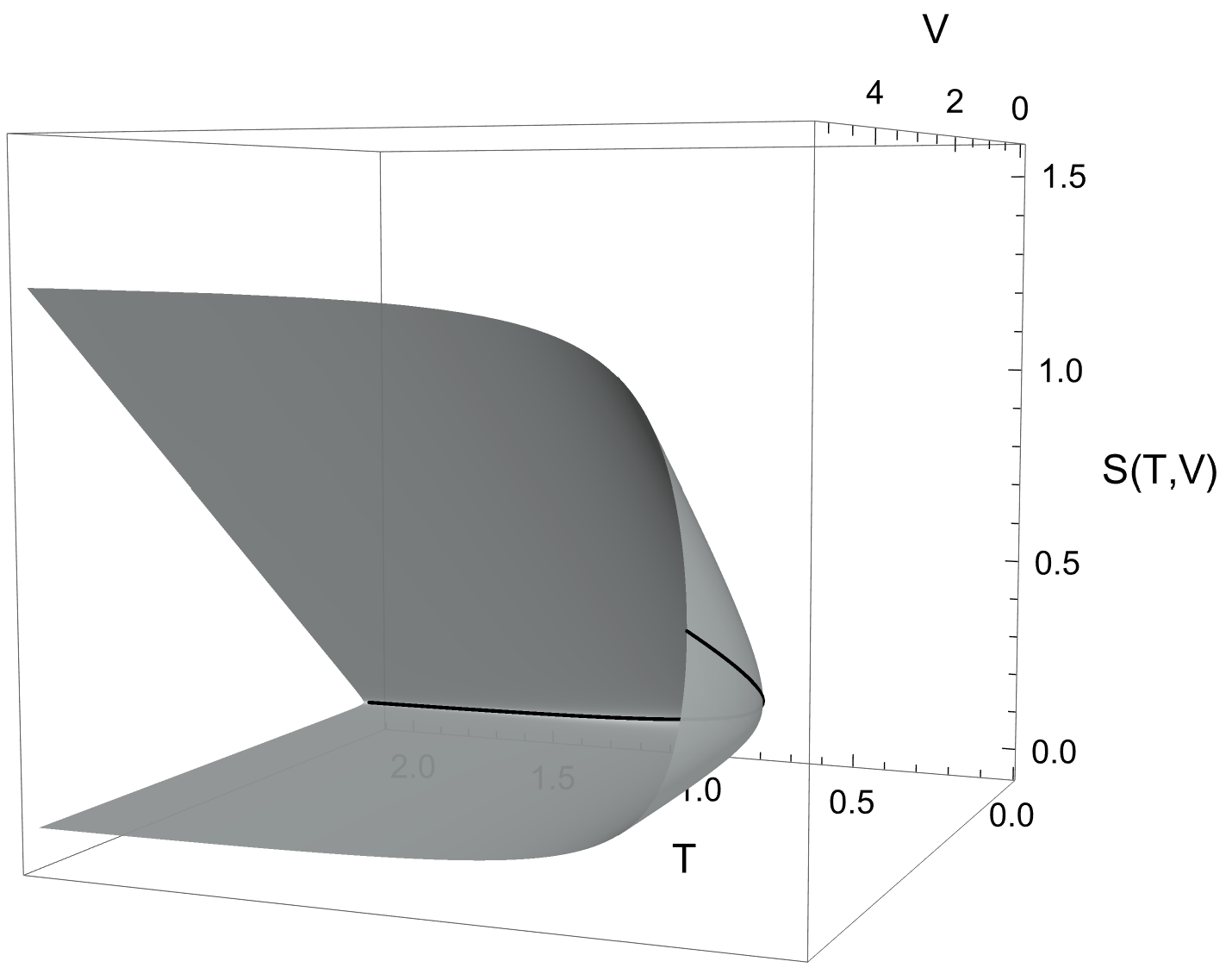}
    \includegraphics[width=0.30\textwidth]{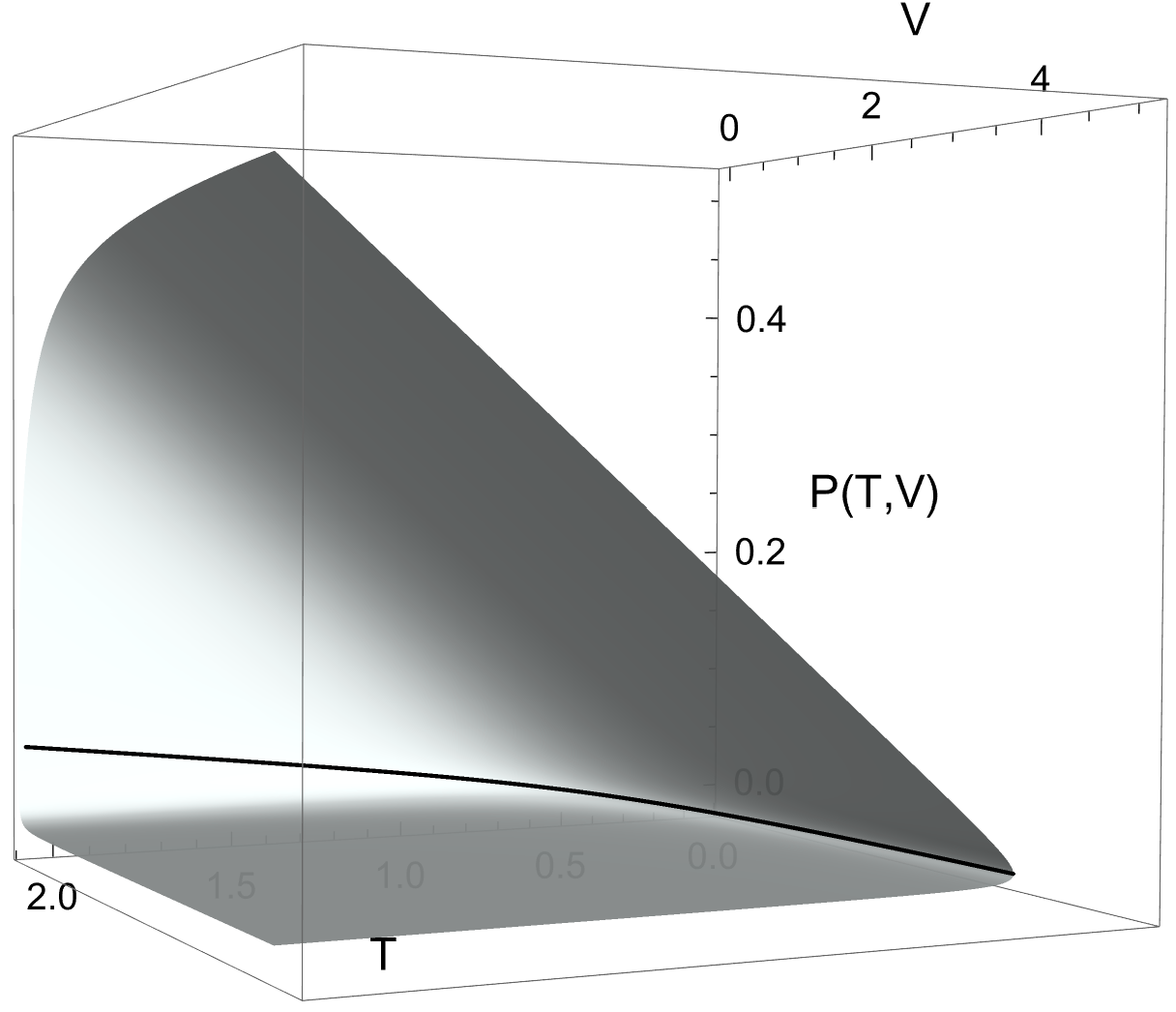}

    \includegraphics[width=0.33\textwidth]{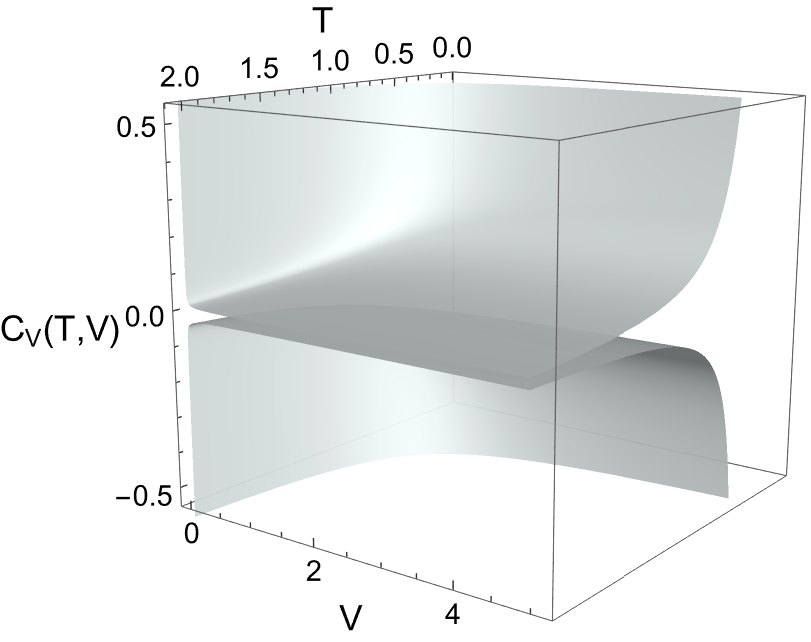}
    \includegraphics[width=0.32\textwidth]{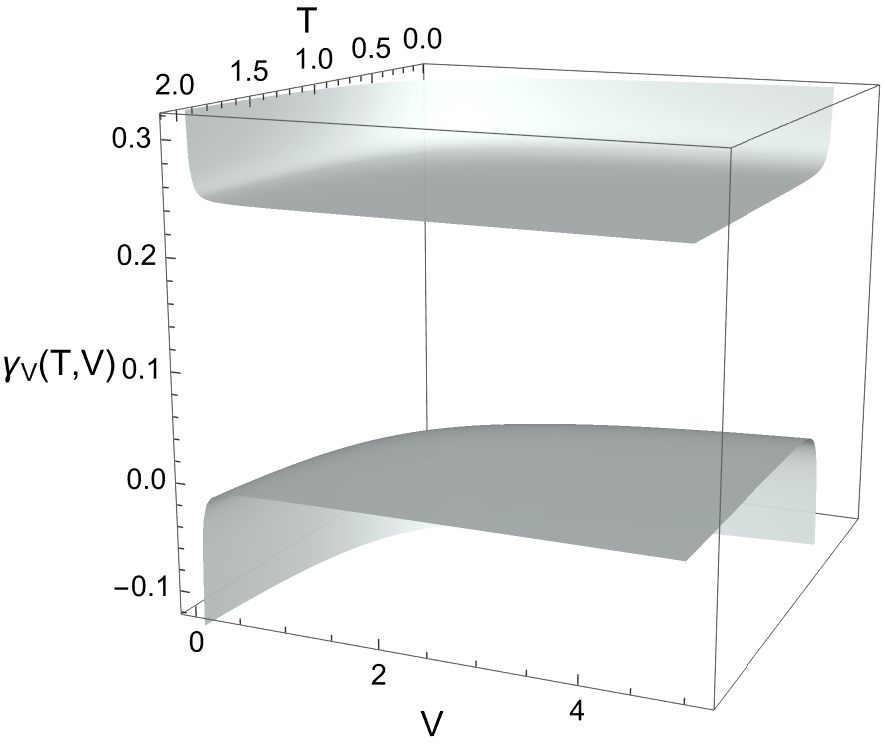}
    \includegraphics[width=0.33\textwidth]{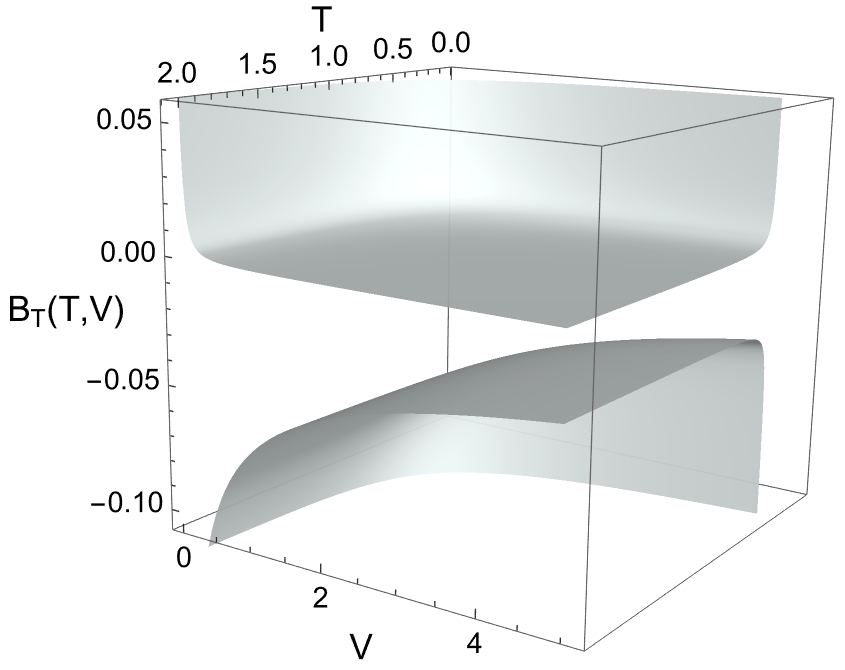}
    
    \caption{
\emph{Helmholtz representation.}
Top row from left to right: the fundamental relation $F(T,V)$ and the equations of state $S(T,V)$ and $P(T,V)$. \textcolor{black}{Bottom row: the isochoric heat capacity $C_V$, the thermal pressure coefficient $\gamma_V$, and the isothermal bulk modulus $B_T$. The physical domain is $V>0$ and $T\geq T_{\min}(V)=\sqrt{27}/(4\sqrt{\pi V})$. The black curve marks the locus $x=2/3$, where the small and large black
hole branches merge and the projection onto the Helmholtz variables
$(T,V)$ has a fold singularity. There $F$ remains finite, whereas
$C_V$, $\gamma_V$, and $B_T$ all diverge. The first two response coefficients are negative on the small branch and positive on the large branch, while $B_T$ has the opposite sign.}}
\label{fig:schwarzschild-canonical-representation}
\end{figure}

All Helmholtz response functions diverge at
$ 
x=2/3,
$ 
corresponding to the curve
\begin{equation}
T= T_\text{min}=\frac{\sqrt{27}}{4 \sqrt{\pi V}}\,,
\end{equation}
where the small and large black hole branches merge. As is well known, for
$x<2/3$ one has $C_V<0$, so the small black hole is thermally unstable, while for $x>2/3$ one has $C_V>0$, giving local thermal stability of the large black hole branch. \textcolor{black}{Analogous behavior is exhibited by $\gamma_V$.} By contrast, $B_T$ changes sign in the opposite way: the small branch has positive isothermal bulk modulus, whereas the large branch has negative isothermal bulk modulus. Thus, the large black hole is thermally stable at fixed volume but mechanically unstable under isothermal compression. This  behavior is visualized in the bottom row of Figure \ref{fig:schwarzschild-canonical-representation}, which clearly showcases the two branches.

% \textcolor{black}{The relation between the fold at $x=2/3$ and the behavior of response function is illuminated further by the contact geometric viewpoint, introduced in Appendix \ref{appendixContactGeom}. When the required local invertibility condition  for the transformation $(S,V)\to (T(S,V),V)$ fails, the transformed potential may cease to provide a globally single-valued graph over its chosen independent variables, thus failing to parametrize the Legendre submanifold of equilibrium states as in Eq. \eqref{eq:legendresubmfd}.}

\textcolor{black}{
The relation between the fold at $x=2/3$ and the behavior of the Helmholtz
response functions is further clarified by the contact geometric viewpoint
reviewed in Appendix~\ref{appendixContactGeom}. A Legendre transformation
does not change the underlying equilibrium Legendre submanifold. What can
become singular is its projection onto the independent variables of a
particular thermodynamic representation. If the induced map
$(S,V)\to(T(S,V),V)$ loses local invertibility, the equilibrium Legendre
submanifold can no longer be represented locally as a smooth graph over the
Helmholtz variables $(T,V)$, as in Eq.~\eqref{eq:legendresubmfd}.
}

\textcolor{black}{
For the quasi-local Schwarzschild black hole, the Helmholtz projection
becomes singular when the map $(S,V)\to(T(S,V),V)$ ceases to be locally
invertible. Its Jacobian is
}
\begin{align} \label{jacobian516}
    \frac{\partial (T,V)}{\partial (S,V)}
    =
    \begin{pmatrix}
        \mathcal{H}_{SS}^{(E)} & \mathcal{H}_{SV}^{(E)} \\
        0 & 1
    \end{pmatrix}
    =
    \begin{pmatrix}
        K_T^{(E)} & \chi_{SV}^{(E)} \\
        0 & 1
    \end{pmatrix},
\end{align}
\textcolor{black}{
with determinant $K_T^{(E)}$. Hence the isochoric thermal stiffness
controls the local invertibility of the Helmholtz projection.} \textcolor{black}{\textcolor{black}{
Since $K_T^{(E)}$ vanishes at $x=2/3$, the Helmholtz projection becomes
singular there. For a map of the form
$(S,V)\mapsto(T(S,V),V)$, such a singularity is a \emph{fold} \cite{GolubitskyGuillemin1973,gilmore1993catastrophe} if the turning
point in the $S$-direction is non-degenerate, i.e.
$(\partial T/\partial S)_V=0$ while
$(\partial^2T/\partial S^2)_V\neq0$. In the present case,
$(\partial T/\partial S)_V=K_T^{(E)}$, and
Eq.~\eqref{firstequation51} shows that $K_T^{(E)}$ has a simple zero at
$x=2/3$. Since $(\partial x/\partial S)_V=2G/(Vx)\neq0$, it follows that
$(\partial K_T^{(E)}/\partial S)_V
=(\partial^2T/\partial S^2)_V\neq0$ there. The singularity of the
Helmholtz projection at $x=2/3$ is therefore a fold.
}}
% Since
% $K_T^{(E)}$ vanishes at $x=2/3$, this projection becomes singular there.
% As discussed in Sec.~\ref{sec:helmholtzrepr}, the resulting singularity is a fold. 

\textcolor{black}{
When the potential coordinate $F$ is retained, the two Helmholtz sheets
coalesce along the curve corresponding to the fold locus in the plotted
$(T,V,F)$ surface, as shown in Figure~\ref{fig:schwarzschild-canonical-representation}. The Helmholtz
potential therefore ceases to furnish a globally single-valued smooth
graph over $(T,V)$. At the fold, $F$ and its first derivatives remain
finite, while its second derivatives diverge. All in all the fold is a
singularity of the Helmholtz projection, rather than of the equilibrium
Legendre submanifold itself.}

\textcolor{black}{
The loss of local invertibility also explains why the three Helmholtz
response functions diverge along the same fold. Away from the fold,
differentiation with respect to $T$ at fixed $V$ can be written as
}
\begin{equation}
    \left(\frac{\partial}{\partial T}\right)_V
    =
    \frac{1}{K_T^{(E)}}
    \left(\frac{\partial}{\partial S}\right)_V .
\end{equation}
\textcolor{black}{
Consequently, the Helmholtz response functions can be expressed directly
in terms of derivatives of the energy as
}
\begin{equation}
    C_V=\frac{T}{K_T^{(E)}},
    \qquad
    \gamma_V=-\frac{\chi_{SV}^{(E)}}{K_T^{(E)}},
    \qquad
    B_T=\frac{V\det\mathcal{H}^{(E)}}{K_T^{(E)}} .
\end{equation}
\textcolor{black}{
At $x=2/3$, the isochoric thermal stiffness
$K_T^{(E)}=\mathcal{H}_{SS}^{(E)}$, which appears in the denominator of
all three expressions, vanishes. By contrast, $T$,
$\chi_{SV}^{(E)}=\mathcal{H}_{SV}^{(E)}$, and
$\det\mathcal{H}^{(E)}$ remain finite and non-zero. Hence the divergences
of $C_V$, $\gamma_V$, and $B_T$ have the same origin: the loss
of local invertibility of the projection onto the Helmholtz variables
$(T,V)$.
}

\subsection{Gibbs representation}

\begin{figure}[t]
    \centering
    \includegraphics[width=0.32\textwidth]{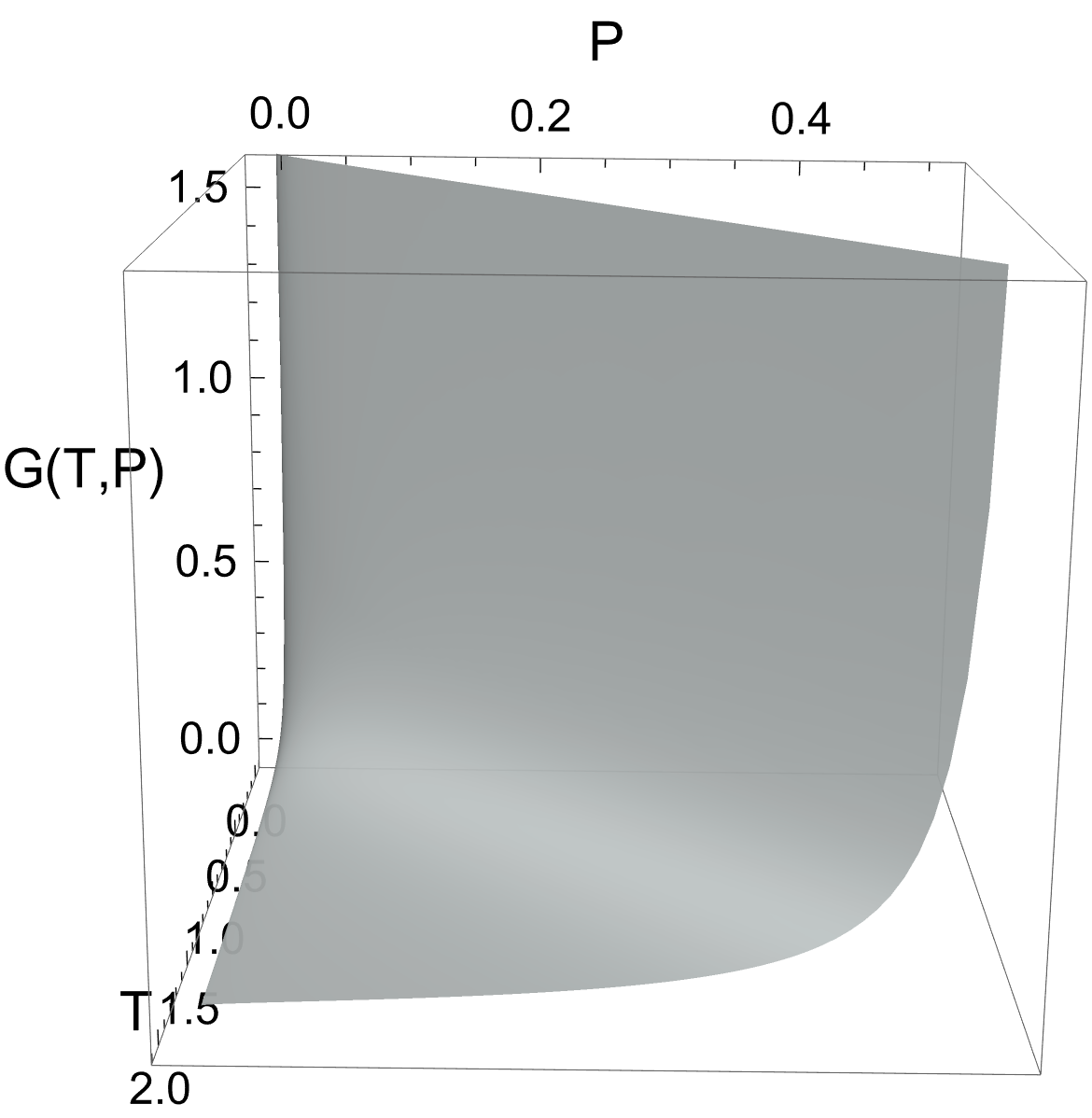}
    \includegraphics[width=0.35\textwidth]{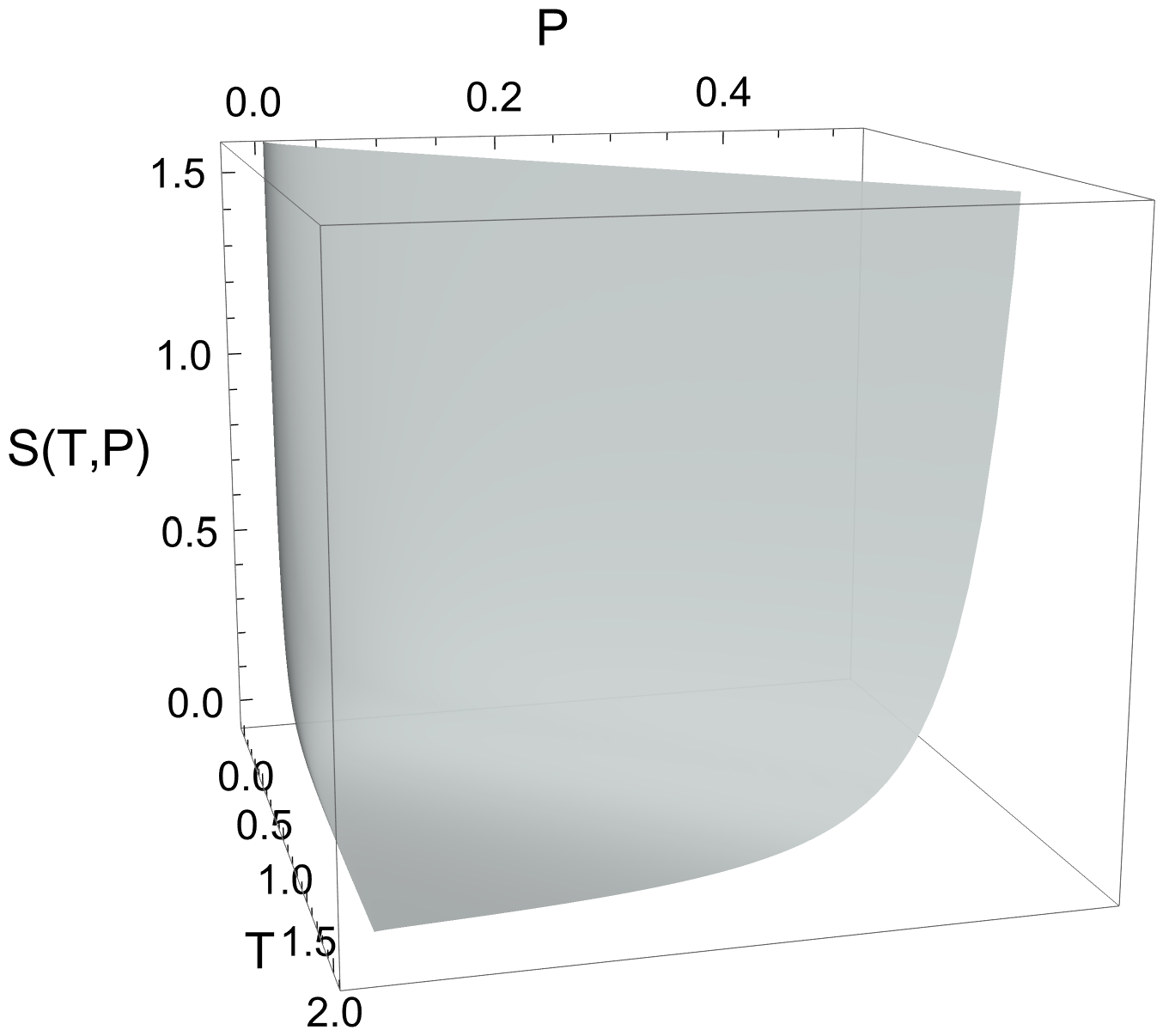}
    \includegraphics[width=0.31\textwidth]{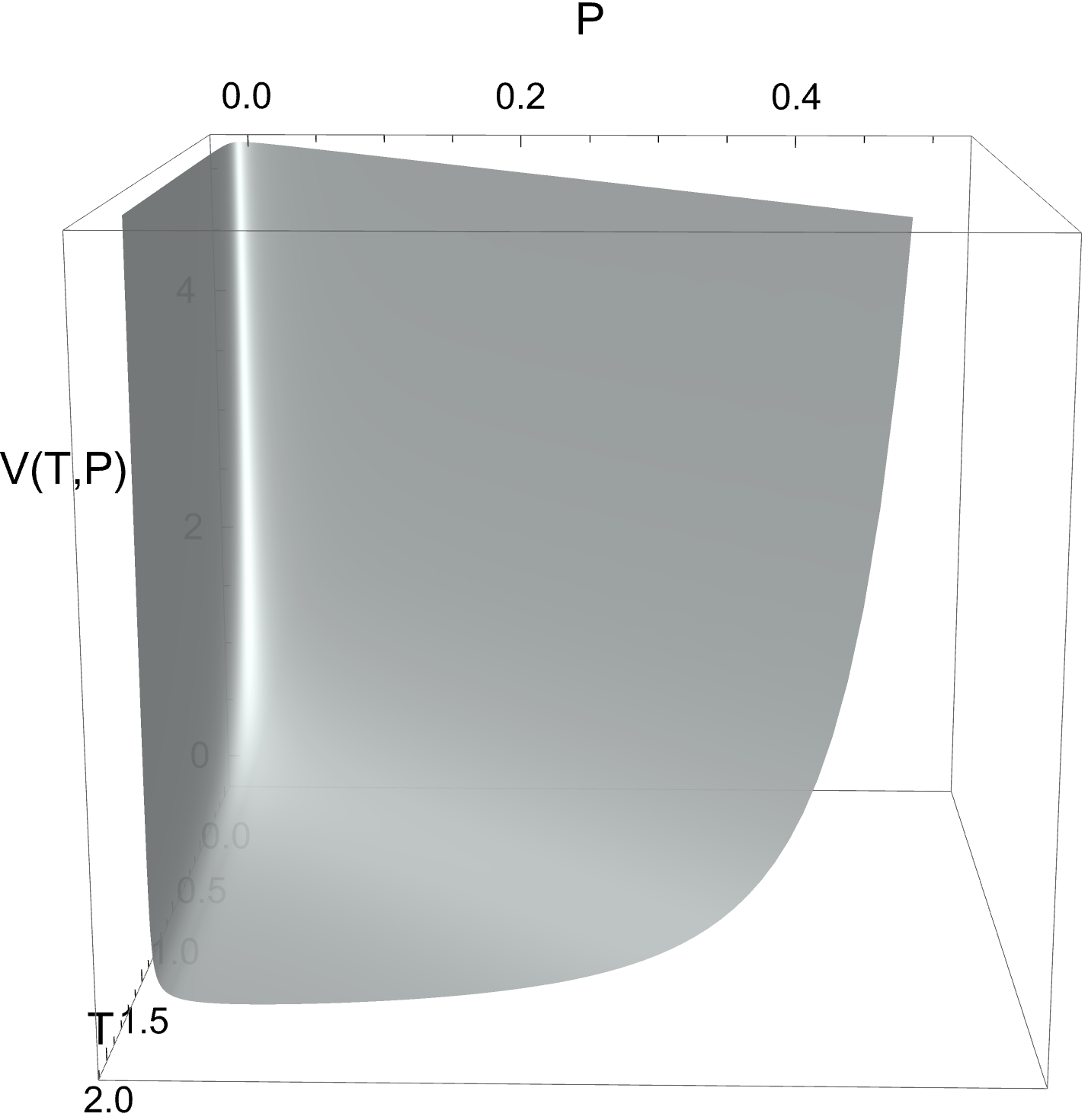}

    \includegraphics[width=0.32\textwidth]{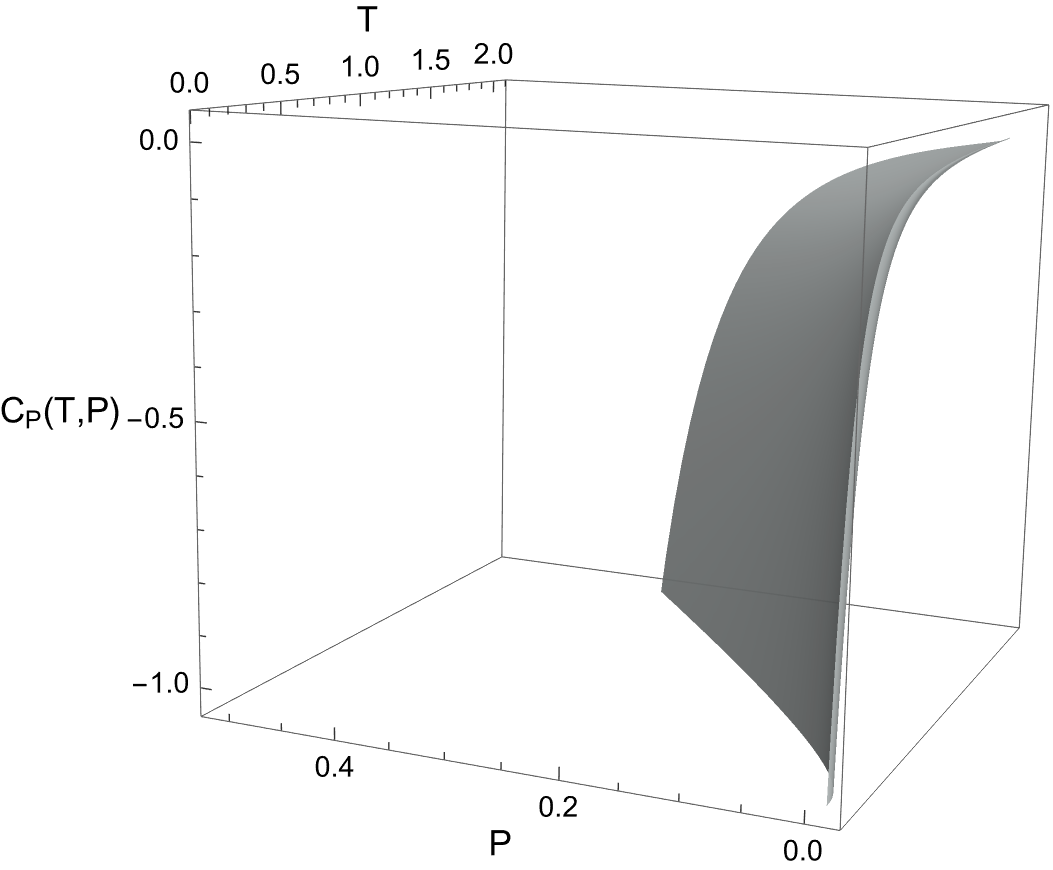}
    \includegraphics[width=0.33\textwidth]{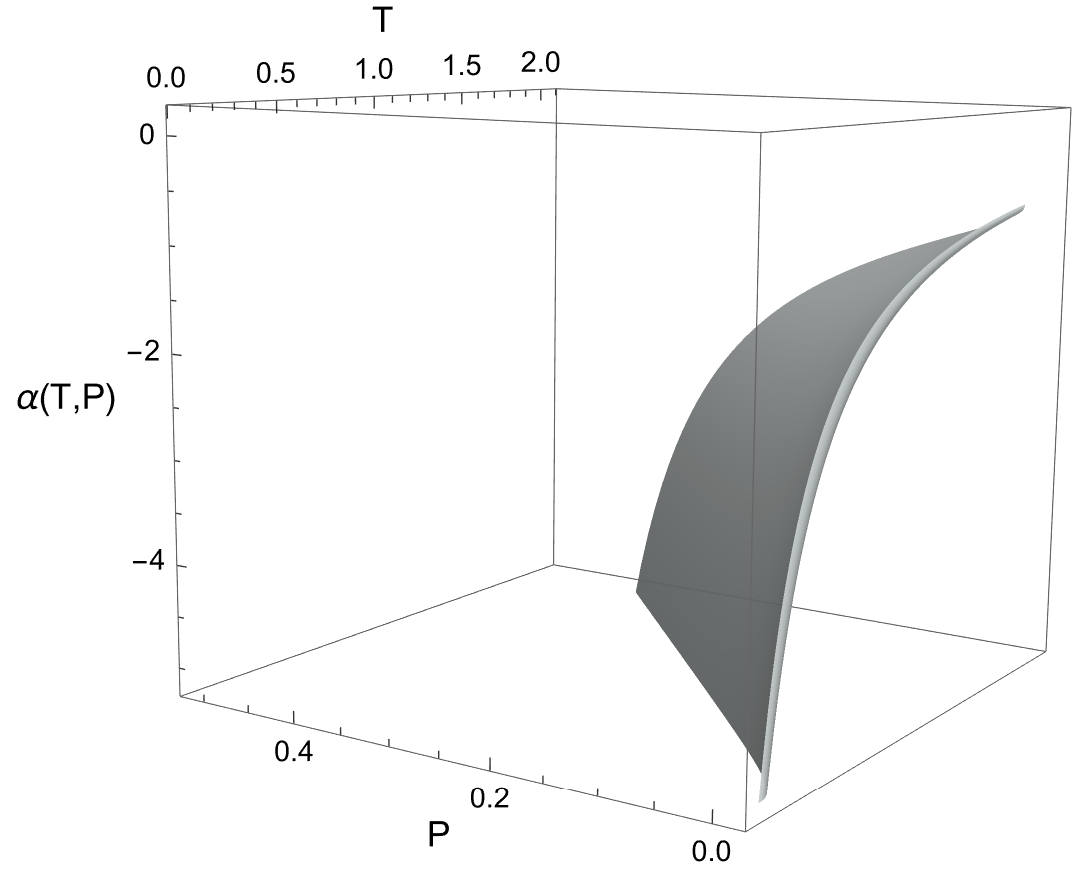}
    \includegraphics[width=0.33\textwidth]{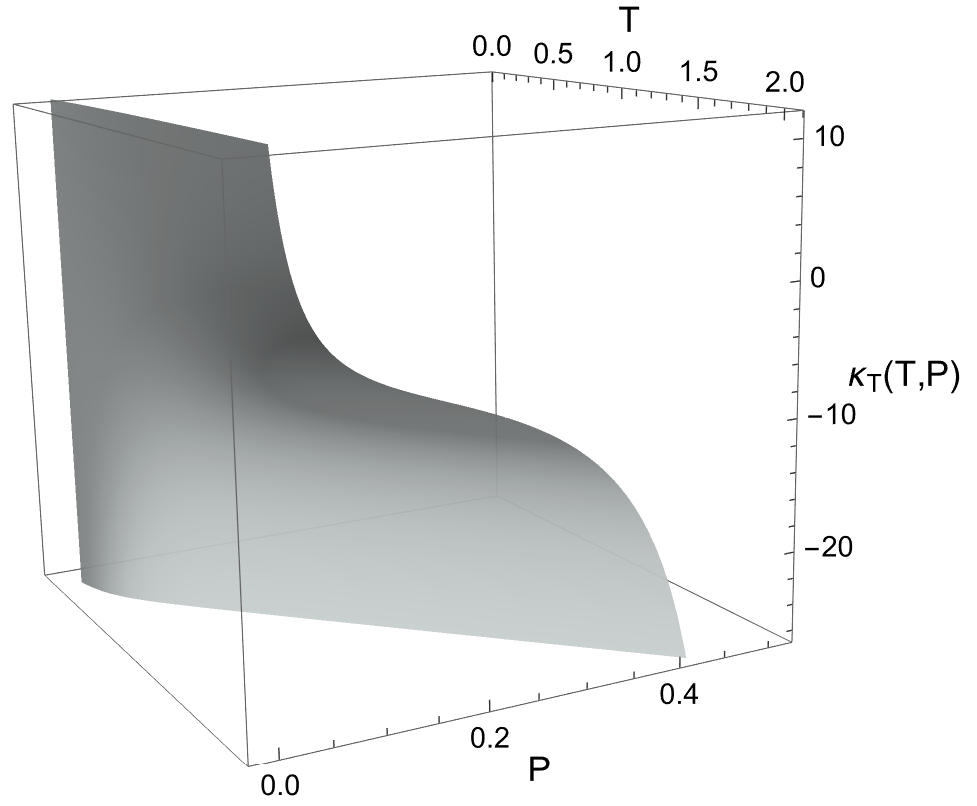}
    
    \caption{
    \emph{Gibbs representation.}
Top row from left to right: the fundamental relation \(G(T,P)\) and the equations of state \(S(T,P)\) and \(V(T,P)\).
    The physical domain is \(0<P<T/4G\). \textcolor{black}{Bottom row: the isobaric heat capacity $C_P$, the thermal expansion coefficient $\alpha$, and the isothermal compressibility $\kappa_T$.   Both $C_P$ and $\alpha$ are negative throughout the physical domain. The compressibility $\kappa_T$ is positive for   the small black hole branch, vanishes at the merger $x=2/3$, and is negative for the large black hole branch.
}
    }
    \label{fig:schwarzschild-gibbs-representation}
\end{figure}

In the Gibbs representation the independent variables are $(T,P)$. In terms of  the function  
$y(T,P)$ in \eqref{eq:yexplicit}
the   Hessian components of the Gibbs free energy are
\begin{align}
\mathcal{H}^{(G)}_{TT}
&=
-\left(\frac{\partial S}{\partial T}\right)_P
=
\frac{3y^2+2y+1}
{16\pi G\,T^3 y^3(1+2y)}\,,
\\
\mathcal{H}^{(G)}_{TP}
=
\mathcal{H}^{(G)}_{PT}
&=
\left(\frac{\partial V}{\partial T}\right)_P
=
-\left(\frac{\partial S}{\partial P}\right)_T
=
-\frac{1}
{4\pi T^3 y^3(1-y)^2(1+2y)}\,,
\\
\mathcal{H}^{(G)}_{PP}
&=
\left(\frac{\partial V}{\partial P}\right)_T
=
-\frac{3y^2-1}
{4\pi P\,T^2 y^3(1-y^2)^2(1+2y)}\,,
\end{align}
where $S$ and $V$ are functions of $T$ and $P$, cf. \eqref{eq:gibbseos}. 
The corresponding response functions are therefore
\begin{align}
C_P
&=
-T\mathcal{H}^{(G)}_{TT}
=
-\frac{3y^2+2y+1}
{16\pi G\,T^2 y^3(1+2y)}\,,
\\
\alpha
&=
\frac{1}{V}\mathcal{H}^{(G)}_{TP}
=
-\,\frac{(1+y)^2}
{T\,y(1+2y)}\,,
\\
\kappa_T
&=
-\frac{1}{V}\mathcal{H}^{(G)}_{PP}
=
\frac{3y^2-1}
{P\,y(1+2y)}\,.
\end{align}
Since $3y^2+2y+1>0$ and $y(1+2y)>0$ for $y\in(0,1)$, one has $C_P<0$ for all allowed values in Eq.~\eqref{allowedregimenow2}, i.e. $0<P<T/4G$. \textcolor{black}{Hence the Schwarzschild black hole is thermally unstable at fixed pressure
throughout the Gibbs representation, even though the large black hole
branch is thermally stable at fixed volume in the Helmholtz representation.
This agrees with the instability of the canonical pressure ensemble found
by Comer~\cite{Comer:1992pc}.} It illustrates that thermodynamic stability depends on the chosen representation and therefore on the allowed equilibrium fluctuations.

The thermal expansion coefficient also clearly satisfies $\alpha<0$ everywhere. Hence, at fixed pressure, increasing the temperature decreases the cavity volume. Further, the isothermal compressibility  $\kappa_T$ changes sign at $y=1/\sqrt{3}$,
which coincides with the photon sphere $r_h=2r_B/3$. For $y>1/\sqrt{3}$ (small black hole) one has $\kappa_T>0$, whereas for $y<1/\sqrt{3}$ (large black hole) one finds $\kappa_T<0$, which is the opposite sign pattern to that of $C_V$ in the Helmholtz representation, with the difference that there $C_V$ diverges at the critical value. This sign change was already identified by York \cite{york1986}, who analyzed \(\kappa_T\) in terms of the cavity area and thus within the Helmholtz representation, whereas it is more naturally interpreted as a Gibbs-representation response function. \textcolor{black}{
The response functions are shown in the bottom row of
Figure~\ref{fig:schwarzschild-gibbs-representation}, with the isothermal
compressibility $\kappa_T$ on the right, clearly exhibiting this smooth
change of sign. Since
$\kappa_T=-V^{-1}(\partial V/\partial P)_T$, the negative compressibility
of the large black hole states implies
$(\partial V/\partial P)_T>0$, or equivalently
$(\partial P/\partial V)_T>0$. An isothermal compression therefore lowers
the system pressure, so that an external pressure maintaining mechanical
equilibrium drives further compression rather than restoring the original
volume. These states are thus mechanically unstable under isothermal
volume perturbations.
}

\textcolor{black}{The relation to Comer's stability analysis in \cite{Comer:1992pc} can now
be made explicit. In the canonical pressure ensemble, both the quasi-local
energy and the cavity area are allowed to fluctuate. Comer analyzes local stability by expanding the reduced Euclidean action
to second order around the equilibrium configuration. Positivity of this
second variation for all allowed fluctuations is the condition that the
equilibrium be a local minimum rather than a saddle. The first condition
obtained in this way is the fixed-area canonical stability condition,
corresponding to $C_V>0$. Once this condition is satisfied, stability under
the additional area fluctuations yields a second condition, namely positive isothermal mechanical
stiffness 
$K_P^{(F)}=-(\partial P/\partial V)_T>0$. Away from the singular locus,
this is equivalent to $\kappa_T>0$. The small black hole is already
thermally unstable under fixed-area fluctuations, since $C_V<0$. The large
black hole instead satisfies $C_V>0$, but becomes unstable once area
fluctuations are admitted, because $K_P^{(F)}<0$, equivalently
$\kappa_T<0$. Our response function analysis thus provides a direct
thermodynamic interpretation of Comer's result that the canonical pressure
ensemble is nowhere locally stable. In particular, for the large black
hole the instability corresponds to a negative fluctuation direction of
the reduced Euclidean action once boundary-area fluctuations are admitted.
The equilibrium is therefore a saddle rather than a local minimum in the
canonical pressure ensemble.
}

% \textcolor{black}{
% The relation to Comer's stability analysis in \cite{Comer:1992pc} can now
% be made explicit. In the canonical pressure ensemble, both the quasi-local
% energy and the cavity area are allowed to fluctuate. The first condition
% obtained from the second variation of the reduced Euclidean action is
% precisely the fixed-area canonical stability condition, corresponding  
% to $C_V>0$. Once this condition is satisfied, stability under the
% additional area fluctuations requires positive isothermal mechanical
% stiffness, $-(\partial P/\partial V)_T>0$, equivalently
% $\kappa_T>0$ wherever the inverse relation is regular. The small black hole
% therefore fails the thermal condition, $C_V<0$, whereas the large black
% hole satisfies $C_V>0$ but fails the mechanical condition,
% $\kappa_T<0$. Our response function analysis thus provides a direct
% thermodynamic interpretation of Comer's result that the canonical
% pressure ensemble is nowhere locally stable. In particular, for the large
% black hole the instability corresponds to a negative fluctuation direction
% of the reduced Euclidean action that appears once boundary-area
% fluctuations are admitted. The equilibrium is therefore a locally unstable saddle in the
% canonical pressure ensemble rather than a metastable local minimum.  
% }

\subsection{Enthalpy representation}\label{sec:enthalpyresponse}

\textcolor{black}{
Like the Gibbs representation but unlike the Helmholtz representation,
the enthalpy representation is single-valued for the Schwarzschild black
hole. The odd status of the Helmholtz representation can be seen by
comparing its projection with those onto the enthalpy and Gibbs variables.
The Jacobian determinants of the latter two are
}
\begin{equation}
    \det\frac{\partial(S,P)}{\partial(S,V)}
    =
    \left(\frac{\partial P}{\partial V}\right)_S
    =
    -\mathcal{H}_{VV}^{(E)}
    \neq 0\,,
    \qquad
    \det\frac{\partial(T,P)}{\partial(S,V)}
    =
    -\det\mathcal{H}^{(E)}
    \neq 0 \, .
\end{equation}
\textcolor{black}{
Since $\mathcal{H}_{VV}^{(E)}>0$ and
$\det\mathcal{H}^{(E)}<0$ throughout the interior of the physical state
space, the projections onto both $(S,P)$ and $(T,P)$ remain locally regular. Thus, unlike the Helmholtz projection, neither develops a fold. This local regularity is consistent with the global single-valuedness of the enthalpy and Gibbs representations, and confirms that the Helmholtz fold is a property of the chosen projection rather than of the equilibrium Legendre submanifold (cf. Appendix \ref{appendixContactGeom})
itself.
}

\begin{figure}[t]
    \centering
    \includegraphics[width=0.3\textwidth]{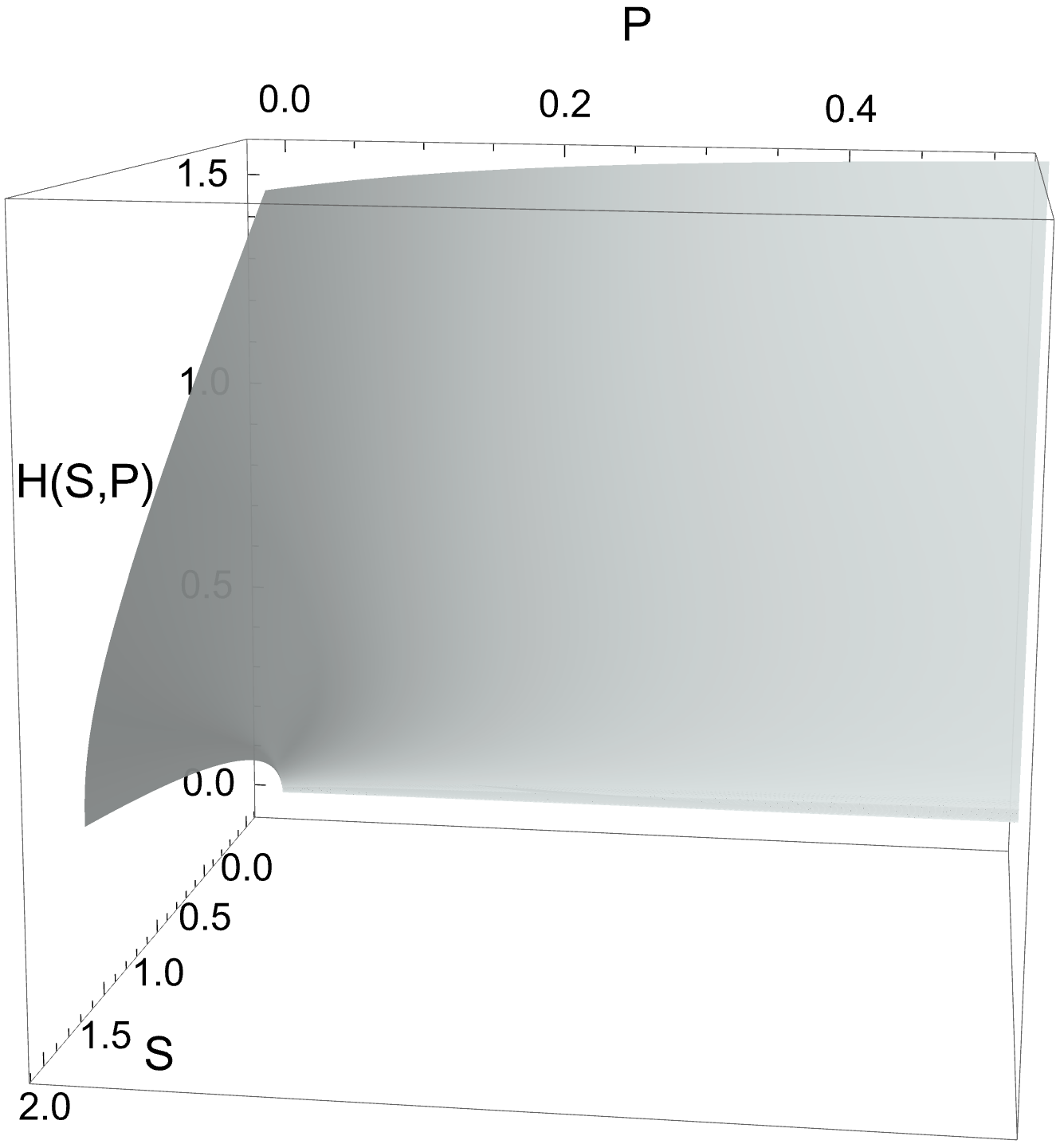}
    \includegraphics[width=0.34\textwidth]{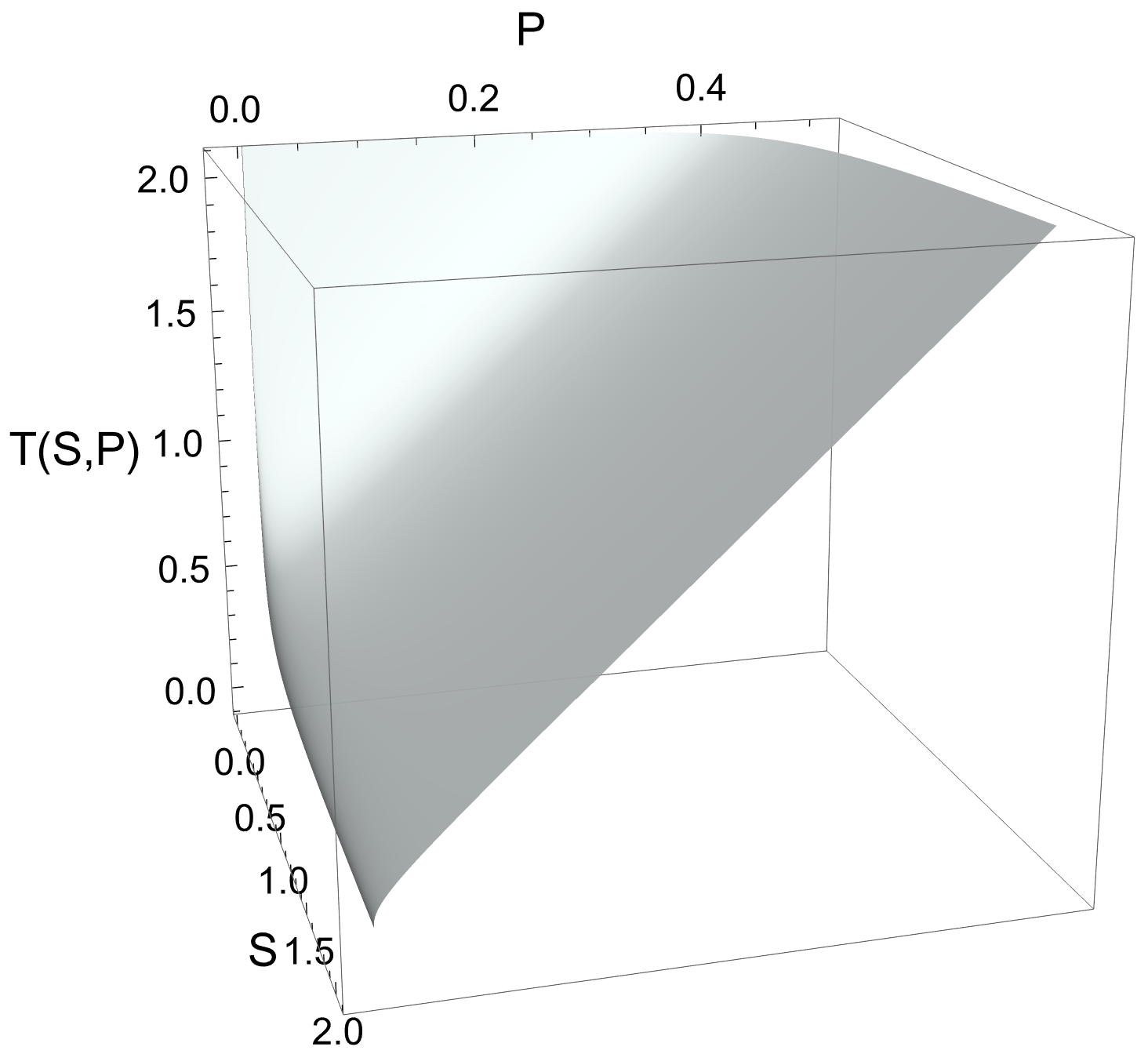}
    \includegraphics[width=0.34\textwidth]{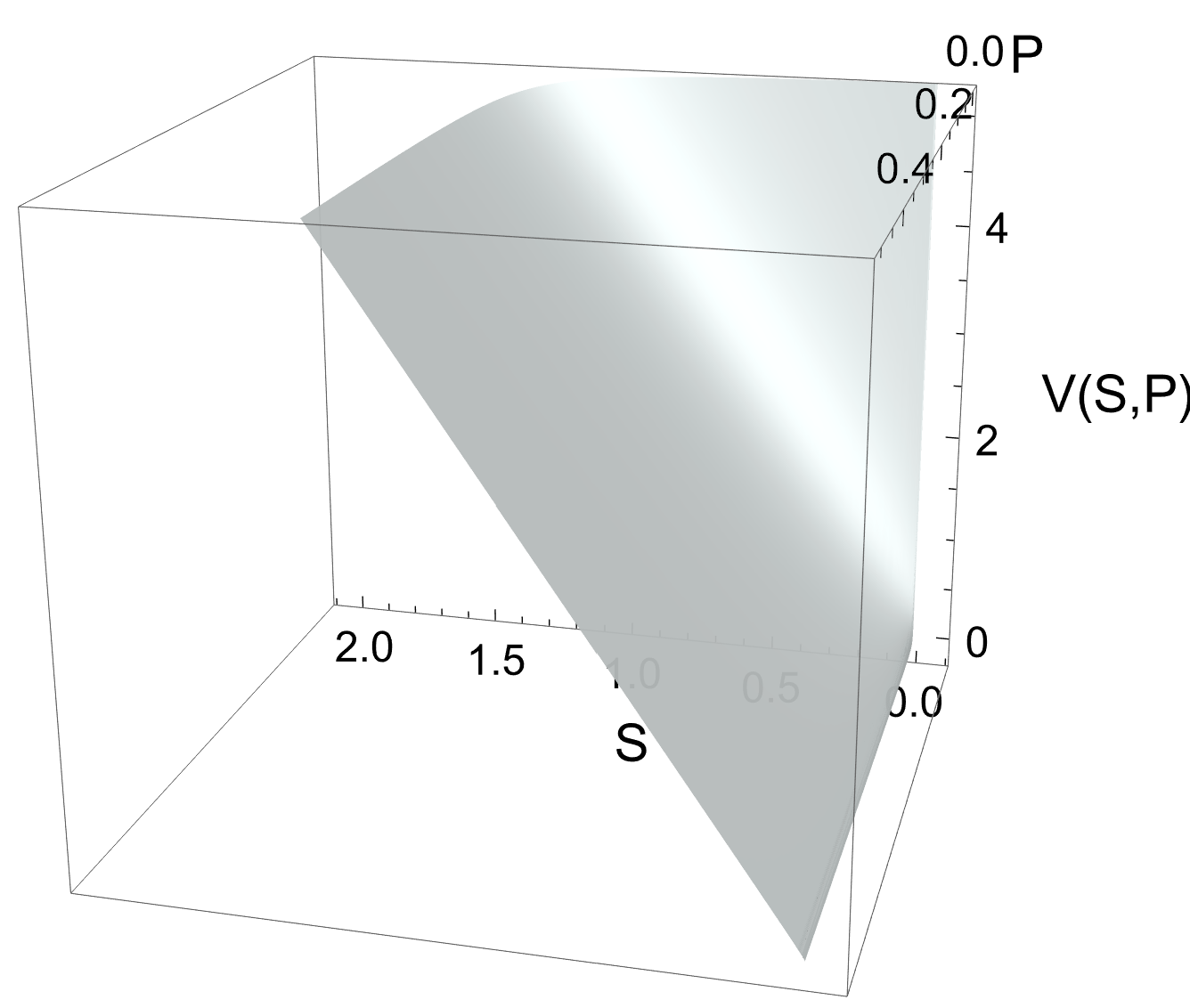}

    \includegraphics[width=0.35\textwidth]{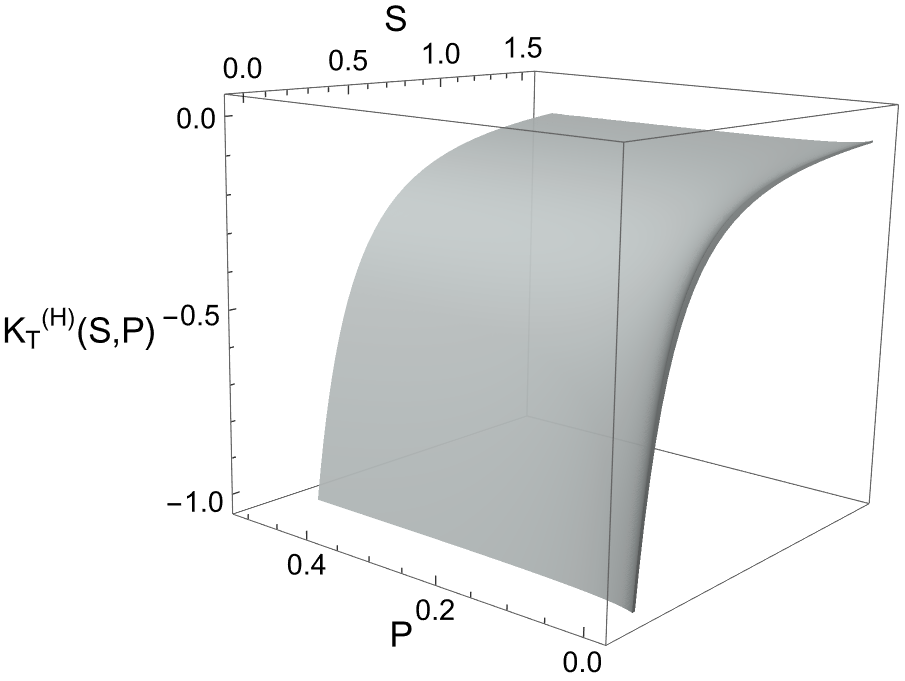}
    \includegraphics[width=0.31\textwidth]{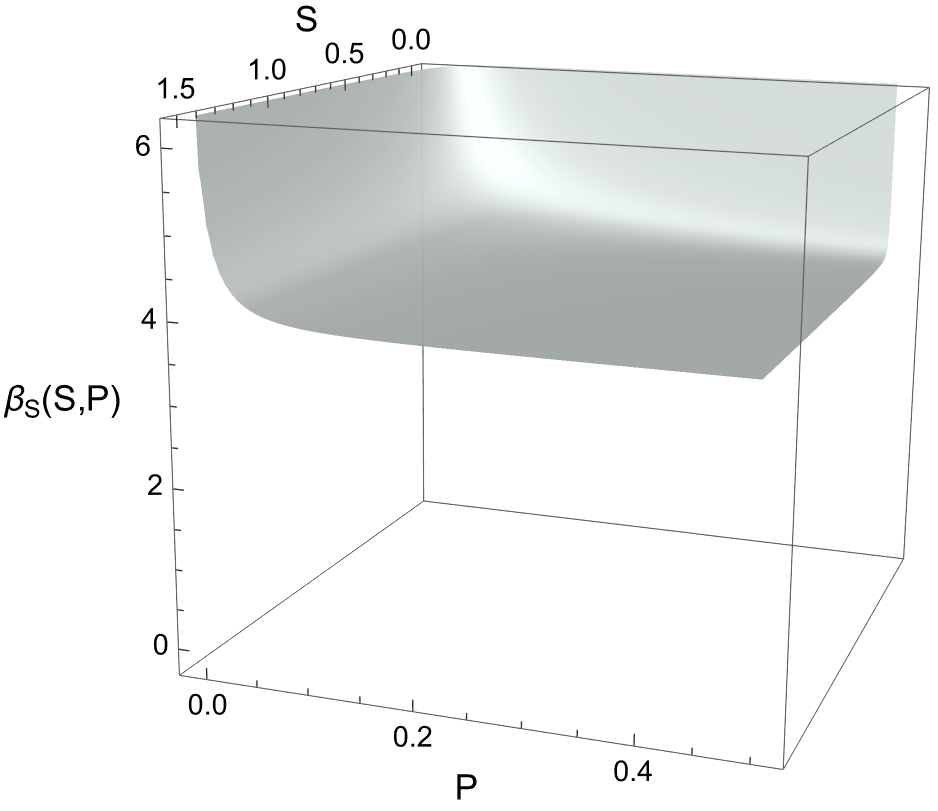}
    \includegraphics[width=0.32\textwidth]{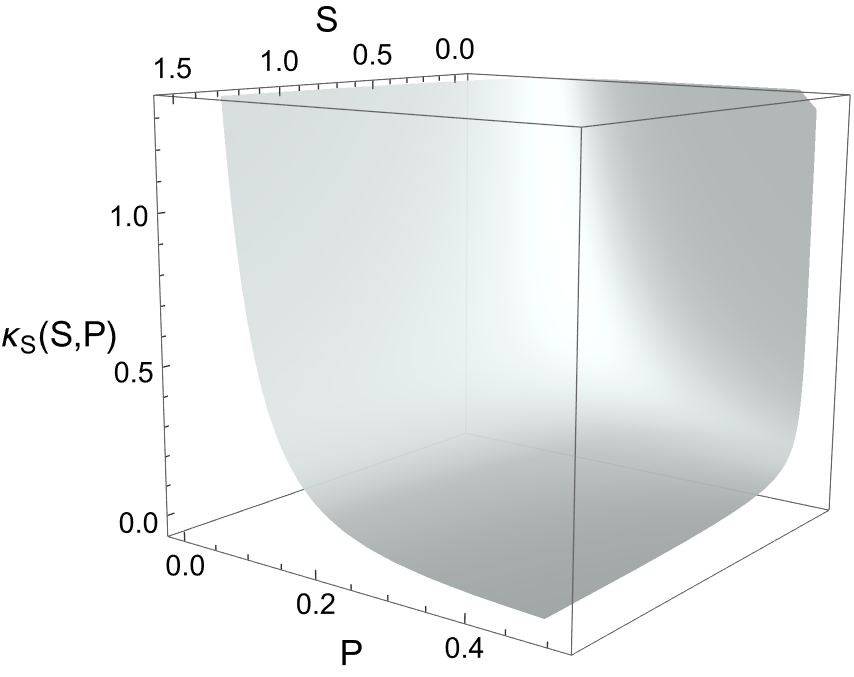}
    
    \caption{
    \emph{Enthalpy representation.}
    Top row from left to right: the fundamental relation~\(H(S,P)\) and equations of state \(T(S,P)\) and \(V(S,P)\).
    The physical domain is~\(S>0\) and \( P>0\).
    The zero-pressure edge corresponds to the limit \(V\to\infty\) at fixed entropy. Along this edge \(H\) approaches the asymptotically flat Schwarzschild mass and therefore remains non-vanishing, as seen in the left plot. \textcolor{black}{Bottom row: the isobaric thermal stiffness $K_T^{(H)}$, the isentropic temperature-pressure coefficient $\beta_S$, and the adiabatic compressibility $\kappa_S$. The physical domain is $S>0$ and $P>0$. One has $K_T^{(H)}<0$, while $\beta_S>0$ and $\kappa_S>0$ throughout the domain, showing thermal instability at fixed pressure but mechanical stability under adiabatic compression. As $P\to0$ at fixed entropy,   $H$ approaches the asymptotically flat Schwarzschild mass, and $\beta_S$ and $\kappa_S$ diverge.}
}
    \label{fig:schwarzschild-enthalpy-representation}
\end{figure}

In the enthalpy representation the three independent Hessian components in terms of $y(S,P)$ are
\begin{align}
\mathcal{H}^{(H)}_{SS}
&=
\left(\frac{\partial T}{\partial S}\right)_P
=
-\frac{1+2y}
{4S\sqrt{\pi G S}\left(3y^2+2y+1\right)}\,,
\\
\mathcal{H}^{(H)}_{SP}
=
\mathcal{H}^{(H)}_{PS}
&=
\left(\frac{\partial T}{\partial P}\right)_S
=
\left(\frac{\partial V}{\partial S}\right)_P
=
\frac{1-y^2}
{4P\sqrt{\pi G S}\,y\left(3y^2+2y+1\right)}\,,
\\
\mathcal{H}^{(H)}_{PP}
&=
\left(\frac{\partial V}{\partial P}\right)_S
=
-\frac{16GS\,y^2}
{P\left(1-y^2\right)^2
\left(3y^2+2y+1\right)} \,.
\end{align}
The isobaric thermal stiffness in the enthalpy representation,   $
K_T^{(H)} =T / C_P=
\mathcal{H}^{(H)}_{SS}
$, 
is negative everywhere, so the Schwarzschild black hole is thermally unstable at fixed pressure.  The mechanical response is encoded in the adiabatic compressibility
\begin{equation}
\kappa_S
=
-\frac{1}{V}
\left(\frac{\partial V}{\partial P}\right)_S
=
-\frac{1}{V}\mathcal{H}^{(H)}_{PP}=
\frac{4y^2}
{P(3y^2+2y+1)} \,.
\end{equation}
Hence $\kappa_S>0$ everywhere.
Thus the system is mechanically stable under adiabatic compression.
\textcolor{black}{
Finally, the isentropic temperature-pressure coefficient
$\beta_S=\mathcal{H}_{SP}^{(H)}$ is positive everywhere. The three
response functions are visualized in
Figure~\ref{fig:schwarzschild-enthalpy-representation}.
}

\subsection{Gr\"{u}neisen parameter,  Joule-Thomson coefficient and speed of sound}\label{sec:gruneisenetc}

\begin{figure}[t]
    \centering
    \includegraphics[width=0.33\textwidth]{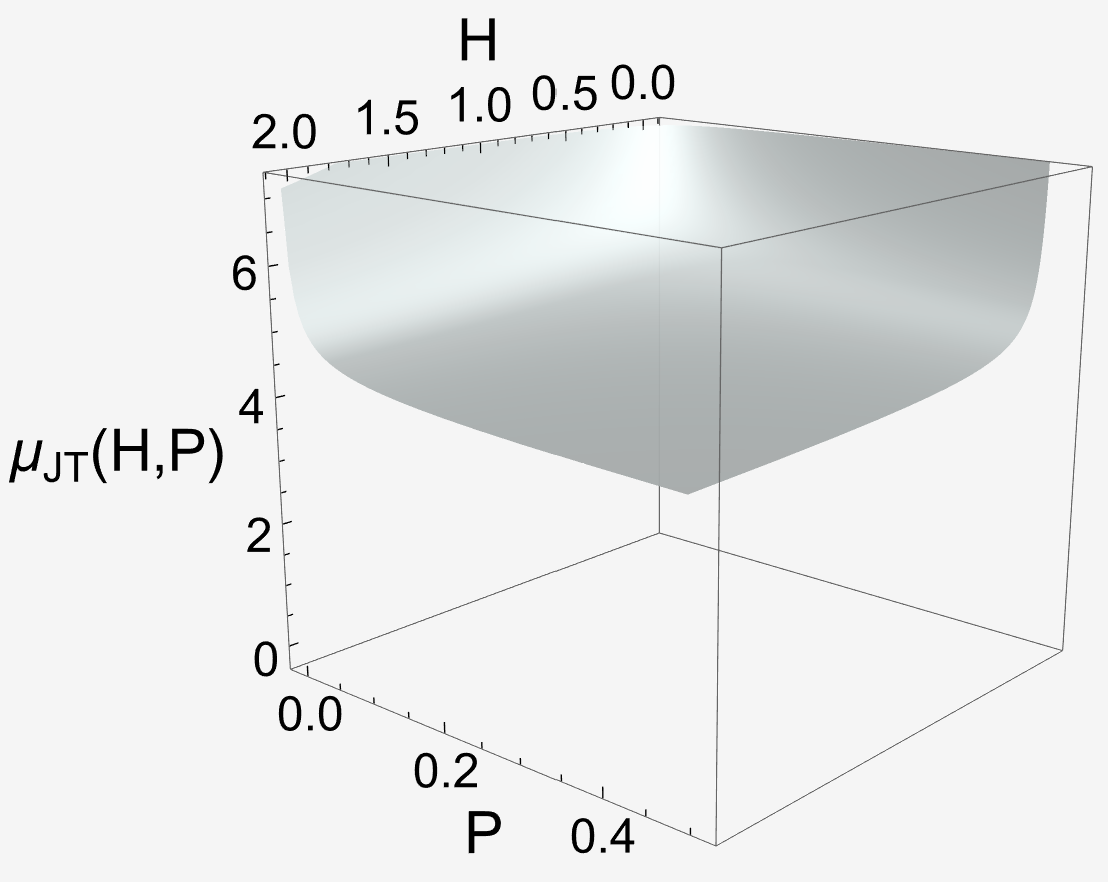}
    \includegraphics[width=0.325\textwidth]{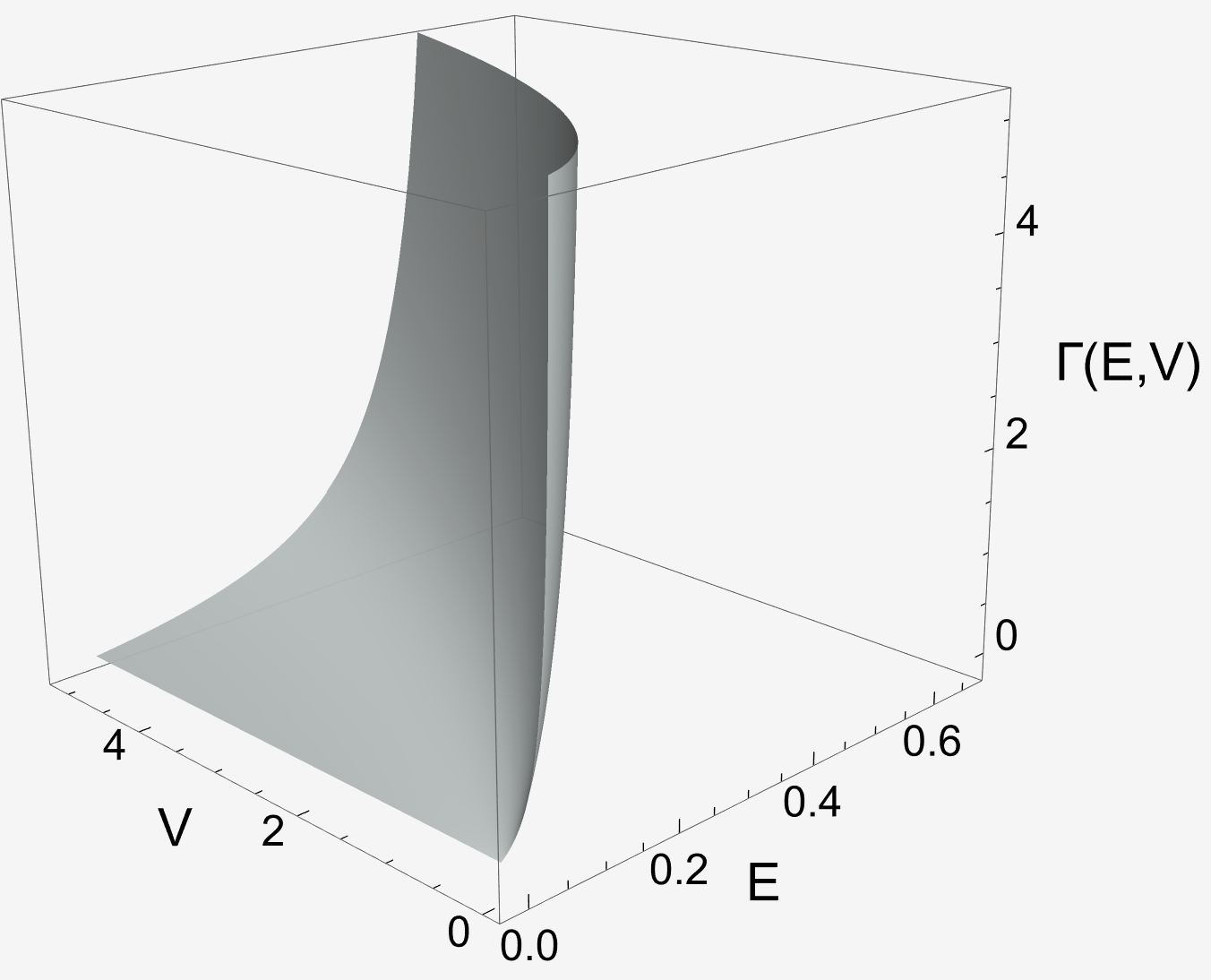}
    \includegraphics[width=0.325\textwidth]{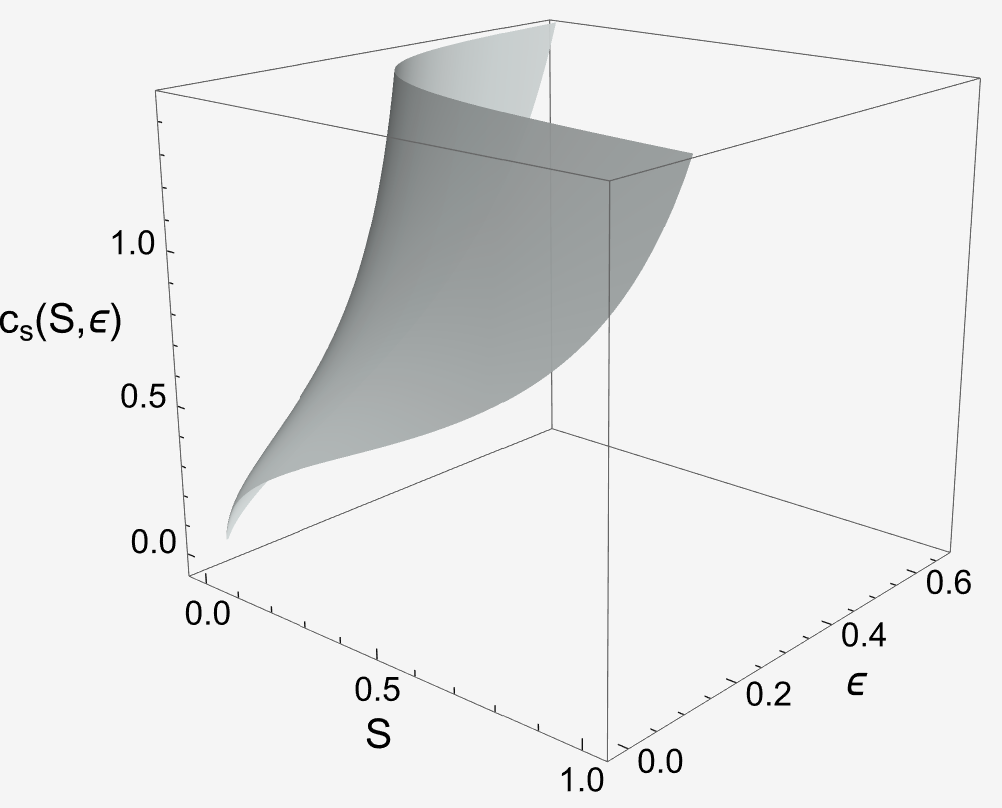}
    \caption{
    Derived thermodynamic response functions of the Schwarzschild black hole.
    From left to right: \(\mu_{\mathrm{JT}}(H,P)\), \(\Gamma(E,V)\) and \(c_s(S,\varepsilon)\).
    \textcolor{black}{The Joule-Thomson coefficient is positive throughout the physical state space, so isenthalpic expansion always cools the black hole and there is no finite inversion curve. Both $\Gamma$ and $c_s$ vanish in the small-black-hole limit $x\to0$ and diverge as the horizon approaches the cavity wall, $x\to1$. Moreover, $c_s>1$ for $x\gtrsim0.84$.}
    }
    \label{fig:schwarzschild-derived-response-functions}
\end{figure}

Some response functions are not tied to a single thermodynamic representation, but are obtained by combining the response coefficients derived above. Three useful examples are the Gr\"uneisen parameter, the Joule-Thomson coefficient and the speed of sound. The Joule-Thomson coefficient has been computed before for charged Anti-de Sitter black holes in the framework of black hole chemistry \cite{Okcu:2016tgt}, but as far as we are aware the Gr\"uneisen parameter has not been considered before in the context of black hole thermodynamics. The speed of sound has previously been studied in black hole chemistry \cite{Dolan:2011jm,Dolan:2012jh}, and in finite-cutoff AdS/CFT, where it governs signal propagation in the dual \(T\bar T\)-deformed boundary theory \cite{McGough:2016lol}. \textcolor{black}{These derived response functions are plotted in Figure \ref{fig:schwarzschild-derived-response-functions}.}

The Gr\"uneisen parameter measures the pressure response to an energy perturbation at fixed volume,
\begin{equation}
\Gamma
=
V\left(\frac{\partial P}{\partial E}\right)_V
=
\frac{\alpha V}{C_V\kappa_T}=-\frac{V}{T}\chi_{SV}^{(E)}\,.
\end{equation}
For the Schwarzschild black hole in a cavity this gives the simple expression
\begin{equation}
\Gamma
=
\frac{x}{4(1-x)}\,.
\label{eq:GammaSchwarzschild}
\end{equation}
where $x$ should be regarded as the function \(x=x(E,V)\).  Thus \(\Gamma>0\) throughout the state space: at fixed cavity size, increasing the quasi-local energy increases the pressure. Moreover, \(\Gamma\) diverges as \(x\to1\), where the horizon approaches the cavity wall.

The Joule-Thomson coefficient describes the temperature response during an isenthalpic pressure variation,
\begin{equation} \label{generalformulamujt}
\mu_{\rm JT}
=
\left(\frac{\partial T}{\partial P}\right)_H
=
\frac{V}{C_P}\left(T\alpha-1\right)\,.
\end{equation}
Its sign determines whether the system cools or heats during isenthalpic expansion. Since an
expansion corresponds to $dP<0$, a positive Joule-Thomson coefficient implies
cooling, whereas a negative coefficient implies heating.
The Joule-Thomson coefficient is naturally expressed in terms of the enthalpy variables \((H,P)\). Using the redshift parameter \(y=y(H,P)\), determined implicitly by
\begin{equation}
64\pi G^2PH
=
\frac{(1-y)^3(1+3y)}{y^2}\,,
\end{equation}
one finds
\begin{equation}
\mu_{\rm JT}
=
\frac{4G(1+3y+3y^2)}
{(1-y^2)^2(3y^2+2y+1)} \,.
\label{eq:JTSchwarzschild-y}
\end{equation}
All factors in \eqref{eq:JTSchwarzschild-y} are positive for \(0<y<1\), and therefore
\begin{equation}
\mu_{\rm JT}>0 \,.
\end{equation}
Hence an isenthalpic expansion always cools the quasi-local Schwarzschild black hole. There is no finite inversion curve, i.e.\ no curve in the \((P,T)\) plane separating regions of heating and cooling under isenthalpic expansion: the entire state space lies in the cooling region.

\textcolor{black}{
The absence of an inversion curve can also be understood directly from
Eq.~\eqref{generalformulamujt}. Since $C_P<0$ and $\alpha<0$ everywhere, one has $T\alpha-1<0$, and hence $\mu_{JT}>0$.
Equivalently, the inversion condition $T\alpha=1$ can never be satisfied.
The difference from the charged AdS result of~\cite{Okcu:2016tgt} should not be attributed simply to the absence
of a cosmological constant. In that case the charge introduces an
additional scale and modifies the equation of state in such a way that
$\mu_{JT}$ can change sign. Indeed, their inversion condition has no
positive horizon radius solution in the neutral limit. It would therefore be interesting to extend the present quasi-local
analysis to charged black holes and investigate whether the additional
charge scale can generate a finite inversion curve.
}

As a final derived response function, one may define the adiabatic speed of sound by
\begin{equation}
c_s^2=\left(\frac{\partial P}{\partial \varepsilon}\right)_S\,,
\qquad\text{with}\qquad
\varepsilon=\frac{E}{V}\,\label{speedofsound1}.
\end{equation}
\textcolor{black}{
In the holographic interpretation, $c_s$ is   interpreted as the
adiabatic sound speed of the boundary system living on the cavity wall,
characterizing the adiabatic pressure response to changes in the quasi-local
energy density. This interpretation is supported by the closely related
finite-cutoff AdS/CFT analysis of \cite{McGough:2016lol}, where the
same thermodynamic derivative was shown, through an independent dynamical
analysis, to coincide with the propagation speed of boundary excitations.
We therefore expect the same interpretation to apply here, although
establishing this equivalence directly for the asymptotically flat
Schwarzschild cavity would require a separate dynamical analysis.
}

The derivative in \eqref{speedofsound1} can be related to the adiabatic bulk modulus $B_S$ by viewing both the pressure~\(P\) and energy density \(\varepsilon\) as functions of \(V\) at fixed entropy. Since the energy changes under an adiabatic change of volume according to \((\partial E/\partial V)_S=-P\), one has
\begin{equation}
\left(\frac{\partial \varepsilon}{\partial V}\right)_S
=
-\frac{\varepsilon+P}{V}\,.
\end{equation}
Therefore
\begin{equation}
c_s^2
=
\frac{B_S}{\varepsilon+P}\,,
\qquad
B_S=-V\left(\frac{\partial P}{\partial V}\right)_S \,.
\end{equation}
For the Schwarzschild black hole in a cavity this gives
\begin{equation}
c_s^2
=
\frac{\left(1-\sqrt{1-x}\right)
\left(4-3x+2\sqrt{1-x}\right)}
{4(1-x)\left(1+3\sqrt{1-x}\right)}\,,
\end{equation}
where $x=r_h/r_B$ should be viewed as the function $x=x(\varepsilon,S)$.
Thus $c_s^2$ is positive throughout the physical state space, vanishes in the small black hole limit $x\to0$, \textcolor{black}{exceeds unity for $x\gtrsim0.84$,} 
and diverges as the horizon approaches the cavity wall.
This divergence means that an adiabatic perturbation produces an increasingly
large pressure response for a given change in quasi-local energy density as
the cavity boundary approaches the black hole horizon.
\textcolor{black}{
It indicates that the boundary system becomes increasingly stiff as
$x\to1$. This is also a singular limit of the quasi-local setup, since
the timelike cavity wall approaches the horizon and the local Tolman
temperature diverges.
}

\textcolor{black}{
The occurrence of $c_s^2>1$ deserves some further discussion, since if
$c_s$ is interpreted as a physical propagation speed it may appear to
conflict with relativistic causality. However, a superluminal
thermodynamic sound speed does not necessarily imply acausal signal
propagation. In the finite-cutoff AdS/CFT setting, McGough, Mezei and
Verlinde~\cite{McGough:2016lol} found precisely such a superluminal sound
speed and showed, through an independent dynamical analysis, that it
coincides with the propagation speed of boundary excitations. In their
description, these excitations propagate superluminally relative to the
fixed background metric on the cutoff surface, but remain causal with
respect to the fluctuating metric that governs their propagation. The
light cone of this fluctuating metric is modified by the stress tensor of
the state, and therefore need not coincide with that of the fixed
background metric. On the gravity side, the corresponding excitations
are boundary graviton modes rather than localized particles propagating
outside the relevant causal cone. Their analysis thus provides a closely
related example in which $c_s^2>1$ is compatible with causality.
}

\textcolor{black}{
We expect that an analogous interpretation may apply to the
  Schwarzschild cavity considered here. In particular,
nothing in the equilibrium thermodynamic analysis singles out the point
$c_s^2=1$, or equivalently $x\simeq0.84$, as a boundary of the physical
state space. Nevertheless, demonstrating explicitly that the
thermodynamic sound speed in Eq.~\eqref{speedofsound1} coincides with the
characteristic propagation speed of boundary perturbations, and
determining the corresponding effective causal structure, would require
a linearized dynamical analysis beyond the scope of the present work.
}

\section{Conclusion}
\label{sec:conclusion}

\textcolor{black}{
In thermodynamics, a choice of thermodynamic potential and independent
variables determines the equations of state: the conjugate quantities are
obtained as first derivatives of the potential and thereby become functions
of the chosen independent variables. The corresponding response functions
are obtained from second derivatives of the chosen potential. A non-trivial
thermal-mechanical response theory therefore requires an independent
mechanical degree of freedom in addition to the entropy. Standard
Schwarzschild black hole thermodynamics does not provide such a degree of
freedom, since it has no independently variable thermodynamic volume.
Extended black hole thermodynamics introduces a pressure by varying the
cosmological constant, but for static black holes the thermodynamic volume
is fixed by the entropy, so the entropy and volume do not constitute
independent thermodynamic variables \cite{Borsboom:2026ash}. The
finite cavity construction of York \cite{york1986} supplies the missing
degree of freedom: the entropy associated with the horizon and the area of
the cavity boundary can be varied independently. For this quasi-local
Schwarzschild black hole, the resulting response structure has not been
systematically investigated. In this article we therefore used the four
standard thermodynamic representations (energy, Helmholtz, Gibbs, and
enthalpy) to organize the equations of state and response functions of the
quasi-local Schwarzschild black hole, with the ideal gas serving as a
benchmark.
}

For the quasi-local Schwarzschild black hole, with the cavity area and
surface pressure interpreted as the thermodynamic volume and pressure of a
holographically dual system, the resulting stability properties are highly
representation dependent. \textcolor{black}{Each thermodynamic representation selects a
different set of independent thermodynamic variables, and hence a different
set of natural response coefficients, which measure the response of the
conjugate quantities to variations of one independent variable while the
others are held fixed.} This is reflected in the different sign behavior of the response functions: some are uniformly positive, some are uniformly negative, and others change sign between the small and large black hole branches. The named response functions calculated in this article, together with their natural representations and sign behavior, are summarized in Table~\ref{tab:schwarzschild-response-summary}. In particular, the sign of $C_V$ is reversed relative to both $B_T$ and $\kappa_T$: the large black hole branch is thermally stable at fixed volume, but mechanically unstable under isothermal compression. Thus, black hole stability depends not only on the equilibrium branch under consideration, but also on the thermodynamic representation and the corresponding variables held fixed.

% The independent pressure-volume sector is essential for this richer structure. In the standard asymptotically flat Schwarzschild description, the thermodynamic state space is effectively one-dimensional, so there is no independent notion of mechanical response. By contrast, the quasi-local description in a cavity makes it meaningful to distinguish thermal and mechanical stability. The resulting response theory also differs sharply from that of an ordinary ideal gas: the thermal expansion coefficient is negative rather than positive, isenthalpic expansion cools the black hole rather than leaving the temperature unchanged, and the adiabatic speed of sound diverges as the cavity boundary approaches the horizon rather than remaining finite. These features show that quasi-local black holes can display equilibrium response behavior that is qualitatively different from familiar thermodynamic  systems.

\textcolor{black}{
Beyond this representation-dependent stability structure, the quasi-local
response theory also differs sharply from that of an ordinary ideal gas.}
The thermal expansion coefficient is negative rather than positive,
isenthalpic expansion cools the black hole rather than leaving the
temperature unchanged, and the adiabatic speed of sound diverges as the
cavity boundary approaches the horizon rather than remaining finite. These
features illustrate that quasi-local black holes can exhibit equilibrium
response behavior qualitatively different from that of familiar
thermodynamic systems.

More broadly, the framework developed here provides a systematic route to deriving black hole equations of state and response functions using quasi-local gravitational thermodynamics. Although we focused on four-dimensional asymptotically flat Schwarzschild black holes, the same approach can be applied to charged \cite{Braden:1990hw} and rotating black holes \cite{Martinez:1994ja,Dehghani:2001af}, black holes coupled to matter \cite{Creighton:1995au,Astefanesei:2024wfj}, higher-dimensional solutions \cite{Andre:2020czm,Andre:2021ctu}, other theories of gravity \cite{Neri:2024qgb}, spacetimes with de Sitter or anti-de Sitter asymptotics \cite{Carlip2003,Simovic:2018tdy} and other horizon topologies \cite{Astefanesei:2005ad}. These extensions can also be combined, leading to a large class of quasi-local thermodynamic systems with different state spaces and stability structures. Charged static black holes in a cavity provide a particularly natural next step: the charge and electrostatic potential introduce an additional conjugate pair, while the charged solution has a richer branch structure than the Schwarzschild case \cite{Braden:1990hw,Carlip2003,Lundgren2008,Fernandes:2023byx,Fernandes:2025vzb}. Consequently, one obtains more thermodynamic representations, equations of state, and response functions, including coefficients that measure the response to variations of the charge or electrostatic potential.

\textcolor{black}{
A further direction would be to use the quasi-local response functions in
thermodynamic fluctuation theory and study the gravitational backreaction
of the resulting energy and entropy fluctuations. This may provide a
semiclassical point of contact with proposals by Sorkin \cite{Sorkin:1996sr} and Marolf \cite{Marolf:2003bb} that
quantum fluctuations give the black hole horizon a finite effective
width.\\
}

\textbf{Acknowledgements.}
SGAB thanks the audiences of the Qubits and Spacetime Unit group meeting at the Okinawa Institute of Science and Technology and of the Oxford Thursday seminar in philosophy of physics for their insightful questions. SGAB also acknowledges the Julian Schwinger Foundation for financial support during the 2026 Peyresq Spacetime Meeting, where  the final work on this article was carried out. MRV thanks the audiences and organizers of the Gravitational Thermodynamics and Beyond meeting at the Dublin Institute for Advanced Studies, TedFest 2026 at the University of Maryland, and the physics colloquium at Bilkent University, where this work was presented. This project was supported by the Spinoza Grant of the Dutch Research Council (NWO), awarded to Klaas Landsman.
We developed the main ideas in this paper ourselves, but utilized ChatGPT (OpenAI) for routine tasks throughout this work, especially for doing calculations, writing Mathematica code and rewriting text. We remain fully responsible for the entire content of the paper.

\begin{table}[H]
\centering
\small
\setlength{\tabcolsep}{2pt}
\renewcommand{\arraystretch}{1.25}
\begin{tabular}{|l|l|l|l|}
\hline
\textbf{Representation} & \textbf{Response function} & \textbf{Formula} & \textbf{Sign/behavior} \\
\hline

\multirow{3}{*}{Energy \(E(S,V)\)}
&
Isochoric thermal stiffness
&
\(K_T^{(E)}=\left(\frac{\partial T}{\partial S}\right)_V\)
&
\begin{tabular}[c]{@{}l@{}}
\(<0\) small BH,\\
zero at merger,\\
\(>0\) large BH.
\end{tabular}
\\
\cline{2-4}

&
Thermo-mechanical coupling
&
\(\chi_{SV}^{(E)}=\left(\frac{\partial T}{\partial V}\right)_S\)
&
\begin{tabular}[c]{@{}l@{}}
\(<0\); diverges as horizon\\
approaches cavity wall.
\end{tabular}
\\
\cline{2-4}

&
Adiabatic bulk modulus
&
\(B_S=-V\left(\frac{\partial P}{\partial V}\right)_S\)
&
\(>0\)
\\
\hline

\multirow{3}{*}{Helmholtz \(F(T,V)\)}
&
Isochoric heat capacity
&
\(C_V=T\left(\frac{\partial S}{\partial T}\right)_V\)
&
\begin{tabular}[c]{@{}l@{}}
\(<0\) small BH,\\
\(>0\) large BH,\\
diverges at merger.
\end{tabular}
\\
\cline{2-4}

&
Thermal pressure coefficient
&
\(\gamma_V=\left(\frac{\partial P}{\partial T}\right)_V\)
&
\begin{tabular}[c]{@{}l@{}}
\(<0\) small BH,\\
\(>0\) large BH,\\
diverges at merger.
\end{tabular}
\\
\cline{2-4}

&
Isothermal bulk modulus
&
\(B_T=-V\left(\frac{\partial P}{\partial V}\right)_T\)
&
\begin{tabular}[c]{@{}l@{}}
\(>0\) small BH,\\
\(<0\) large BH,\\
diverges at merger.
\end{tabular}
\\
\hline

\multirow{3}{*}{Gibbs \(G(T,P)\)}
&
Isobaric heat capacity
&
\(C_P=T\left(\frac{\partial S}{\partial T}\right)_P\)
&
\(<0\)
\\
\cline{2-4}

&
Thermal expansion coefficient
&
\(\alpha=\frac{1}{V}\left(\frac{\partial V}{\partial T}\right)_P\)
&
\(<0\)
\\
\cline{2-4}

&
Isothermal compressibility
&
\(\kappa_T=-\frac{1}{V}\left(\frac{\partial V}{\partial P}\right)_T\)
&
\begin{tabular}[c]{@{}l@{}}
\(>0\) small BH,\\
\(<0\) large BH,\\
zero at merger.
\end{tabular}
\\
\hline

\multirow{3}{*}{Enthalpy \(H(S,P)\)}
&
Isobaric thermal stiffness
&
\(K_T^{(H)}=\left(\frac{\partial T}{\partial S}\right)_P\)
&
\(<0\)
\\
\cline{2-4}

&
Temperature-pressure coefficient
&
\(\beta_S=\left(\frac{\partial T}{\partial P}\right)_S\)
&
\(>0\)
\\
\cline{2-4}

&
Adiabatic compressibility
&
\(\kappa_S=-\frac{1}{V}\left(\frac{\partial V}{\partial P}\right)_S\)
&
\(>0\)
\\
\hline

\multirow{3}{*}{Derived}
&
Gr\"uneisen parameter
&
\(\Gamma=V\left(\frac{\partial P}{\partial E}\right)_V\)
&
\begin{tabular}[c]{@{}l@{}}
\(>0\); diverges as horizon\\
approaches cavity wall.
\end{tabular}
\\
\cline{2-4}

&
Joule-Thomson coefficient
&
\(\mu_{\rm JT}=\left(\frac{\partial T}{\partial P}\right)_H\)
&
\begin{tabular}[c]{@{}l@{}}
\(>0\); no finite\\
inversion curve.
\end{tabular}
\\
\cline{2-4}

&
Speed of sound
&
\(c_s=\sqrt{\left(\frac{\partial P}{\partial\varepsilon}\right)_S}\)
&
\begin{tabular}[c]{@{}l@{}}
\(>0\); diverges as horizon\\
approaches cavity wall.
\end{tabular}
\\
\hline

\end{tabular}
\caption{Summary of twelve response functions associated with independent
Hessian components in the four thermodynamic representations, together
with three derived response functions, for the quasi-local Schwarzschild
black hole. The merger denotes the locus where the small and large black
hole branches in the Helmholtz representation coincide, corresponding to
the minimum temperature.}
\label{tab:schwarzschild-response-summary}
\end{table}

\appendix

\newpage

 \section{Standard thermodynamic representations}
\label{appA}

Here we summarize several standard thermodynamic representations, their
equations of state, and some physically relevant    response functions. We denote each representation by its corresponding potential. \\

\noindent $(E)$ In the \emph{energy representation} the fundamental relation and its differential are
\begin{equation}
E=E(S,V,N)\,,
\qquad
dE=TdS-PdV+\mu dN\,,
\end{equation}
and the equations of state are
\begin{equation}
T
=
\left(\frac{\partial E}{\partial S}\right)_{V,N},
\qquad
P
=
-
\left(\frac{\partial E}{\partial V}\right)_{S,N},
\qquad
\mu
=
\left(\frac{\partial E}{\partial N}\right)_{S,V}\,.
\end{equation}
The natural response coefficients in this representation are the thermal,
mechanical, and chemical stiffness coefficients
\begin{equation}
K_T^{(E)}
=
\left(\frac{\partial T}{\partial S}\right)_{V,N},
\qquad
K_P^{(E)}
=
-
\left(\frac{\partial P}{\partial V}\right)_{S,N},
\qquad
K_\mu^{(E)}
=
\left(\frac{\partial \mu}{\partial N}\right)_{S,V},
\end{equation}
together with mixed thermo-mechanical and thermo-chemical response
coefficients such as
\begin{equation}
\left(\frac{\partial T}{\partial V}\right)_{S,N},
\qquad
\left(\frac{\partial T}{\partial N}\right)_{S,V},
\qquad
\left(\frac{\partial \mu}{\partial V}\right)_{S,N},
\end{equation}
\textcolor{black}{the first of which we have denoted $\chi_{SV}^{(E)}$ in the main text.} The mechanical stiffness coefficient determines the adiabatic bulk modulus,
\begin{equation}
B_S = V K_P^{(E)}
=
-
V
\left(\frac{\partial P}{\partial V}\right)_{S,N}.\\
\end{equation}

\noindent $(S)$ In the \emph{entropy representation},
\begin{equation}
S=S(E,V,N)\,,
\qquad
dS=\beta dE+\beta PdV-\beta\mu dN\,,
\end{equation}
where $\beta=1/T$, with equations of state
\begin{equation}
\beta
=
\left(\frac{\partial S}{\partial E}\right)_{V,N},
\qquad
\beta P
=
\left(\frac{\partial S}{\partial V}\right)_{E,N},
\qquad
-\beta\mu
=
\left(\frac{\partial S}{\partial N}\right)_{E,V}.\\
\end{equation}

\noindent $(F)$ In the \emph{Helmholtz} or \emph{canonical representation},
\begin{equation}
F=E-TS\,,
\qquad
F=F(T,V,N)\,,
\qquad
dF=-S dT-P dV+\mu dN\,,
\end{equation}
with equations of state
\begin{equation}
S
=
-
\left(\frac{\partial F}{\partial T}\right)_{V,N},
\qquad
P
=
-
\left(\frac{\partial F}{\partial V}\right)_{T,N},
\qquad
\mu
=
\left(\frac{\partial F}{\partial N}\right)_{T,V}.
\end{equation}
This representation is naturally adapted to isothermal  and isochoric processes. The corresponding response functions include the constant-volume heat
capacity,
\begin{equation}
C_{V}
=
T\left(\frac{\partial S}{\partial T}\right)_{V,N},
\end{equation}
\textcolor{black}{the thermal pressure coefficient}
\begin{equation} \label{thermalpressurecoef}
    \gamma_V=\left(\frac{\partial P}{\partial T}\right)_{V,N}
\end{equation}
and the isothermal bulk modulus
\begin{equation}
B_T
=
-
V
\left(\frac{\partial P}{\partial V}\right)_{T,N}.
\end{equation}
The mixed response coefficients satisfy Maxwell relations such as
\begin{equation}
\left(\frac{\partial S}{\partial V}\right)_{T,N}
=
\left(\frac{\partial P}{\partial T}\right)_{V,N}.
\end{equation}

\noindent $(G)$ In the \emph{Gibbs representation},
\begin{equation}
G=E-TS+PV\,,
\qquad
G=G(T,P,N)\,,
\qquad
dG=-S dT+V dP+\mu dN\,,
\end{equation}
with equations of state
\begin{equation}
S
=
-
\left(\frac{\partial G}{\partial T}\right)_{P,N},
\qquad
V
=
\left(\frac{\partial G}{\partial P}\right)_{T,N},
\qquad
\mu
=
\left(\frac{\partial G}{\partial N}\right)_{T,P}.
\end{equation}
This representation is naturally adapted to systems coupled to thermal and
mechanical reservoirs at fixed temperature and pressure. The corresponding
response functions include the constant-pressure heat capacity,
\begin{equation}
C_{P}
=
T\left(\frac{\partial S}{\partial T}\right)_{P,N},
\end{equation}
the thermal expansion coefficient,
\begin{equation}
\alpha
=
\frac1V
\left(\frac{\partial V}{\partial T}\right)_{P,N},
\end{equation}
the isothermal compressibility,
\begin{equation}
\kappa_T
=
-
\frac1V
\left(\frac{\partial V}{\partial P}\right)_{T,N},
\end{equation}
and mixed thermo-chemical response coefficients such as
\begin{equation}
\left(\frac{\partial \mu}{\partial T}\right)_{P,N},
\qquad
\left(\frac{\partial \mu}{\partial P}\right)_{T,N}.
\end{equation}

\noindent $(H)$ In the \emph{enthalpy representation},
\begin{equation}
H=E+PV\,,
\qquad
H=H(S,P,N)\,,
\qquad
dH=T dS+V dP+\mu dN\,,
\end{equation}
the equations of state are
\begin{equation}
T
=
\left(\frac{\partial H}{\partial S}\right)_{P,N},
\qquad
V
=
\left(\frac{\partial H}{\partial P}\right)_{S,N},
\qquad
\mu
=
\left(\frac{\partial H}{\partial N}\right)_{S,P}.
\end{equation}
This representation is naturally adapted to adiabatic and isobaric processes. The associated
response coefficients include
the adiabatic compressibility
\begin{equation}
\kappa_S
=
-
\frac1V
\left(\frac{\partial V}{\partial P}\right)_{S,N},
\end{equation}
the thermal 
  and chemical stiffness coefficients
\begin{equation}
K_T^{(H)}
=
\left(\frac{\partial T}{\partial S}\right)_{P,N},
\qquad
K_\mu^{(H)}
=
\left(\frac{\partial \mu}{\partial N}\right)_{S,P},
\end{equation}
and mixed thermo-mechanical and thermo-chemical response coefficients such as
\begin{equation}
\left(\frac{\partial T}{\partial P}\right)_{S,N},
\qquad
\left(\frac{\partial V}{\partial S}\right)_{P,N},
\qquad
\left(\frac{\partial T}{\partial N}\right)_{S,P},
\qquad
\left(\frac{\partial V}{\partial N}\right)_{S,P},
\end{equation}
\textcolor{black}{the first of which is called the isentropic temperature-pressure coefficient (also called adiabatic lapse rate) and was denoted $\beta_S$ in the main text.}\\

\noindent $(\Omega)$ Further, exchanging the particle number for the chemical potential yields the
\emph{grand canonical representation},
\begin{equation}
\Omega=E-TS-\mu N\,,
\qquad
\Omega=\Omega(T,V,\mu)\,,
\qquad
d\Omega=-S dT-P dV-N d\mu\,,
\end{equation}
with equations of state
\begin{equation}
S
=
-
\left(\frac{\partial \Omega}{\partial T}\right)_{V,\mu},
\qquad
P
=
-
\left(\frac{\partial \Omega}{\partial V}\right)_{T,\mu},
\qquad
N
=
-
\left(\frac{\partial \Omega}{\partial \mu}\right)_{T,V}.
\end{equation}
This representation naturally describes open systems in contact with particle
reservoirs. The corresponding response functions include the particle-number
susceptibility
\begin{equation}
\chi_N
=
\left(\frac{\partial N}{\partial \mu}\right)_{T,V},
\end{equation}
together with mixed thermo-chemical response coefficients
\begin{equation}
\left(\frac{\partial N}{\partial T}\right)_{V,\mu},
\qquad
\left(\frac{\partial S}{\partial \mu}\right)_{T,V}.
\end{equation}

Moreover, some important response functions are derived constrained response
coefficients constructed from the thermodynamic variables and   the fundamental
response functions introduced above. An   example is the
\emph{Gr\"uneisen parameter},
\begin{equation}
\Gamma
=
V
\left(
\frac{\partial P}{\partial E}
\right)_{V,N}
=
\frac{\alpha V}{C_{V}\kappa_T}\,,
\end{equation}
which characterizes thermo-mechanical coupling by measuring how the pressure
changes under variations of the internal energy at fixed volume. Another
important example is the \emph{Joule-Thomson coefficient},
\begin{equation}
\mu_{\rm JT}
=
\left(\frac{\partial T}{\partial P}\right)_{H,N} =
\frac{V}{C_{P}}
\left(
T\alpha-1
\right)\,,
\end{equation}
which characterizes cooling or heating during isenthalpic expansion. Additionally, there is the \emph{adiabatic speed of sound}
\begin{equation}
c_s^2=\left(\frac{\partial P}{\partial \varepsilon}\right)_{S,N}\,,
\qquad\text{with}\qquad
\varepsilon=\frac{E}{V}\,,
\end{equation}
which quantifies the pressure response to a change in energy density under
adiabatic variations.

Finally, the response functions in the different representations are not independent but satisfy the standard
thermodynamic identities \cite{Landau:1980mil,callen1985thermodynamics} 
\begin{align}
\kappa_T = \frac{1}{B_T}\,,& \quad \kappa_S = \frac{1}{B_S}\,,\\ 
C_V = \frac{T}{K_T^{(E)}}\,,& \quad C_P = \frac{T}{K_T^{(H)}}, \\
C_{P}-C_{V}
&=
\frac{TV\alpha^2}{\kappa_T}\,,
\\
\kappa_T-\kappa_S
&=
\frac{TV\alpha^2}{C_{P}}\,,
\\
\gamma=\frac{C_{P}}{C_{V}}
&=
\frac{\kappa_T}{\kappa_S} = \frac{B_S}{B_T}\,,
\\
\Gamma&=-\frac{V}{T}\chi_{SV}^{(E)},
\\
c_s^2
&=
\frac{B_S}{\varepsilon+P}\,.
\end{align}
\textcolor{black}{Here, $\gamma$ should not be confused with $\gamma_V$ in Eq.~\eqref{thermalpressurecoef}.} Thus the bulk moduli are the inverse of the compressibilities, and the stiffness coefficients are the inverse (up to a factor of $T$) of the heat capacities.  For thermodynamically stable systems one   has
\begin{equation}
C_{P}\ge C_{V}\,,
\qquad
\kappa_T\ge\kappa_S\,,
\qquad
B_S\ge B_T\,.
\end{equation}
These inequalities follow directly from the thermodynamic identities above
together with the positivity conditions
\textcolor{black}{$C_{V}>0$}
 and 
$\kappa_T>0$ (together with $T>0$ and $V>0$).

\section{The contact geometry of equilibrium thermodynamics}
\label{appendixContactGeom}

In the contact-geometric axiomatization of thermodynamics
\cite{hermann1973geometry,MRUGALAgeometricformulation,1997haslachnonequil,bravettiContactGeometryThermodynamics2019},
one considers a $(2n+1)$-dimensional smooth manifold $M$ together with a contact
1-form $\alpha\in\Omega^1(M)$ satisfying \begin{equation}\alpha\wedge(d\alpha)^n\neq 0\,.\end{equation}
Here $n$ is the same number as in Section \ref{sec:eqnofstateinTD}, i.e. the number of
independent thermodynamic variables. The kernels
$\mathcal{D}_p=\ker(\alpha_p)$ define a distribution
$\mathcal{D}=\bigsqcup_{p\in M}\mathcal{D}_p\subset TM$, called the
\textit{contact structure}. The condition $\alpha\wedge(d\alpha)^n\neq 0$ is known as
\textit{maximal non-integrability}: it states that the hyperplane distribution
$\mathcal D=\ker(\alpha)$ is as far as possible from being integrable. Recall that a
distribution $\mathcal D$ is integrable if through every point there passes a
submanifold $N\subset M$ whose tangent spaces coincide with the distribution,
$T_qN=\mathcal D_q$ for all $q\in N$. For a codimension-one distribution
$\mathcal D=\ker(\alpha)$, the Frobenius theorem states that integrability is
equivalent to $\alpha\wedge d\alpha=0$. Thus non-integrability is satisfied already if
$\alpha\wedge d\alpha\neq 0$, but the contact condition additionally implies that
$d\alpha$ restricts to a non-degenerate 2-form on each~$\mathcal{D}_p$, \textcolor{black}{and is in this sense as far away from integrability as possible.}

%\textcolor{black}{By Darboux's theorem, this locally allows the contact form to be
%written in terms of $n$ pairs of conjugate variables.}

\textcolor{black}{By Darboux's theorem, around every point there exist local coordinates
$(x^i,y_i,z)$, with $i=1,\ldots,n$, in which the contact form takes the
canonical form $\alpha=dz-\sum_i y_i dx^i$. In the thermodynamic
interpretation, the pairs $(x^i,y_i)$ are identified with conjugate
variables.
The simplest example is $dz-y dx$ on $\mathbb{R}^3$, visualized as a
non-integrable distribution of planes. In thermodynamics one may, for
example, identify $z=E$, $x=S$, and $y=T$, in which case the contact
1-form becomes $dE-TdS$.
Its kernel defines the contact distribution, from which the
differential fundamental relation $dE=TdS$ is recovered upon restriction
to the equilibrium thermodynamic state space.}

\indent \textcolor{black}{Now, the contact manifold $(M,\alpha)$ is
$(2n+1)$-dimensional, whereas the equilibrium thermodynamic state space is
$n$-dimensional.} Thus one restricts to a \textit{Legendre submanifold}, which
is an integral submanifold of maximal dimension of the contact distribution
$\mathcal{D}\subset TM$. It can straightforwardly be derived that such a
Legendre submanifold is $n$-dimensional and is therefore an appropriate
candidate for the thermodynamic state space. When the contact 1-form $\alpha$
is pulled back to a Legendre submanifold, it automatically vanishes, ensuring
that the differential fundamental relation
\eqref{eq:firstlaw-general-representation} holds everywhere on the
thermodynamic state space.

\textcolor{black}{A standard example of a contact manifold is the \emph{1-jet bundle}
\cite{bravettiContactGeometryThermodynamics2019}:}
\\
\\
\textbf{Example: the 1-jet bundle.} For any smooth manifold $Q$, let
$J^1 Q=J^1(Q,\mathbb{R})$ be \textcolor{black}{the space of 1-jets of real-valued
functions on $Q$}, with $j^1 f(x)=\left(x,f(x),\mathrm{d}_x f\right)$ for
$f\colon Q\to\mathbb{R}$, where $\mathrm{d}_x f$ stands for the differential of
$f$ at $x\in Q$. Then $J^1 Q$ is isomorphic to $T^*Q\times\mathbb{R}$ and it has
a canonical contact structure defined by
$\mathcal{D}=\operatorname{ker}\left(\mathrm{d}z-y_i dx^i\right)$, where $x^i$
are coordinates on $Q$, and $y_i$ are coordinates on the fiber $T_x^*Q$.\\

\textcolor{black}{
This example turns out to be useful in equilibrium thermodynamics because
it provides a natural geometric setting for describing thermodynamic
potentials and their equations of state.
To see more explicitly how equations of state enter the formalism, the
following result is useful}  \cite[Theorem 2.2]{bravettiContactGeometryThermodynamics2019}:
\\
\\
\textbf{Theorem.} Consider a disjoint partition $I \cup J$ of the set of indices $\{1, \ldots, n\}$ and a function of $n$ variables $f\left(y_i, x^j\right)$, with $i \in I$ and $j \in J$. The $n+1$ equations

\begin{equation}
x^i=-\frac{\partial f}{\partial y_i}\,, \quad y_j=\frac{\partial f}{\partial x^j}\,, \quad z=f-y_i \frac{\partial f}{\partial y_i}\,,
\end{equation}
define a Legendre submanifold $\mathcal{L}$ of $(M,\alpha)$. Conversely, any Legendre submanifold is locally defined by these equations for at least one of the $2^n$ possible choices of the partition of the set $\{1, \ldots, n\}$.
\\
\\
\indent The thermodynamic interpretation of this theorem is that, locally, there will always exist a fundamental relation (here $z=f-y_i\partial f/\partial y_i$) and equations of state (here $x^i=-\partial f/\partial y_i$ and $y_j=\partial f/\partial x^j$). Thus the contact geometric interpretation of equations of state is that they are the relations parametrizing the thermodynamic Legendre submanifold.

% Yet the above theorem only establishes the existence of local equations of state. In the main body of this article we assumed that equations of state are globally defined. To understand this better, consider the following crucial example.
\textcolor{black}{The theorem above is local: a chosen set of thermodynamic variables need not
provide a global parametrization of the equilibrium Legendre submanifold. The prime example of a globally parametrized Legendre submanifold is the following.}
\\
\\
\textbf{Example.} Given the 1-jet bundle $J^1 Q$ with the contact structure defined by the 1-form $\alpha=dz-y_i dx^i$, for every function $f\colon Q\to\mathbb{R}$, the \emph{1-graph} of $f$ defines the following Legendre submanifold 
\begin{equation}\label{eq:legendrianasgraph}
\mathcal{L}_f:=\left\{\left(z, x^i, y_i\right) \in J^1 Q \mid z=f\left(x^i\right), x^i=x^i, y_i=\frac{\partial f}{\partial x^i}\right\} .
\end{equation}
% This is the prototypical example of the Legendre submanifolds that we encounter in thermodynamics, and it is indeed the type of Legendre submanifold that we considered in this paper, where $f$ is the thermodynamic potential and the \textit{globally defined} equations of state are $y^i=\partial f/\partial x^i$.\\
\textcolor{black}{
This is the prototypical graph description encountered in thermodynamics,
with $f$ playing the role of the thermodynamic potential and
$y_i=\partial f/\partial x^i$ giving the corresponding equations of state \cite{1997haslachnonequil}.
A given thermodynamic representation need not, however, provide a single
such graph globally. If one performs a coordinate transformation that is not globally one-to-one from the coordinates in \eqref{eq:legendrianasgraph} to differently chosen independent
variables, then several branches are required. If
the transformation loses local invertibility, the resulting 1-graph description itself
becomes singular. The Helmholtz fold discussed in Sec.~\ref{sec:helmholtzresponse} provides an
explicit example.}\\

In the contact-geometric discussion above, $f$ denotes a generic generating function. We now switch to the notation of Section \ref{sec:eqnofstateinTD}, where the thermodynamic potential is denoted by~$\Phi$. We can make precise how Legendre transformations of thermodynamic representation arise in the contact-geometric description. 

Let \(X^A\) denote the chosen independent variables in a representation and write the conjugate variables with the signs absorbed as
\begin{equation}
    Y_A:=\frac{\partial \Phi}{\partial X^A}\,.
\end{equation}
Thus, if \(X^A=x^i\), then \(Y_A=y_i\), while if \(X^A=y^j\), then \(Y_A=-x_j\). In these variables the contact 1-form \textcolor{black}{written in representation-adapted Darboux coordinates} is
\begin{equation}
    \alpha_\Phi=d\Phi-Y_A dX^A\,,
\end{equation}
% and the equilibrium Legendre submanifold is the graph
\textcolor{black}{and, wherever the transformation into the variables $X^A$ is regular, the
equilibrium Legendre submanifold is locally represented as the graph}
\begin{equation}\label{eq:legendresubmfd}
    \mathcal{L}_\Phi
    =
    \left\{(\Phi,X^A,Y_A)\mid
    \Phi=\Phi(X^A),\ Y_A=\frac{\partial \Phi}{\partial X^A}
    \right\}.
\end{equation}
This is the same 1-graph construction as above, only written in the notation of general thermodynamic representations. \textcolor{black}{The subscript $\Phi$ labels the choice of generating function, not a
physically different thermodynamic system. Different thermodynamic potentials describe the same underlying equilibrium relation in different contact coordinates, whenever the required graph projection is regular.}

% This is the same 1-graph construction as above, written in the notation of
% general thermodynamic representations. The subscript $\Phi$ labels the
% generating-function presentation, not a physically different
% thermodynamic system. Different thermodynamic potentials therefore provide
% different generating-function descriptions of the same underlying
% equilibrium Legendre submanifold wherever the corresponding graph
% projection is regular.

Following \cite[Definition 2.9]{bravettiContactGeometryThermodynamics2019}, a Legendre transformation is a \textcolor{black}{strict} contact transformation (i.e. a diffeomorphism whose pullback preserves the contact 1-form) which exchanges some coordinates with their conjugates. Consider a disjoint partition \(I\cup J\) of the set of indices \(\{1,\ldots,n\}\), where the variables with \(i\in I\) are transformed and the variables with \(j\in J\) are left unchanged. With the sign convention used here this transformation is
\begin{equation}
    \widetilde{\Phi}
    =
    \Phi-Y_i X^i\,.
\end{equation}
The corresponding conjugate variables are determined by the differential of the transformed potential:
\begin{equation}
    d\widetilde{\Phi}
    =
    -X_i dY^i+Y_j dX^j \,,  \qquad i\in I,\quad j\in J \,.
\end{equation}
Hence the equations of state in the transformed representation are
\begin{equation}
    \frac{\partial \widetilde{\Phi}}{\partial Y^i}
    =
    -X_i\,,
    \qquad
    \frac{\partial \widetilde{\Phi}}{\partial X^j}
    =
    Y_j \,.
\end{equation}
Returning to the original \((y_a,x^a)\) notation, this is precisely the sign pattern in Eq.~\eqref{eq:yasdifferentialofx} and \eqref{eq:xasdifferentialofy}: after a Legendre transformation, the variable which was previously independent becomes an equation of state in the new representation.

% Since Legendre transforms are contact transformations, they map Legendre submanifolds to Legendre submanifolds \cite{bravettiContactGeometryThermodynamics2019}. \textcolor{black}{Regarded passively as a change of thermodynamic coordinates, they provide a different generating-function presentation of the same underlying equilibrium relation, but do not create a physically different equilibrium system.} In coordinates, this says that the image of \(\mathcal{L}_\Phi\) is again described by a fundamental relation and its equations of state, namely \(\mathcal{L}_{\widetilde{\Phi}}\). The only extra assumption needed for this to be a single-valued change of representation is that the expressions for the new variables in terms of the old variables can be inverted. Locally, this is equivalent to the corresponding Jacobian, or
% equivalently the relevant Hessian block of $\Phi$, being non-degenerate. \textcolor{black}{When every conjugate pair is exchanged, the Legendre transformation is called total. Otherwise it is partial. The passage from \(E(S,V)\) to \(F(T,V)\) is thus a partial Legendre transformation, whose local graph description requires the relation \(T=T(S,V)\) to be invertible with respect to \(S\) at fixed \(V\). A total Legendre transform of a convex or concave potential is automatically a diffeomorphism of Legendre submanifolds \cite{bravettiContactGeometryThermodynamics2019}, but this need not hold for partial Legendre transforms or potentials which are not convex/concave.} 

\textcolor{black}{
Since Legendre transformations are contact diffeomorphisms, they map
Legendre submanifolds to Legendre submanifolds \cite{bravettiContactGeometryThermodynamics2019}. Regarded passively as changes of thermodynamic coordinates, Legendre
transformations therefore do not change the underlying equilibrium system,
but provide different generating function descriptions of the same
equilibrium Legendre submanifold. The contact transformation itself remains regular. What may
fail is the representation of the equilibrium Legendre submanifold as a
graph over the transformed independent variables.}

\textcolor{black}{For the partition $I\cup J$ above, such a local graph description requires
the map from $(X^i,X^j)$ to $(Y_i,X^j)$ to be locally invertible.
Equivalently, the corresponding Hessian block
$\partial^2\Phi/\partial X^i\partial X^k$, with $i,k\in I$, must be
non-degenerate. When every conjugate pair is exchanged, the Legendre
transformation is called total; otherwise it is partial. The passage from
$E(S,V)$ to $F(T,V)$ is a partial Legendre transformation, whose local
graph description requires
$\mathcal{H}_{SS}^{(E)}=(\partial T/\partial S)_V\neq0$. For the Schwarzschild cavity system this quantity vanishes at $x=2/3$,
so the Helmholtz projection becomes singular there. As shown explicitly
in Sec.~\ref{sec:helmholtzresponse}, \textcolor{black}{this singularity is a fold,} even though the
underlying equilibrium Legendre submanifold remains regular.}
\\
\\
Response functions and fluctuation metrics have a natural contact-geometric interpretation. On the Legendre submanifold $\mathcal{L}_\Phi$, the equations of state are the functions
\begin{equation}
    Y_A=Y_A(X^B)=\frac{\partial\Phi}{\partial X^A}\,.
\end{equation}
Differentiating these equations of state gives
\begin{equation}
    \textcolor{black}{\mathcal{H}}^{(\Phi)}_{AB}
    :=
    \frac{\partial Y_A}{\partial X^B}
    =
    \frac{\partial^2\Phi}{\partial X^A\partial X^B}\,.
\end{equation}
Hence the fundamental response coefficients are precisely the components of the Hessian of the thermodynamic potential, i.e. the Hessian matrix introduced in Section \ref{sec:eqnofstateinTD}. 
\textcolor{black}{Thus, the Hessian data play two related roles: their components determine
the response coefficients, while the non-degeneracy of the appropriate
Hessian block determines whether a change of thermodynamic representation
provides a regular local graph.}

% \textcolor{black}{These Hessians can also be interpreted as fluctuation metrics, although this requires additional metric structure that is not determined by the contact form alone. In the construction of \cite[Proposition V and Result II]{bravettiContactGeometryThermodynamics2019}, thermodynamic fluctuation theory equips the thermodynamic phase space with a particular pseudo-Riemannian
% metric which, together with the contact form, Reeb vector field, and
% associated almost para-contact structure, defines a \emph{para-Sasakian
% manifold}. The pullback of the phase space metric to an equilibrium Legendre submanifold yields the corresponding Hessian fluctuation metric. In the standard energy and entropy representations this reproduces the Weinhold and Ruppeiner metrics, respectively. A contact manifold does not become para-Sasakian merely by being supplied with an arbitrary compatible metric: additional normality or integrability conditions are required. We will not treat this further here.}

 \textcolor{black}{These Hessians can also be interpreted as fluctuation metrics, although this
requires additional metric structure that is not determined by the contact
form alone. In the construction of
\cite[Proposition~V and Result~II]{bravettiContactGeometryThermodynamics2019},
thermodynamic fluctuation theory equips the thermodynamic phase space with
a particular pseudo-Riemannian metric whose pullback to an equilibrium
Legendre submanifold yields the corresponding Hessian fluctuation metric.
In the standard energy and entropy representations this reproduces the
Weinhold and Ruppeiner metrics, respectively. We will not pursue this
additional geometric structure further here.}

In our earlier work \cite{Borsboom:2026ash}, we briefly used contact geometric language to describe black hole thermodynamics with a holographic pressure-volume pair.  For other applications of contact geometry in the context of black hole thermodynamics see  \cite{Quevedo:2007mj,Alvarez:2008wa,ghosh2019,ghoshContactMetricStructures2023}.

% \label{appc}

\bibliographystyle{JHEP}
\bibliography{bibliography}

\end{document}